\documentclass{article}
\usepackage{amsmath} 
\usepackage{PRIMEarxiv}
\usepackage{natbib}
\usepackage[utf8]{inputenc} 
\usepackage[T1]{fontenc}    
\usepackage{hyperref}       
\usepackage{url}            
\usepackage{booktabs}       
\usepackage{amsfonts}       
\usepackage{nicefrac}       
\usepackage{microtype}      
\usepackage{lipsum}
\usepackage{fancyhdr}       
\usepackage{graphicx}       
\graphicspath{{media/}}     
\title{A physics-assisted deep neural network-based closure framework for velocity gradient dynamics in compressible flows with vibrational non-equilibrium}

\author{
  Deep Shikha, Sawan S Sinha*\\
  Department of Applied Mechanics \\
  Institute of Technology Delhi, New Delhi, India \\
  New Delhi\\
  \texttt{*Email address for correspondence:}sawan@am.iitd.ac.in}

\begin{document}
\maketitle

\begin{abstract}
In this study, we propose a dynamical model for the evolution of velocity gradients in compressible turbulent flows with vibrational non-equilibrium effects, using physics-assisted deep neural networks. Such models provide a powerful framework for understanding the nonlinear physics associated with small-scale structures. In compressible flows, the influence of thermodynamic fields on velocity-gradient dynamics is represented through the thermodynamic gradient field (TGF) tensor. The TGF tensor is one of the primary unclosed terms in the velocity-gradient evolution equations. The TGF tensor comprises contributions from the pressure-Hessian tensor, $\rho\boldsymbol{H}$, and the baroclinic tensor, $\boldsymbol{B}$. Accordingly, the proposed framework incorporates closures for both $\boldsymbol{H}$ and $\boldsymbol{B}$ tensor dynamics. Building upon existing phenomenological closures for the $\boldsymbol{H}$ tensor governing mechanisms, we develop a neural-network-based closure for the inviscid mechanism responsible for generating the $\boldsymbol{B}$ tensor. 
Unlike the other recently used tensor bases, the presented work employs a novel tensor basis allowing for the inclusion of non-symmetric features in the model. The framework also incorporates a data-driven closure for vibrational non-equilibrium effects. 
The resulting framework combines phenomenological and data-driven representations of various $\boldsymbol{H}$ and $\boldsymbol{B}$ tensors governing mechanisms, termed as the \textit{hybrid enhanced homogenized Euler equation} (H-EHEE) model. Model predictions are evaluated across a range of turbulent Mach numbers and compared against direct numerical simulation (DNS) data and existing compressible velocity-gradient models. The H-EHEE model exhibits close agreement with DNS statistics and provides significant improvements over existing models, particularly in highly compressible flow regimes.
\end{abstract}


\section{Introduction}
Understanding the physics of small-scale structures is essential for developing accurate and generalizable turbulence models \citep{ashurst1987alignment, ohkitani1993eigenvalue, pumir1994numerical, soria1994study}. The velocity gradient tensor plays a central role in shaping the dynamics of small-scale structures \citep{meneveau2011lagrangian, johnson2024multiscale}. The evolution of the velocity gradient tensor following a fluid particle can be studied through: (i) experiments, (ii) direct numerical simulations (DNS), and (iii) a closed set of ordinary differential equations (ODEs) describing velocity gradient dynamics. Performing experimental measurements of velocity gradient dynamics is extremely challenging. To the best of our knowledge, the experimental studies of measuring velocity gradient tensors following a fluid particle are very limited \citep{xu2011pirouette}. Using DNS to access the dynamics of the velocity gradient tensor requires implementing a Lagrangian particle tracker within the simulation, which makes the approach computationally very expensive \citep{yeung1989lagrangian, danish2014direct}.

Having a closed set of ordinary differential equations (ODEs) that describes the time evolution of the velocity gradient tensor following the motion of a fluid particle offers three main advantages. First, solving these ODEs is computationally cheaper than solving the full set of partial differential equations. Second, their solution directly provides the Lagrangian evolution of the velocity gradient tensor, which is physically more intuitive. Third, such an ODEs system can serve as closures in Lagrangian probability density function (PDF) methods. The Lagrangian PDF methods offer a closure approach for turbulence modeling in inhomogeneous flows \citep{pope1985pdf}.

In incompressible flows, the time evolution of the velocity gradient tensor contains two unclosed terms: the pressure Hessian tensor and the viscous Laplacian tensor. Given the central role of the velocity gradient tensor in turbulence dynamics, modeling these unclosed terms has been a major research focus. A key milestone in this line of research is the work of \cite{vieillefosse1982local}. In his study, the author has derived the velocity gradient evolution equation for incompressible flows. In his formulation, the anisotropic part of the pressure Hessian and the viscous effects are neglected. The resulting closed system is known as the \textit{restricted Euler equation} (REE) model. Despite these simplifications, the REE system can reproduce several important statistical features of turbulence. However, because of the imbalance between the self-stretching term and the isotropic pressure Hessian term, the REE system leads to a finite-time singularity. To address these shortcomings, several modifications have been proposed to incorporate the effects of the anisotropic part of the pressure Hessian tensor and the viscous Laplacian tensor~\citep{cantwell1992exact,jeong2003velocity,  martin1997inverse, chevillard2008modeling}. Among these, the \textit{recent fluid deformation} (RFD) model is regarded as one of the most successful phenomenological models for velocity gradient dynamics in incompressible turbulence.

Along a similar line of research, there is a growing need for velocity-gradient closure models in compressible turbulence. In compressible flows, the influence of pressure fields on the velocity-gradient dynamics is manifested through two unclosed contributions: the pressure-Hessian tensor, which is inherently symmetric in nature, and the tensor arising from the interaction between the density-gradient and pressure-gradient fields, referred to herein as the baroclinic tensor. The combined effect of these two contributions is represented by the \emph{thermodynamic gradient field} (TGF) tensor. In earlier studies, the TGF tensor itself was referred to as the \emph{pressure-Hessian tensor}.  Since the earlier terminology was somewhat misleading and may lead to some confusion about the tensor symmetry/non-symmetry,  the term TGF tensor is now adopted consistently in the present work. In an incompressible flow field, wherein the density variation is absent, the baroclinic tensor vanishes leading to the TGF tensor becoming the same as  symmetric pressure Hessian tensor- with density being a constant multiplicative factor.

Unlike incompressible flows, where pressure is governed by the pressure Poisson equation, pressure in compressible flows acts as a thermodynamic variable. Consequently, compressible flows require a separate evolution equation for the thermodynamic gradient field (TGF) tensor as well. The first velocity gradient dynamical model for compressible flows was proposed by \cite{suman2009homogenized}, referred to as the \textit{homogenized Euler equation} (HEE) model. This framework demonstrated reasonable agreement with several features of compressible turbulence dynamics; however, the model is unable to capture the influence of varying Mach numbers. To overcome this limitation, \cite{suman2011dynamical} in their next study derived an explicit evolution equation for the TGF tensor. The authors proposed a coupled system of modeled velocity gradient and TGF tensor evolution equations for compressible flows. 
At higher turbulent Mach numbers, the baroclinic tensor contributions becomes more significant. However, the model of \cite{suman2011dynamical} does not include the closures for such baroclinic effects in the TGF tensor evolution equation . To address this,~\cite{danish2014direct} proposed a velocity gradient model that incorporates the influence of the inviscid mechanisms responsible for the generation and evolution of the baroclinic tensor into the proposed TGF tensor evolution equation. Note that, the mechanisms responsible for the generation and evolution of the baroclinic tensor were originally referred to as the non-symmetric mechanisms by \citet{danish2014direct}. In the present work, these mechanisms are referred to as baroclinic-tensor-generating mechanisms to better reflect their physical role. The model proposed by \cite{danish2014direct} is referred to as the \textit{new enhanced homogenized Euler equation} (N-EHEE) model. The N-EHEE model demonstrated significant improvement over the existing models. Nonetheless, the model relied on the simplified assumption that pressure and density fluctuations remain small compared to their mean values. This assumption does not hold in high-turbulence Mach number regimes. These limitations highlight the need for more advanced models capable of accurately capturing the dynamics of the baroclinic tensor contributions in compressible turbulence. 

At sufficiently high Mach numbers, turbulent flows start exhibiting various non- equilibrium effects. In particular, in high-temperature regions ($>800\,\mathrm{K}$), such as those encountered in hypersonic flows and across strong shock fronts, vibrational modes of air molecules become excited. Although these vibrationally excited molecules tend to relax toward equilibrium with the evolving translational–rotational temperature field, the associated relaxation timescale, referred to as the vibrational relaxation time, can be comparable to the characteristic timescale of the turbulent velocity field. This timescale interaction introduces additional complexity into the flow dynamics. Hence, modeling such flows becomes even more challenging. In such regimes, the vibrational energy must be treated as an independent thermodynamic variable and hence evolves separately from the internal energy associated with the temperature field.

Recently, \cite{shikha2024modeling} proposed a neural-network-based framework to model vibrational non-equilibrium effects in the TGF tensor evolution equation. Their model was evaluated at a single time instant (non-dynamical framework), with all input features extracted directly from DNS data. To the best of our knowledge, however, no dynamical model of velocity-gradient evolution (using neural networks or otherwise) currently exists that explicitly incorporates the effects of vibrational non-equilibrium. In this study, we aim to model vibrational non-equilibrium effects within a dynamical framework for the evolution of velocity gradients.

In recent years, data-driven approaches have gained significant attention in turbulence modeling, offering a promising scope towards more accurate and generalizable closure models \citep{brunton2019machine}. Within the context of velocity-gradient dynamics, the first neural-network-based model was introduced by \cite{parashar2020modeling}, who employed a tensor-basis neural network (TBNN) framework to model the pressure-Hessian tensor in incompressible turbulence. Subsequent studies have focused on improving neural network architectures and incorporating additional physics-based constraints to enhance the modeling of the pressure-Hessian tensor for incompressible flows \citep{tian2021physics, shikha2023augmented}. 

In contrast, machine-learning-based closures for compressible turbulence remain limited. Recently, \cite{shikha2024modeling} proposed a neural network framework to capture vibrational non-equilibrium effects in the TGF tensor evolution. In the models proposed by~\cite{shikha2023augmented,shikha2024modeling}, a combination of two distinct neural network architectures is employed to represent two different aspects of the output tensor: its self-normalized form (using the TBNN architecture), which exhibits fairly universal behavior across different flow fields, and its magnitude, which is comparatively more sensitive across varying flow conditions. Subsequently, \cite{wu2025two} have used a similar approach to model the subgrid stress tensor for forced homogeneous isotropic turbulent flows. In this study, we adopt a similar dual neural network-based architecture to develop closure for the inviscid mechanism which is contributing in the generation and the evolution of the baroclinic tensor. In general, such mechanisms are asymmetric in nature. A tensor-basis neural network (TBNN) designed to represent a second-order tensor requires an appropriate set of tensor bases and scalar invariants. To the best of our knowledge, no prior work has derived a complete set of tensor bases and invariants specifically for modeling a general second-order tensor that is not constrained to be symmetric. To this end, we aim to derive a complete set of tensor bases and scalar invariants suitable for representing an arbitrary second-order tensors without symmetry constraints.

Furthermore, all existing neural-network-based models for velocity-gradient closures (both incompressible and compressible) have been assessed solely through \textit{a priori} evaluations, wherein the input quantities are extracted directly from DNS databases at a given time instant and the models are evaluated in a non-dynamical setting. As a result, these studies do not provide a dynamical closure of velocity-gradient evolution, in which both the velocity-gradient tensor and the neural-network inputs evolve simultaneously through a coupled system of ODEs. To the best of our knowledge, no existing study has developed a dynamical velocity-gradient framework in which neural-network-predicted closures are embedded directly into the governing evolution equations and and integrated over time. The present work addresses this gap by constructing a fully dynamical velocity-gradient framework that incorporates neural-network predictions within a closed system of ODEs, thereby providing access to the continuous evolution of the velocity-gradient tensor along fluid-particle trajectories.
Modeling the velocity-gradient dynamics of compressible flows with vibrational non-equilibrium effects using a single neural-network architecture can be particularly challenging because multiple physical mechanisms contribute simultaneously to the flow evolution. Therefore, the present framework employs separate neural-network closures for distinct physical mechanisms governing the velocity-gradient and TGF tensor dynamics. In particular, vibrational non-equilibrium effects are represented through a dedicated neural-network closure, while the mechanism contributing in the baroclinic tensor evolution is modeled independently. This modular strategy reduces the complexity of the learning task, allows each network to focus on a specific physical mechanism, and improves the interpretability and robustness of the resulting dynamical framework.
Motivated by the identified research gaps and the associated challenges, the present study pursues the following novel objectives:
\begin{itemize}
    \item To develop an apt tensor-basis, and invariants-based neural-network closure for the inviscid mechanism contributing to the baroclinic tensor generation and evolution, thereby overcoming the limitations of the existing phenomenological model proposed by \cite{danish2014direct}.

    \item To incorporate recently developed neural-network-based closures for the inviscid baroclinic tensor generating mechanism and vibrational non-equilibrium mechanisms into a dynamical velocity-gradient framework.

    \item To formulate a novel \textit{hybrid} dynamical velocity-gradient model (a closed set of ODEs) for compressible flows, in which unclosed terms are represented using a combination of phenomenological modeling and neural-network-based closures.

    \item To assess the performance of the proposed model across a wide range of flow regimes, spanning low- to high-turbulent-Mach-number conditions, including regions with active vibrational non-equilibrium effects by making comparisons with DNS data as well as with existing models of compressible velocity gradient dynamics.
\end{itemize}

This paper is organized into ten sections. Section~\ref{II} introduces the evolution equation of the velocity gradient and pressure Hessian tensors for compressible flows. Section~\ref{III}
gives a brief review of the existing model of velocity gradient dynamics in compressible flows. Sections~\ref{IV} to \ref{VII} discuss the enhancements and modeling strategies employed in the current study. Section \ref{VII} gives a summary of the newly proposed model of velocity gradient tensor in compressible flows. Section \ref{VIII} presents the non-dimensionalized form of the model. Sections~\ref{IX} provide an extensive performance assessment of the model across different flow conditions. Finally, Section~\ref{X} concludes the paper with a brief summary.
\section{\label{II}{Evolution equations of velocity gradient dynamics in compressible turbulence}}
For a calorically perfect compressible fluid, the Navier-Stokes equation set can be written as follows:

\begin{equation}
\label{eq:continuity}
\frac{\partial{\rho}}{\partial{t}} + {V_k} \frac{\partial{\rho}}{\partial{x_k}}= -\rho \frac{\partial{V_{k}}}{\partial{x_k}},
\end{equation}

\begin{equation}
\label{eq:momentum}
\frac{\partial{V_i}}{\partial{t}} + {V_k} \frac{\partial{V_i}}{\partial{x_k}}= -\frac{1}{\rho} \frac{\partial{p}}{\partial{x_i}} + \frac{1}{\rho}\frac{\partial{\sigma_{ik}}}{\partial{x_k}},
\end{equation}
where $V_i$, $t$, and $x_i$ denote the velocity, time, and spatial coordinates in the Eulerian frame of reference, respectively. The symbol $p$ represents pressure, $\rho$ represents density, and $\sigma_{ik}$ represents the viscous stress tensor. For a Newtonian fluid, the viscous stress is given by
\begin{equation}
    \label{eq:sigma}
    \sigma_{ij} = \rho\, \mu \left( A_{ij} + A_{ji} - \frac{2}{3} A_{kk} \delta_{ij} \right),
\end{equation}
where $\delta_{ij}$ is the Kronecker delta. The evolution of the velocity gradient tensor $(A_{ij}=\frac{\partial V_{j}}{\partial x_{i}})$ is derived by taking the material derivative ($\frac{D}{Dt}\equiv \frac{\partial}{\partial t} + u_{i} \frac{\partial}{\partial x_{i}}$), of the momentum equation (\ref{eq:momentum})~\citep{suman2011dynamical}:
\begin{align}
\label{eq:velGrad}
\frac{D A_{ij}}{D t} 
= - A_{ik}A_{kj}
-\underbrace{\frac{\partial}{\partial x_{j}} \left( \frac{1}{\rho}\frac{\partial p}{\partial x_{i}} \right)}_{P_{ij}}
+ \underbrace{\mu\frac{\partial ^{2} A_{ij}}{\partial x_{k} \partial x_{k}} + \frac{\mu}{3}\frac{\partial ^{2} A_{kk}}{\partial x_{i} \partial x_{j}}-\frac{\mu}{\rho}\frac{\partial \rho}{\partial x_{j}} \left ( \frac{\partial A_{ik}}{\partial x_{k}} + \frac{1}{3}\frac{\partial A_{kk}}{\partial x_{i}}\right)}_{\Gamma_{ij}},
\end{align}
where $\mu$ denotes the kinematic viscosity. The left-hand side (LHS) of (\ref{eq:velGrad}) represents the Lagrangian evolution of the velocity gradient tensor, i.e., its rate of change following a fluid particle. The right-hand side (RHS) contains three terms that govern this evolution:  
The first term, $-A_{ik}A_{kj}$, is the \textit{self-stretching term}. It depends solely on the velocity gradient tensor itself and is therefore mathematically closed. The second term, $P_{ij}$, denotes the \textit{thermodynamic gradient field tensor} (TGF tensor). In earlier studies, this term was referred to as the \textit{pressure-Hessian tensor}. However, since the tensor incorporates the effects of density and pressure gradients in addition to the pressure Hessian itself, the term \textit{thermodynamic gradient field tensor} is adopted in the present work to more accurately reflect its physical behavior. 
The third term, $\Gamma_{ij}$, is the \textit{viscous Laplacian tensor}. Unlike the self-stretching term, the TGF tensor ($P_{ij}$) and the viscous Laplacian tensors $\Gamma_{ij}$ are mathematically unclosed and inherently non-local. Consequently, closing (\ref{eq:velGrad}) requires either deriving separate evolution equations for these terms or modeling them explicitly. In compressible flows, the term $P_{ij}$ needs its own evolution equation, which can be obtained from the evolution of the internal energy, $e = C_{v} T$, where $C_{v}$ is the specific heat at constant volume and $T$ is the thermodynamic temperature. \cite{suman2011dynamical} derived the evolution equation for the $\boldsymbol{P}$ tensor in compressible flows without accounting for vibrational non-equilibrium effects. Later, \cite{srivastava2024influence} incorporated vibrational non-equilibrium effects into the pressure-Hessian evolution equation
\begin{multline}
    \label{eq:Pressure_Hessian}
    \frac{DP_{ij}}{Dt} = \underbrace{-A_{kj}P_{ik}-A_{ki}P_{kj}-(n-1) A_{kk}P_{ij}}_{(I_{ij})} \\
     \underbrace{-\frac{n}{\rho}\left (\frac{\partial A_{kk}}{\partial x_j}\ \frac{\partial p}{\partial x_i} + \frac{\partial A_{kk}}{\partial x_i}  \frac{\partial p}{\partial x_j}\right )}_{II_{ij}}
     \underbrace{-\frac{1}{\rho} \frac{\partial A_{ki}}{\partial x_{j}} \frac{\partial P} {\partial x_{k}}}_{III_{ij}}
     \underbrace{-\frac{np}{\rho}\frac{\partial^2A_{kk}}{\partial x_{i}\partial x_{j}}}_{IV_{ij}}\\
     \underbrace{+\frac{1}{\rho}\frac{\partial A_{kk}}{\partial x_j}\frac{\ \partial p}{\partial x_i} +\frac{np}{\rho^{2}}\frac{\partial A_{kk}}{\partial x_i}\frac{\ \partial\rho}{\partial x_j}}_{(V_{ij})}\\
      \underbrace{+\left(\frac{n-1}{R}\right)\ \frac{\partial}{\partial x_j}\left\{\frac{1}{\rho}\frac{\partial^2}{\partial x_i\partial x_k}\left[k\ \frac{\partial}{\partial x_k}\left(\frac{p}{\rho}\right)\right]\right\}}_{(VI_{ij})} \\
      +\underbrace{(n-1) \frac{\partial}{\partial x_j}\left[\frac{1}{\rho}\ \frac{\partial\sigma_{km}}{\partial x_i} A_{km}+\frac{1}{\rho}\sigma_{km}\ \frac{\partial A_{km}}{\partial x_i}\right]}_{(VII_{ij})} \\+  \underbrace{(n-1)\frac{\partial}{\partial x_j}\left[\frac{1}{\rho}\frac{\partial}{\partial x_i}\left(\frac{\rho{e}_{v}}{\tau_{vib}}\right)\right]}_{(VIII_{ij})}
    - \underbrace{(n-1)\frac{\partial}{\partial x_j}\left[\frac{1}{\rho}\frac{\partial}{\partial x_i}\left(\frac{\rho{e}_{v}^{*}}{\tau_{vib}}\right)\right]}_{(IX_{ij})}, \\
    \end{multline}
here, $k$ denotes the thermal conductivity. The symbol $e_{\nu}$ denotes the vibrational non-equilibrium energy per unit mass, while $e_{\nu}^{*}$ denotes the vibrational energy per unit mass in equilibrium with the local temperature $T$. The vibrational non-equilibrium energy relaxes toward $e_{\nu}^{*}$ on a characteristic timescale $\tau_{vib}$, which is referred to as the vibrational relaxation time. Following \cite{vincenti1966introduction}, the equilibrium vibrational energy per unit mass is expressed as
\begin{equation}
    \label{eq:vib_eqlb_energy}
    e_{\nu}^{*} = \frac{\frac{h \nu}{k_{B} T}}{\exp\!\left( \frac{h \nu}{k_{B} T} \right) - 1}\, RT,
\end{equation}
where $R$ is the gas constant, $h$ is Planck's constant, $\nu$ is the molecular vibrational frequency, and $k_{B}$ is Boltzmann's constant.

The terms written on the right-hand side (RHS) of (\ref{eq:Pressure_Hessian}) are grouped according to their physical significance, and each group corresponds to a distinct mechanism governing the evolution of the $P_{ij}$ tensor. The group $I_{ij}$ mechanisms represent the direct interaction of the $P_{ij}$ with the $A_{ij}$ tensor and are therefore mathematically closed. The mechanism groups $II_{ij}$--$V_{ij}$ are inviscid and non-local in nature, capturing interactions involving the gradients of $A_{kk}$ and pressure $(p)$. In particular, the term $IV_{ij}$ contains the factor $\tfrac{n p}{\rho}$, which highlights the influence of acoustic time scales or Mach number effects~\citep{suman2011dynamical}. Mechanisms $VI_{ij}$ and $VII_{ij}$ describe non-local contributions arising from thermal conduction and viscous heating, respectively. Mechanisms $VIII_{ij}$ and $IX_{ij}$ account for the influence of vibrational non-equilibrium on the evolution of the pressure Hessian. Due to the presence of higher-order gradients of the velocity gradient and density fields, the mechanisms $II_{ij}$--$IX_{ij}$ are mathematically unclosed. Consequently, closing the $P_{ij}$ evolution equation requires appropriate modeling of these mechanisms ($II_{ij}$--$IX_{ij}$).

The $P_{ij}$ tensor can be decomposed into two parts as
\begin{equation}
    \label{eq:PH}
    P_{ij} 
    \equiv 
    \frac{\partial}{\partial x_{j}}\left( \frac{1}{\rho}\frac{\partial p}{\partial x_{i}} \right)
    =
    \underbrace{\frac{1}{\rho}\frac{\partial^{2} p}{\partial x_{j}\partial x_{i}}}_{H_{ij}}
    -
    \underbrace{\frac{1}{\rho^{2}}\frac{\partial \rho}{\partial x_{j}}\,\frac{\partial p}{\partial x_{i}}}_{B_{ij}} ,
\end{equation}
such that,
\begin{equation}
\label{eq:Pij_decomposition}
P_{ij} = H_{ij} - B_{ij}.
\end{equation}
Here, the tensor $\rho H_{ij}$ is indeed  the \emph{pressure-Hessian tensor}. The tensor $B_{ij}$ is referred to as the \emph{baroclinic tensor} because it arises from the interaction between the density-gradient and pressure-gradient fields. The pressure-Hessian tensor is inherently symmetric, whereas the baroclinic tensor is generally asymmetric. In the work of \citet{danish2014direct}, the tensors $H_{ij}$ and $B_{ij}$ were referred to as the symmetric and non-symmetric pressure-Hessian components, respectively.  In  the incompressible-flow limit, density gradients vanish. Consequently, the term $P_{ij}$ is merely $H_{ij}$ and  $B_{ij} =0$.

\section{\label{III}{Existing model for velocity-gradient dynamics in compressible turbulence: a brief review}}

In this section, we provide a brief review of the existing model for compressible velocity-gradient dynamics proposed by \cite{danish2014direct}. In this framework, a phenomenological closure strategy is employed to model the unclosed terms appearing in the evolution equations of the $A_{ij}$ (\ref{eq:velGrad}) and the $P_{ij}$ tensors (\ref{eq:Pressure_Hessian}). The resulting formulation is summarized below:
\begin{equation}
\label{eq:DanishAij}
\frac{dA_{ij}}{dt}
=
-A_{ik}A_{kj}
-P_{ij}
-\frac{C_{pq}C_{pq}}{3\tau_{v}}
\left(
A_{ij}
+ 
\frac{A_{kk}}{3C_{pq}C_{pq}}C_{ri}C_{rj}
\right),
\end{equation}
\begin{equation}
\label{eq:DanishPij}
\frac{dP_{ij}}{dt}
=
-\left(
P_{ij}
+
\frac{A_{lm}A_{ml}}{C_{pq}C_{pq}}C_{ri}C_{rj}
\right)\frac{c}{L}
-\frac{C_{pq}C_{pq}}{3\tau_{p}}P_{ij}
+A_{kk}\sqrt{P_{mn}P_{mn}}
\frac{C_{ij}}{\sqrt{C_{qr}C_{qr}}},
\end{equation}
\begin{equation}
\label{eq:M_danish}
\mathbf{M} \equiv \mathbf{C}^{-1},
\end{equation}
\begin{equation}
\label{eq:dmij_danish}
\frac{dM_{ij}}{dt}
=
A_{ik}M_{kj},
\end{equation}
\begin{equation}
\label{eq:L_danish}
L
=
\max\left[
\frac{L_{0}}{\sqrt{C_{pq}C_{pq}/3}},
\frac{L_{0}}{\sqrt{\tau_{\nu}A}}
\right],
\qquad 
\text{where } A=\sqrt{A_{ij}A_{ij}}.
\end{equation}
Equations~(\ref{eq:DanishAij})--(\ref{eq:L_danish}) constitute a closed system of 27 ordinary differential equations (ODEs) governing the Lagrangian evolution of the velocity-gradient tensor and the $P_{ij}$ tensor in compressible turbulence. This ODE system is referred to as the \textit{new enhanced homogenized Euler equation} (N-EHEE) model. Here, $c$ denotes the speed of sound and $L$ represents a characteristic fluid length scale, with $L_{0}\equiv L(t=0)$. The term $C_{ij}$ represents the deformation-gradient tensor, which is defined as
\begin{equation}
\label{eq:deformation_gradient_tensor}
C_{ij} \equiv \frac{\partial X_{i}}{\partial x_{j}},
\end{equation}
where $x$ and $X$ denote the Eulerian and Lagrangian coordinates of a fluid particle, respectively. A detailed derivation and physical interpretation of various terms can be found in the original work of \cite{danish2014direct}.

\section{\label{IV}Enhancements and new modeling suggestions}
The N-EHEE model relies on two key simplifying assumptions: (i) thermodynamic fluctuations are sufficiently small compared to their mean values, and (ii) the thermodynamic variables follow a polytropic process. Previous DNS studies \citep{sarkar1991analysis,blaisdell1993compressibility} indicate that these assumptions remain valid up to a turbulent Mach number of approximately $0.5$. However, at higher Mach numbers, thermodynamic fluctuations become significant and deviations from polytropic behavior increase, thereby limiting the applicability of the N-EHEE model. Moreover, the model does not incorporate any closure for vibrational non-equilibrium effects. Being aware of the limitations of the existing model (\ref{eq:DanishAij})--(\ref{eq:L_danish}), the present study aims to develop an improved formulation that adapts to varying flow conditions, including high turbulent Mach numbers, and accounts for the effects of vibrational non-equilibrium.

In the current modeling framework, separate evolution equations are formulated for the $\boldsymbol{H}$ and $\boldsymbol{B}$ tensors (\ref{eq:PH}). Having these two separate governing equations allows the already-modeled component of the $\boldsymbol{P}$ tensor to serve as an input for predicting the unknown component.
On the right-hand side of~(\ref{eq:Pressure_Hessian}), after identifying the distinct effects governing mechanisms, we develop their respective closure forms and incorporate them into the modeled evolution equations for the $\boldsymbol{H}$ and $\boldsymbol{B}$ tensor, consistent with their underlying physical nature.

We first aim to modify the already modeled mechanisms as per the new modeling framework. In this, we first focus on modifying the closure of 
mechanism $IV_{ij}$ proposed by \cite{danish2014direct}. 
In their formulation (\ref{eq:DanishPij}), the mechanism $IV_{ij}$ (\ref{eq:Pressure_Hessian}) is modeled such that the $P_{ij}$ tensor relaxes toward its incompressible counterpart over the acoustic time scale: \(L/c\). 
In the present work, owing to the availability of separate evolution equations for the $\boldsymbol{H}$ and $\boldsymbol{B}$ tensors, the existing model of the \(IV_{ij}\) (\ref{eq:Pressure_Hessian}) mechanism is modified as:
\begin{align}
\label{eq:PH_IV}
-\frac{np}{\rho}\frac{\partial^2 A_{kk}}{\partial x_i \partial x_j}
\;\equiv\;
-\left(
P_{ij}
+
\frac{A_{lm}A_{ml}}{C_{pq}C_{pq}}C_{ri}C_{rj}
\right)\frac{c}{L} \\ \equiv
-\left(
H_{ij}
+
\frac{A_{lm}A_{ml}}{C_{pq}C_{pq}}\, C_{ri} C_{rj}
\right)
\frac{c}{L}+B_{ij}\frac{c}{L}.
\end{align}

Next, we modify the closure of the thermal conduction processes denoted by $VI_{ij}$ mechanism in (\ref{eq:Pressure_Hessian}), contributing to the $\boldsymbol{P}$ tensor \citep{danish2014direct}. 
Upon decomposing the existing model of the thermal heating mechanisms explicitly into their $\boldsymbol{H}$ and $\boldsymbol{B}$ contributing parts , the resulting expression is  written as:
\begin{multline}
    \label{eq:PH_VI_VII}
    \left(\frac{n-1}{R}\right)
    \frac{\partial}{\partial x_j}
    \left\{
    \frac{1}{\rho}
    \frac{\partial^2}{\partial x_i\partial x_k}
    \left[
    k\frac{\partial}{\partial x_k}\left(\frac{p}{\rho}\right)
    \right]
    \right\}
    \equiv
   - \frac{C_{pq}C_{pq}}{3\tau_{p}}P_{ij}
    =
   - \frac{C_{pq}C_{pq}}{3\tau_{p}}(H_{ij}-B_{ij}).
\end{multline}

The mechanism $I_{ij}$ (\ref{eq:Pressure_Hessian}) represents the direct interaction between the $\boldsymbol{A}$ and $\boldsymbol{P}$ tensors, and hence it is mathematically closed. We directly utilize the various terms involved in the mechanism $I_{ij}$ (\ref{eq:Pressure_Hessian}) in our model. For this, we decompose the exact form of the mechanism $I_{ij}$ (\ref{eq:Pressure_Hessian})  using (\ref{eq:Pij_decomposition}). Accordingly,  the mechanism \(I_{ij}\) (\ref{eq:Pressure_Hessian}) can be expressed as the sum total of two parts
\begin{align}
\label{eq:split_I}
 -A_{kj} P_{ik}
 -A_{ki} P_{kj}
 -(n-1) A_{kk} P_{ij}&=&
 \\\nonumber
 \left[-A_{kj} H_{ik} 
 -A_{ki} H_{kj} 
 -(n-1) A_{kk} H_{ij} \right]
  &+&\left[A_{kj} B_{ik} 
 +A_{ki} B_{kj} 
 +(n-1) A_{kk} B_{ij} \right]
\end{align}
It is desirable to retain both the $\boldsymbol{H}$ and $\boldsymbol{B}$ tensors contributions arising from the decomposition of these exact terms (mechanism $I_{ij}$, as evident in \ref{eq:split_I}) . However, prior modeling experience suggests that including the $\boldsymbol{H}$ contributions of the mechanism $I_{ij}$ (\ref{eq:Pressure_Hessian}) in the evolution equation for the $\boldsymbol{H}$ tensor compromises the model’s ability to maintain the correct behavior of the velocity divergence in low Mach regions (wherein, $A_{ii}$ must be negligible  at all times for all fluid elements). Consequently, the $\boldsymbol{H}$ contributions of these terms are neglected, and only the $\boldsymbol{B}$ of the mechanism $I_{ij}$ is retained and incorporated into the evolution equation of the $\boldsymbol{B}$ tensor. The resulting expression for the $I_{ij}$ (\ref{eq:Pressure_Hessian}) tensor closure is hence approximated as:
\begin{equation}
\label{eq:I}
 -A_{kj} P_{ik}
 -A_{ki} P_{kj}
 -(n-1) A_{kk} P_{ij}
\approx
 A_{kj} B_{ik}
 +A_{ki} B_{kj}
 +(n-1) A_{kk}B_{ij}.
\end{equation}

With the closures for the various mechanisms proposed in this section (as given in equations: \ref{eq:I}, \ref{eq:PH_IV}, \ref{eq:PH_VI_VII}), the evolution equations governing the $\boldsymbol{H}$ and $\boldsymbol{B}$ tensors can be written as:

\begin{multline}
\label{eq:Hij_modeled_without_vibnon}
    \frac{dH_{ij}}{dt}=-\left( H_{ij} +\frac{A_{lm}A_{ml}}{C_{pq}C_{pq}} C_{ri}C_{rj}\right ) \frac{c}{L} - \frac{C_{pq}C_{pq}}{3\tau_{p}} H_{ij}+M_{ij}^{vib},  
\end{multline}

\begin{multline}
\label{eq:Bij_modeled_without_nonsym}
\frac{dB_{ij}}{dt}= -A_{kj}B_{ik} -A_{ki}B_{kj} -(n-1)A_{kk}B_{ij}-B_{ij}\frac{c}{L} -V_{M_{ij}}  -\frac{C_{pq}C_{pq}}{3\tau_{p}} B_{ij}.
\end{multline}
The term $M_{ij}^{\mathrm{vib}}$, appearing in~(\ref{eq:Hij_modeled_without_vibnon}), represents the modeled contribution of the \emph{symmetric} component of the vibrational non-equilibrium mechanism denoted by $VIII_{ij}+IX_{ij}$ (\ref{eq:Pressure_Hessian}) . Although the vibrational non-equilibrium mechanism is inherently asymmetric in nature, DNS investigations by~\cite{shikha2024modeling} indicate that the magnitude of the antisymmetric component of the vibrational non-equilibrium tensor remains negligible for flow fields with turbulent Mach numbers ($M_t$) up to $0.5$. 
In the present modeling framework, only the symmetric component of the vibrational non-equilibrium mechanism is incorporated into the evolution equation of the $\boldsymbol{H}$ tensor. No explicit representation of the vibrational non-equilibrium mechanism effects, either symmetric or antisymmetric, is included in the evolution of the $\boldsymbol{B}$ tensor. The development of a neural-network-based closure for the antisymmetric vibrational non-equilibrium mechanism, which may require addressing further modeling complexities,  is left for future work. Addressing these complexities may require performing detailed analysis employing  high Mach number DNS database before any modeling attempts can be made, and is out of the scope of this study.

The term $V_{M_{ij}}$, used in (\ref{eq:Bij_modeled_without_nonsym}), denotes the modeled representation of the $V_{ij} $ mechanism identified in (\ref{eq:Pressure_Hessian}). The mechanism $V_{ij}$ (\ref{eq:Pressure_Hessian}) is the inviscid mechanism which contribute to the generation of $\boldsymbol{B}$ tensor even when the initial velocity-gradient and TGF tensors do not contain density-gradient effects (i.e.,\ \(B_{ij}=0, ~$at$~ t=0\)). This behavior arises from the explicit involvement of density-gradient-related terms within the \(V_{ij}\) mechanism. To close (\ref{eq:Hij_modeled_without_vibnon}) and~(\ref{eq:Bij_modeled_without_nonsym}), appropriate closure models are required for the remaining unclosed mechanisms, namely \( M_{ij}^{vib} \) and \( V_{M_{ij}} \). In the following two sections, we develop closure models for the vibrational non-equilibrium mechanism and the $V_{ij}$ mechanism, respectively.

\section{\label{V} Model of \texorpdfstring{$M_{ij}^{vib}$}{Mij vib}}
For the first time, in this work, we incorporate a model of vibrational non-equilibrium mechanisms into a dynamical velocity-gradient framework. To this end, we build upon the modeling strategy proposed by \cite{shikha2024modeling}. In that work, a physics-assisted neural-network model was developed to represent the non-dimensional form of the vibrational non-equilibrium mechanisms (\(VIII_{ij}\)+\(IX_{ij}\)) appearing in the $P_{ij}$ evolution (\ref{eq:Pressure_Hessian}). The non-dimensionalization was carried out using local magnitudes of the $A_{ij}$ and the $P_{ij}$ tensors.
\begin{equation}
    \label{eq:vib_mechanism}
    \Pi_{ij}= \frac{1}{\epsilon \sqrt{P_{mn}P_{mn}}}
    \left[
    (n-1)\frac{\partial}{\partial x_{j}}
    \left\{
    \frac{1}{\rho}\frac{\partial}{\partial x_{i}}
    \left( \frac{\rho e_{v}}{\tau_{vib}}\right)
    \right\}
    -
    (n-1)\frac{\partial}{\partial x_{j}}
    \left\{
    \frac{1}{\rho}\frac{\partial}{\partial x_{i}}
    \left( \frac{\rho e_{v}^{*}}{\tau_{vib}}\right)
    \right\}
    \right],
\end{equation}
where $\epsilon=\sqrt{A_{ij}A_{ij}}$. By definition, the tensor $\boldsymbol{\Pi}$ (\ref{eq:vib_mechanism}) is asymmetric. To simplify the modeling strategy, \cite{shikha2024modeling} decomposed $\boldsymbol{\Pi}$ into its symmetric and antisymmetric components as
\begin{equation}
\label{eq:decomp_vibnoneqlb}
\boldsymbol{\Pi}= \boldsymbol{^{s}\Pi} + \boldsymbol{^{a}\Pi}.
\end{equation}
DNS-based analyses performed by \cite{shikha2024modeling} demonstrated that, for the range of parameters considered in their study (initial turbulent Mach number up to $0.5$), the magnitude of the antisymmetric component ($\boldsymbol{^{a}\Pi}$) is negligible compared to that of the symmetric component ($\boldsymbol{^{s}\Pi}$). Based on this observation, the following approximation was adopted:
\begin{equation}
\label{eq:approximation}
\boldsymbol{\Pi} \approx \boldsymbol{^{s}\Pi}.
\end{equation}
Under this approximation, the vibrational non-equilibrium contribution is modeled as a purely symmetric tensor. The resulting closure for $\boldsymbol{\Pi}$ is referred to as the \textit{invariants-based vibrational non-equilibrium model (IBVM)}.

To model $\Pi_{ij}$, the IBVM employs two neural networks. The first network predicts the self-normalized tensor, denoted by $\widehat{\Pi}_{ij}$. The second network predicts the scalar magnitude, denoted by $\Phi$, of the non-dimensional vibrational non-equilibrium tensor. The self-normalized tensor $\widehat{\Pi}_{ij}$ and the magnitude $\Phi$ are defined as follows:
\begin{equation}
    \label{eq:magnitude_ibvm}    \widehat{\Pi}_{ij} =\frac{\Pi_{ij}}{\sqrt{\Pi_{mn}\Pi_{mn}}},~~~\Phi=\sqrt{\Pi_{ij}\Pi_{ij}},
\end{equation}
where $\sqrt{\widehat{\Pi}_{ij}\widehat{\Pi}_{ij}}=1$. 

The IBVM adopts the \textit{tensor-basis neural network} (TBNN) framework to predict the self-normalized vibrational tensor $\widehat{\Pi}_{ij}$. In this approach, tensor bases and scalar invariants are constructed using the normalized pressure-Hessian tensor $\boldsymbol{\widehat{H}}$, the normalized rotation-rate tensor $\boldsymbol{{W}}$, and the locally defined Damk\"ohler number $D_{m}$. The normalized symmetric pressure-Hessian tensor is defined as
\begin{equation}
    \label{eq:norm_hij}
    \boldsymbol{\widehat{H}}=\frac{\boldsymbol{H}}{|\boldsymbol{H}|},
\end{equation}
where $|\boldsymbol{H}|=\sqrt{H_{ij}H_{ij}}$.
The normalized rotation-rate tensor is obtained from the self-normalized velocity-gradient tensor ($\boldsymbol{a}=\boldsymbol{A}/|\boldsymbol{A}|$, where $|\boldsymbol{A}|=\sqrt{A_{ij}A_{ij}}$))  as
\begin{equation}
    \label{eq:rotation_rate}    
    \boldsymbol{W}=\frac{\boldsymbol{a}-\boldsymbol{a}^{T}}{2}.
\end{equation}
The Damk\"ohler number is a key nondimensional parameter characterizing vibrational non-equilibrium effects. It is traditionally defined as the ratio of the large-eddy turnover time to the vibrational relaxation time~\citep{khurshid2019decaying, srivastava2024influence}. \cite{shikha2024modeling}, introduced locally defined Damk\"ohler number $D_{m}$ given by
\begin{equation}
    \label{eq:local_damkohler}
    D_{m}=\frac{1}{\tau_{\text{vib}}\sqrt{A_{ij}A_{ij}}},
\end{equation}
where $(A_{ij}A_{ij})^{-1/2}$ represents the characteristic time-scale of the motion of a local fluid element. The symbol $\tau_{\text{vib}}$ represents the vibration relaxation time. The modeled representation of the $\widehat{\Pi}_{ij}$ tensor is given as:
\begin{equation}
    \label{eq:model_pihat}
  \boldsymbol{\widehat{\Pi}}^{IBVM} = \mathcal{G}_{ibvm} (\boldsymbol{\widehat{H}}, \boldsymbol{W}, D),
\end{equation}
where $\mathcal{G}^{ibvm}$ is the modeled mapping, which is accessible in the form of weights and biases saved in the trained neural network. The reader is referred to the original paper by \cite{shikha2024modeling} for further details of the tensor bases and invariants employed in this model.

The second neural network predicts the scalar magnitude $\Phi$ of the nondimensional vibrational tensor. To predict the magnitude $\Phi$,  IBVM model \citep{shikha2024modeling} used a fully connected simple neural network with input quantities in the form of the first and third invariants ($I_{r1}, ~ \&,~ I_{r3}$) of the $\boldsymbol{\widehat{H}}$ tensor (\ref{eq:norm_hij}), the locally defined model Damk\"ohler number $D_{m}$ (\ref{eq:local_damkohler}) and the local compressibility parameters such as: (i) the dilatation rate ($a_{ii}$) and (ii) the rate of change of local dilatation rate ($\delta$). These parameters are mathematically defined as:
\begin{equation}
    \label{eq:delta_aii}
    a_{ii}=\frac{A_{ii}}{\epsilon}, 
    \qquad 
    \delta=\frac{-A_{ij}A_{ji}-H_{ii}+B_{ii}}{\epsilon^{2}},
\end{equation}
where $\epsilon=\sqrt{A_{ij}A_{ij}}$.
The invariants of $\boldsymbol{\widehat{H}}$ are defined as:
\begin{subequations}
\begin{equation}
\label{eq:I1_H}
I_{r1}=\text{Tr}(\boldsymbol{\widehat{H}}),
\end{equation}
\begin{equation}
\label{eq:I2_H}
I_{r2}=\frac{1}{2}\left(I_{r1}^{2}-\text{Tr}(\boldsymbol{\widehat{H}}^{2})\right),
\end{equation}
\begin{equation}
\label{eq:I3_H}
I_{r3}=\frac{1}{3}\left(-I_{r1}^{3}+3I_{r1}I_{r2}-\text{Tr}(\boldsymbol{\widehat{H}}^{3})\right),
\end{equation}
\end{subequations}
where $Tr$ means trace of a tensor. Hence, the closure of $\Phi$ is written as:
\begin{equation}
    \label{eq:general_phi_ibvm}
    \Phi^{IBVM}= \mathcal{F}_{ibvm}(a_{ii}, \delta, D_{m}, I_{r1}, I_{r3}),
\end{equation}
where $\mathcal{F}_{ibvm}$ is the modeled mapping, which is obtained in the form of the weights and biases of the trained neural network.

The final non-dimensionalized vibrational non-equilibrium tensor is obtained by combining the outputs of the two networks (\ref{eq:model_pihat} and \ref{eq:general_phi_ibvm}) as:
\begin{equation}
    \label{eq:model_theta}
    \boldsymbol{\Pi}^{IBVM}=\widehat{\boldsymbol{\Pi}}^{IBVM}\,\Phi^{IBVM}.
\end{equation}
We aim to develop a dynamic model having vibrational non-equilibrium effects that relax to the model of vibrational equilibrium flows, in the limit of 
\begin{equation}
    \label{eq:vib_eqlb_cond}
    e_{\nu}=e_{\nu}^{*}.
\end{equation}
The symbol $e_{\nu}$ denotes the vibrational non-equilibrium energy per unit mass, while $e_{\nu}^{*}$ (\ref{eq:vib_eqlb_energy}) denotes the vibrational energy per unit mass in equilibrium with the local temperature. To bring this feature into our dynamical model, we define a new non-dimensional parameter, $\xi_{m}$ as:
\begin{equation}
    \label{eq:xi_m}
   \xi_{m}= \frac{e_{\nu}}{e_{\nu}^{*}}.
\end{equation}
Note that flow regimes in vibrational equilibrium always satisfy $\xi_{m}=1$. Utilizing the locally available quantities $\sqrt{A_{pq}A_{pq}},~\&~ \sqrt{P_{mn}P_{mn}}$, we write the modeled representation of the vibrational non-equilibrium mechanism ($M_{ij}^{vib}$) as:
\begin{equation}
    \label{eq:model_vibnon}
    M_{ij}^{vib}= \frac{(1-{\xi_{m}}_{0})}{(1+\xi_{m})}\,\boldsymbol{\Pi}^{IBVM}\
\sqrt{A_{pq}A_{pq}}\sqrt{P_{mn}P_{mn}},
\end{equation}
where, ${\xi_{m}}_{0}=\xi_{m}(t=0)$.
We propose to model the evolution process of $\xi_{m}$ inspired by the Landau–Teller model of \cite{vincenti1966introduction} for the source term of the evolution equation of vibrational energy per unit mass. The proposed evolution equation of $\xi_{m}$ is written as:
\begin{equation}
\label{eq:xim}
    \frac{d\xi_{m}}{dt}=-\frac{\xi_{m}-1}{\tau_{\text{vib}}}.
\end{equation}
Following this framework, the proposed model tends to the vibrational equilibrium case when ${\xi_{m}}_{0}$ becomes unity. 
With these extensions, the evolution equation for the $\boldsymbol{H}$ tensor can be written as
\begin{multline}
\label{eq:Hij_modeled}
\frac{dH_{ij}}{dt}
=-
\left(
H_{ij}
+
\frac{A_{lm} A_{ml}}{C_{pq} C_{pq}} \, C_{ri} C_{rj}
\right)
\frac{c}{L}
-
\frac{C_{pq} C_{pq}}{3 \tau_{p}} \, H_{ij}
\\+
\frac{(1-{\xi_{m}}_{0})}{(1+\xi_{m})}
\, \Pi^{\mathrm{IBVM}}_{ij}
\sqrt{P_{mn} P_{mn}}
\sqrt{A_{pq} A_{pq}} .
\end{multline}
Given the evolution equations for \(A_{ij}\) (\ref{eq:velGrad}) and \(\xi_{m}\) (\ref{eq:xim}), (\ref{eq:Hij_modeled}) constitutes a closed evolution equation for the $\boldsymbol{H}$ tensor. 
\section{Model of \texorpdfstring{$V_{M_{ij}}$}{VMij}}
\label{VI}
Following the proposed evolution equation of $B_{ij}$ (\ref{eq:Bij_modeled_without_nonsym}), our next objective is to develop a closure for the mechanism \(V_{{ij}}\) (\ref{eq:Pressure_Hessian}), the only inviscid mechanism which can generate a non-zero baroclinic tensor \(B_{ij}\) even when the initial velocity-gradient and TGF tensors are free of density-gradient effects (i.e.\ \(B_{ij}(t=0)=0\)). Previously,~\cite{danish2014direct} proposed a phenomenological model for the mechanism ($V_{ij}$) (as defined in (\ref{eq:Pressure_Hessian})) under a set of simplifying assumptions. In particular, they assumed that the fluctuations of density and pressure are negligible compared to their respective mean values. However, this assumption is not valid in regions of high turbulent Mach number. Furthermore, the DNS investigations conducted by~\cite{danish2014direct} demonstrated that the contribution of baroclinic tensor in the $P_{ij}$ tensor becomes more significant with increasing turbulent Mach number, especially in the vicinity of shocklets. As a consequence, the N-EHEE model (~\cite{danish2014direct}) fails to accurately reproduce the probability density functions (PDFs) of the components of the antisymmetric part of $P_{ij}$ tensor in high turbulent Mach number flows (\(M_t = 1.2\)). Motivated by these observations, the present work aims to incorporate such physical insights in order to make the model adaptive to varying flow conditions. In particular, we aim to develop a new model for the $V_{ij}$ mechanism without assuming that density and pressure fluctuations are negligible relative to their mean values. In the following subsection, we present the detailed modeling strategy adopted for the $V_{ij}$ mechanism.

\subsection{Overall modeling strategy for \texorpdfstring{$V_{ij}$}{Vij} mechanism}
To model $V_{ij}$ (\ref{eq:Pressure_Hessian}), it is not mandatory to model the tensor in its dimensional form. Instead, we nondimensionalize the mechanism $V_{ij}$ by dividing it by the local magnitudes of $P_{ij}$ and $A_{ij}$ tensors for which the governing equations are available in the modeling framework. We define a nondimensional form of the $V_{ij}$ mechanism using symbol $\varsigma_{ij}$ as:
\begin{equation}
    \label{eq:nondimenV}
    \varsigma_{ij} = 
    \frac{1}{\sqrt{P_{mn}P_{mn}} \, \sqrt{A_{qr}A_{qr}}}
    \left(
        \frac{1}{\rho}\frac{\partial A_{kk}}{\partial x_j}\frac{\partial p}{\partial x_i} 
        + \frac{n p}{\rho^{2}}\frac{\partial A_{kk}}{\partial x_i}\frac{\partial \rho}{\partial x_j}
    \right).
\end{equation}
Given the availability of evolution equations for $A_{ij}$, $B_{ij}$, and $H_{ij}$, the non-dimensional formulation of these tensors does not introduce any additional unknowns. This makes it a suitable approach for developing a model of the non-dimensional form of the $V_{ij}$ mechanism, denoted by $\varsigma_{ij}$. From a modeling perspective, the use of non-dimensional quantities offers several advantages. First, their magnitudes are naturally bounded within certain limits, which enhances the numerical stability of the trained model. Second, non-dimensionalization using locally defined quantities reduces the sensitivity of the resulting non-dimensional variable, thereby improving the model's generalizability across different flow regimes. Consequently, in this work, we focus on modeling the non-dimensional form of the $V_{ij}$ mechanism, $\varsigma_{ij}$.
Our objective is to model the $\boldsymbol{\varsigma}$ tensor using the deep neural network (DNN) framework. Any second-order tensor can be mathematically decomposed into two components:  
(i) a self-normalized tensor, which exhibits universal behavior across a wide range of flow conditions~\citep{das2019reynolds}, and 
(ii) the tensor magnitude, which shows sensitivity towards the underlying flow conditions ~\citep{das2019reynolds}.
The $\boldsymbol{\varsigma}$ tensor characterizes the alignment tendencies of its eigendirections. We define this self-normalized form as $\boldsymbol{\hat{\varsigma}} $:
\begin{equation}
    \label{eq:varsigmahat}
    \widehat{\varsigma}_{ij}= \frac{\varsigma_{ij}}{\sqrt{\varsigma_{mn}\varsigma_{mn}}}.
\end{equation}
In addition, "the magnitude" of the $\boldsymbol{\varsigma}$ tensor is defined as a scalar quantity ($\Psi$), such that
\begin{equation}
    \label{eq:magnitude}
    \Psi = \sqrt{\varsigma_{mn}\varsigma_{mn}}.
\end{equation}
By combining the two parts of the tensor: the eigendirections and the magnitude, we obtain the complete non-dimensional form of the to-be modeled tensor, $\boldsymbol{\varsigma}$ as:
\begin{equation}
    \label{eq:combined_varsigma}
    \boldsymbol{\varsigma}= \boldsymbol{\widehat{\varsigma}} \Psi.
\end{equation}
Having a combined model of $\boldsymbol{\varsigma}$ tensor, one can obtain the mechanism $V_{ij}$ (\ref{eq:Pressure_Hessian}) as:
\begin{equation}
    \label{eq:model_VMij}
    V_{ij}=\varsigma_{ij} \sqrt{P_{mn}P_{mn}}\sqrt{A_{qr}A_{qr}}.
\end{equation}
Note that the quantities $\sqrt{P_{mn}P_{mn}}\sqrt{A_{qr}A_{qr}}$ are locally available in the proposed modeled set of equations. 

We propose to model the two aspects of the $\boldsymbol{\varsigma}$ tensor using two separate deep neural networks (DNNs). In this framework, the tensor decomposition is performed such that the magnitude and the eigendirection alignments are modeled independently, with each DNN trained by optimizing a distinct loss function. These loss functions are formulated based on the underlying physics of the respective output quantities. 


To gain deeper insight into the behavior of the target quantities, we first conducted a DNS-based investigation. This investigation focuses on two key aspects of the expected modeling exercise:  
(i) finding some universal behavior of the quantities of interest across different flow conditions, and  
(ii) identifying an optimized set of input features for training the DNNs.   
The DNNs are trained using the optimized set of input features in one dataset and subsequently validated on other datasets representing different flow conditions, thus ensuring the generalizability of the proposed model.

Previous studies by  \cite{shikha2023augmented} 
have demonstrated that the alignment characteristics and magnitude statistics of the $A_{ij}$ and the $P_{ij}$ tensors in isotropic decaying turbulence exhibit similar trends across different classes of turbulent flows (specifically tested in wall-bounded channel flows). Consequently, modeling the $\boldsymbol{\varsigma}$ tensor using DNS data from isotropic decaying turbulence provides a fundamental framework for extending predictive capabilities across a broader range of turbulent flow configurations. Motivated by this observation, we consider three distinct datasets of compressible isotropic decaying turbulence for our investigations. These datasets, obtained from \cite{parashar2019lagrangian}, are generated using a gas-kinetic method-based solver that has been validated against the compressible decaying turbulence results of \cite{samtaney2001direct}. 

In Table \ref{tab:DNS_cases} we present all details of the DNS datasets employed in this study.
\begin{table}
    \centering
\begin{tabular}{c c c c }
   \hline
   \hline
   Simulation  & $M_{t}$ & $Re_{\lambda}$ & $N^{3}$ \\ \hline
   C1 & 1&  250& $512^{3}$\\
   C2 & 0.8&  250& $512^{3}$\\
   C3 & 1 &  60& $256^{3}$\\
   \hline
   \hline
\end{tabular}
\caption{Various DNS databases used to model the $V_{ij}$ (\ref{eq:Pressure_Hessian}) mechanism.}
\label{tab:DNS_cases}
\end{table}
All simulations are characterized by two non-dimensional parameters:  
(i) the turbulent Mach number, $M_{t}$,  and (ii) the Reynolds number based on the Taylor microscale, $Re_{\lambda}$. These numbers are defined as:
\begin{equation}
    \label{eq:mach_number}
    M_{t}=\sqrt{\frac{2K}{nR\langle T \rangle}},
\end{equation}
\begin{equation}
    \label{eq:reynolds_number}
    Re_{\lambda}=\sqrt{\frac{20}{3\epsilon\mu}}\,K,
\end{equation}
where $\langle\cdot\rangle$ denotes spatial averaging. The symbols $\mu,~\epsilon,~\& ~K$ in (\ref{eq:reynolds_number}) and (\ref{eq:mach_number}) represent kinematic viscosity, dissipation rate, and turbulent kinetic energy, respectively.
Among the three simulation databases mentioned in table (\ref{tab:DNS_cases}), cases C1 and C2 differ in their initial turbulent Mach numbers, with $M_{t} = 1$ for C1 and $M_{t} = 0.8$ for C2, while both have same initial Taylor Reynolds number, $Re_{\lambda} = 250$. In contrast, cases C1 and C3 have identical initial turbulent Mach numbers ($M_{t} = 1$) but differ in their initial Taylor Reynolds numbers, with $Re_{\lambda} = 250$ for C1 and $Re_{\lambda} = 60$ for C3.~\cite{parashar2019lagrangian} demonstrated that such variations in initial $M_{t}$ and $Re_{\lambda}$ are sufficiently large to significantly influence the flow physics. In particular, they reported substantial differences in flow topologies (structures) between two simulations that differed by only $\Delta M_{t} = 0.2$ and $\Delta Re_{\lambda} = 120$, where the symbol $\Delta$ signifies the difference in two quantities.

In this study, the Mach number of interest is the turbulent Mach number (\ref{eq:mach_number}). \cite{danish2014direct} demonstrated that the baroclinic effects contributes significantly to the evolution of the $P_{ij}$ tensor only in regions of higher $M_{t}$. 
Therefore, our analysis is restricted to DNS datasets corresponding to higher values of $M_{t}$ in $[0.8, 1]$. Earlier,~\cite{sarkar1991analysis} showed that, in practical applications, a turbulent Mach number of $M_{t}=1$ may correspond to a convective Mach number of approximately 10. Consequently, flows with $M_{t}=1$ can be regarded as representative of high convective Mach number flow fields (hypersonic regimes).

Among the three datasets (Table \ref{tab:DNS_cases}), only case C1 is used to train neural networks. The inclusion of additional datasets (C2 and C3) in this study serves two main purposes. First, they help identify the appropriate forms of the input and output quantities, thereby ensuring that the mapping from the inputs to the $\boldsymbol{\varsigma}$ tensor exhibits universality across the simulations C1–C3, which differ significantly in turbulent Mach number and Reynolds number. Second, the diverse DNS datasets (C2 and C3) are extensively used to assess the performance of the proposed model. Since training is performed exclusively on dataset C1, cases C2 and C3 provide independent test platforms to evaluate the model under flow conditions that differ from C1 in both $Re_{\lambda}$ and $M_{t}$.
 
\subsection{Model of \texorpdfstring{$\boldsymbol{\widehat{\varsigma}}$}{varsigma}}
The first neural network is designed to predict the $\boldsymbol{\widehat{\varsigma}}$ tensor. Being a second-order tensor, $\boldsymbol{\widehat{\varsigma}}$ must be independent of the choice of the coordinate system. The tensor property of rotational invariance ensures that the tensor remains unchanged under rotational transformations. To ensure that the $\boldsymbol{\widehat{\varsigma}}$ tensor satisfies rotational invariance, we model it using a well-established deep neural network architecture known as the tensor bases neural network (TBNN). This architecture was originally developed by \cite{ling2016machine}.

The TBNN framework constructs the output tensor using a set of predefined tensor bases together with their corresponding scalar invariants, thereby ensuring that the modeled tensor remains rotationally invariant under changes in the coordinate system. In recent studies \citep{ling2016reynolds, parashar2020modeling, shikha2023augmented, shikha2024modeling, tian2021physics}, this architecture has been employed to model second-order tensors of interest. Notably, all of these works utilized tensor bases specifically derived for symmetric second-order tensors. In contrast, the tensor considered in the present study is asymmetric and possesses a non-zero trace.
To the best of our knowledge, no prior work has derived the tensor bases for such an asymmetric tensor. Therefore, in this study, we derive and employ a novel set of tensor bases to model the $\boldsymbol{\widehat{\varsigma}}$ tensor.

\subsubsection{
Derivation of the tensor bases and invariants for modeling
\texorpdfstring{$\boldsymbol{\widehat{\varsigma}}$}{varsigma}
tensor}
In our derivation, all second-order tensors of interest are represented as $(3 \times 3)$ matrices. Let $\boldsymbol{O}$ denote the matrix form of the tensor $\boldsymbol{\widehat{\varsigma}}$. Our objective is to express $\boldsymbol{O}$ as an infinite polynomial expansion in terms of two matrices: $\boldsymbol{M}$ and $\boldsymbol{N}$, representing an arbitrary symmetric and antisymmetric second-order tensor, respectively. The most general representation of $\boldsymbol{O}$ in terms of $\boldsymbol{M}$ and $\boldsymbol{N}$ can thus be written as:
\begin{equation}
    \label{eq:genV}
    \boldsymbol{O} = \boldsymbol{M}^{a_{1}} \boldsymbol{N}^{b_{1}} \boldsymbol{M}^{a_{2}} \boldsymbol{N}^{b_{2}} \cdots ,
\end{equation}
where $a_{i}$ and $b_{i}$ are positive integers, and $i$ varies from $1$ to $\infty$.
Following the Caley-Hamilton theorem, the infinite polynomial form of the matrix can be written as the linear sum of finite-degree matrix polynomials. The contracted form of $\boldsymbol{O}$ is expressed using the product combinations of $\boldsymbol{M}$ and $\boldsymbol{N}$ having a maximum degree of five and a maximum extension of three. \cite{spencer1958theory} have listed all the sets of products of $\boldsymbol{M ~\& ~N}$ having a total degree $\leq5$ and a total extension of terms $\leq3$. Since the tensor of our interest is asymmetric and possesses a nonzero trace, we retain only those basis elements that also exhibit non-symmetry and a nonzero trace. Based on this analysis, we derive the following set of tensor basis elements: 
\begin{equation}
\begin{aligned}
\label{eq:tensor_basis_general}
\textbf{T}^{(1)}=\boldsymbol{MN}, & \hspace{2cm} 
\textbf{T}^{(2)}=\boldsymbol{NM^{2}},\\
\textbf{T}^{(3)}=\boldsymbol{N^{2}M}, & \hspace{2cm}
\textbf{T}^{(4)}=\boldsymbol{NMN^{2}},\\
\textbf{T}^{(5)}= \boldsymbol{M}\boldsymbol{N}\boldsymbol{M}^{2},
& \hspace{2cm} \textbf{T}^{(6)}=\boldsymbol{N^{2}M^{2}},\\
\textbf{T}^{(7)}=\boldsymbol{N}\boldsymbol{M}^{2}\boldsymbol{N^{2}}, & \hspace{2cm} 
\textbf{T}^{(8)}=\boldsymbol{M}\boldsymbol{N^{2}}\boldsymbol{M^{2}}.\\
\end{aligned}
\end{equation}
The sets of tensors displayed in (\ref{eq:tensor_basis_general}) are complete in the sense that any asymmetric polynomial with non zero trace involving products of $\boldsymbol{M}$ and $\boldsymbol{N}$ can be written as a linear combination of the eight tensors (listed in (\ref{eq:tensor_basis_general})) with the scalar multipliers expressed as polynomials of the scalar invariants. The non-zero scalar invariants are defined as:
\begin{equation}
\begin{aligned}
\label{eq:invariants_general}
\lambda_{1}=Tr(\boldsymbol{N}^{2}),& \hspace{0.5cm}
\lambda_{2}=Tr(\boldsymbol{M}),  & \hspace{0.5cm}
\lambda_{3}=Tr(\boldsymbol{N^{2}M}),& \hspace{0.5cm}
\lambda_{4}=Tr(\boldsymbol{N^{2}M^{2}}),
\end{aligned}
\end{equation}
where $Tr$ represents the trace of the tensor. We write the reduced form of $\boldsymbol{O}$ tensor as:
\begin{equation}    
\label{eq:gen_theta_hat}  
\boldsymbol{O}=\sum_{n=1}^{8}{[g^{(n)}(\lambda_{1},....\lambda_{4})]\textbf{T}^{(n)}}.
\end{equation}
where the symbol $\boldsymbol{g^{(n)}}$ represents the coefficients. These coefficients are the non-linear functions of the invariants ($\lambda_{i}s$) (\ref{eq:invariants_general}).

To model $\boldsymbol{\varsigma}$, it is essential to determine the appropriate forms of the tensors $\boldsymbol{M}$ and $\boldsymbol{N}$. 
Equation~\ref{eq:nondimenV} indicates that $\boldsymbol{\varsigma}$ is a nonlinear function of the gradients of density, pressure, and the velocity divergence ($A_{kk}$). 
Within our modeling framework, the tensors available instantaneously are $A_{ij}$ (\ref{eq:velGrad}), $H_{ij}$ (\ref{eq:Hij_modeled}), and $B_{ij}$ (\ref{eq:Bij_modeled_without_nonsym}). 
Since the tensors $A_{ij}$ and $H_{ij}$ inherently represent the physics of the velocity gradients and pressure gradients. we intend to model the tensor $\boldsymbol{\widehat{\varsigma}}$ using $A_{ij}$ and $H_{ij}$ tensors. The tensor $H_{ij}$ (\ref{eq:PH}) is inherently symmetric. Following~(\ref{eq:velGrad}), for incompressible flows, the $B_{ij}$ (\ref{eq:PH}) component of the pressure Hessian tensor is expected to vanish. To ensure this behavior, we modify the input tensors of the neural network such that the model is guided to predict $B_{ij} \to 0$ when the inputs correspond to the incompressible limit (wherein $A_{ii}=0$).
In incompressible flows, the trace of the $\boldsymbol{H}$ tensor can be expressed directly from the pressure Poisson equation as  
\begin{equation}
    \label{eq:poison_eq}
    H_{ii} = -A_{lm}A_{ml}.
\end{equation}
Consequently, we define a modified tensor $H_{ij}'$ as:  
\begin{equation}
    \label{eq:Hij_prime}
    H_{ij}' = H_{ij} + \frac{A_{lm}A_{ml}}{3}\,\delta_{ij},
\end{equation}
such that $H_{ii}'=0$ in incompressible flow regions. We further nondimensionalize the modified tensor $H_{ij}'$ as: 
\begin{equation}
    \label{eq:nondimhijp}
    \widehat{H_{ij}'} = \frac{H_{ij}'}{\sqrt{H_{lm}' H_{lm}'}}.
\end{equation}

The antisymmetric tensor $\boldsymbol{N}$ can be extracted from another instantaneously available tensor: $A_{ij}$. First, we self-normalize the tensor $A_{ij}$ as
\begin{equation}
\label{eq:selfnormaij}
a_{ij} = \frac{A_{ij}}{\sqrt{A_{mn}A_{mn}}}.
\end{equation}
Next, the antisymmetric part of $a_{ij}$ is computed using (\ref{eq:rotation_rate}).
We further write the tensor $\boldsymbol{W}$ in terms of the local vorticity vector $\boldsymbol{\omega}$:
\begin{equation}
    \label{eq:vorticity}
    \bf{W}=\begin{bmatrix}  0 & 2\omega_{3} & -2\omega_{2}\\-2\omega_{3} & 0 & 2\omega_{1}\\ 2\omega_{2} & -2\omega_{1} & 0 \end{bmatrix} .
\end{equation}
To support the selection of $\boldsymbol{W}$ as the antisymmetric tensor ($\bf{N}$) and $\boldsymbol{\widehat{H^{'}}}$  (\ref{eq:nondimhijp}) as the symmetric tensor ($\bf{M}$) in (\ref{eq:tensor_basis_general}), we evaluate the universal alignment characteristics between the modeled tensor $\boldsymbol{\widehat{\varsigma}}$ and the proposed input tensors $\boldsymbol{\widehat{H^{'}}}$  (\ref{eq:nondimhijp}) and $\boldsymbol{W}$  (\ref{eq:vorticity}). We present the statistics representing alignment tendencies among the modeled tensor ($\boldsymbol{\widehat{\varsigma}}$) and the proposed input tensors ($\boldsymbol{\widehat{H^{'}}}$  (\ref{eq:nondimhijp}) and $\boldsymbol{W}$  (\ref{eq:vorticity})). Particularly, we investigate the probability density functions (PDFs) of the direction cosines between the three eigendirections of the local tensors $\boldsymbol{\widehat{\varsigma}}$ with respect to $\boldsymbol{\widehat{H^{'}}}$  (\ref{eq:nondimhijp}) and $\boldsymbol{W}$  (\ref{eq:vorticity}). 
Since $\boldsymbol{\widehat{\varsigma}}$ is in general an asymmetric tensor, it does not necessarily possess an orthogonal set of eigenvectors. Therefore, we decompose the $\boldsymbol{\widehat{\varsigma}}$ tensor into its symmetric and antisymmetric parts as:
\begin{align}
    \label{eq:varsigma_decomp}
    \boldsymbol{\widehat{\varsigma_{s}}}= \frac{\boldsymbol{\widehat{\varsigma}}+\boldsymbol{\widehat{\varsigma}}^{T}}{2},\\
\boldsymbol{\widehat{\varsigma_{a}}}=
    \frac{\boldsymbol{\widehat{\varsigma}}-\boldsymbol{\widehat{\varsigma}}^{T}}{2},    
\end{align}
where the symbols $\boldsymbol{\widehat{\varsigma}_{s}}$ and $\boldsymbol{\widehat{\varsigma}_{a}}$ represent the symmetric and antisymmetric part of the $\boldsymbol{\widehat{\varsigma}}$ tensor, respectively. 
We express $\boldsymbol{\widehat{\varsigma_{a}}} $ tensor using its local axial vector $\bf{\zeta}$ as:
\begin{equation}
    \label{eq:axial_varsigma}
    \boldsymbol{\widehat{\varsigma_{a}}}=\begin{bmatrix}  0 & 2\zeta_{3} & -2\zeta_{2}\\-2\zeta_{3} & 0 & 2\zeta_{1}\\ 2\zeta_{2} & -2\zeta_{1} & 0 \end{bmatrix} .
\end{equation}
The three eigendirections of $\boldsymbol{\widehat{\varsigma}_{s}}$ tensor are defined as $\mathbf{\hat{e}_{\alpha,\varsigma_{s}}}, \mathbf{\hat{e}_{\beta,\varsigma_{s}}}, \mathbf{\hat{e}_{\gamma,\varsigma_{s}}}$, with the corresponding eigenvalues ordered as $\alpha_{\varsigma_{s}} > \beta_{\varsigma_{s}} > \gamma_{\varsigma_{s}}$. Similarly, we define the three eigenvectors of the tensor $\widehat{\boldsymbol{H^{'}}}$ as $\mathbf{\hat{e}_{\alpha,h}}, \mathbf{\hat{e}_{\beta,h}}, \mathbf{\hat{e}_{\gamma,h}}$,corresponding to the eigenvalues ordered as $\alpha_{h} > \beta_{h} > \gamma_{h}$.

\begin{figure}
    \centering
    \includegraphics[width=0.5\linewidth]{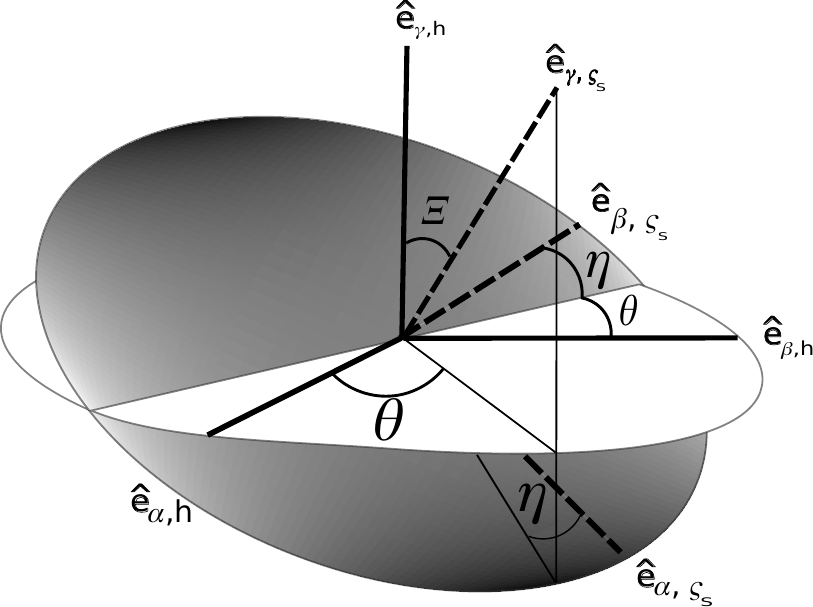}
    \caption{The three Euler angles \citep{rose1995elementary} ($\Xi; \theta; \eta$) to describe the relative orientation of
$\bf{\hat{e}_{\alpha, {\varsigma_{s}}}} , \bf{\hat{e}_{\beta, {\varsigma_{s}}}}, \bf{\hat{e}_{\gamma, {\varsigma_{s}}}}$ with respect to $\bf{\hat{e}_{\alpha,{h}}} , \bf{\hat{e}_{\beta,{h}}}, \bf{\hat{e}_{\gamma, {h}}}$.}
    \label{fig:Euler_angles}
\end{figure}

The relative orientation between the eigenvectors of the two tensors is quantified using three Euler angles: $\Xi$, $\theta$, and $\eta$, following the convention described in the study of \cite{rose1995elementary}. Figure~\ref{fig:Euler_angles} illustrates the definition of these angles with respect to the eigenvector frames of $\boldsymbol{\widehat{\varsigma}}$ and $\boldsymbol{\widehat{H^{'}}}$  (\ref{eq:nondimhijp}). Similarly, the orientation of  the local axial vector $\boldsymbol{\zeta}$ (\ref{eq:axial_varsigma}) and $\boldsymbol{\omega}$ (\ref{eq:vorticity}) is quantified using the angle $\phi$. The relationship between these angles and the various eigenvectors is defined as:
\begin{equation}
\begin{aligned}
\label{eq:cosine}
\cos{\Xi}= \bf{\hat{e}_{\gamma,h}\cdot\bf{\hat{e}_{\gamma,\varsigma_{s}}}},~&~
\cos{\theta}= \bf{\hat{e}_{\alpha,h}} \cdot \bf{\hat{e}_{proj}},~&~
\cos{\eta}= \bf{\hat{e}_{\beta, \varsigma_{s}}} \cdot \bf{\hat{e}_{norm}},~&~
\cos{\phi}= \boldsymbol{\zeta} \cdot \boldsymbol{\omega},
\end{aligned}
\end{equation}
where, 
\begin{equation}
\begin{aligned}
\bf{\hat{e}_{proj}}=\frac{\bf{e^{'}_{proj}}}{|\bf{e^{'}_{proj}}|},~~&~~\bf{e^{'}_{proj}}= \bf{\hat{e}_{\alpha,\varsigma_{s}}} - (\bf{\hat{e}_{\alpha,\varsigma_{s}}} \cdot \bf{\hat{e}_{\gamma, h}})\bf{\hat{e}_{\gamma, h}},~~&~~
\bf{\hat{e}_{norm}}= \bf{\hat{e}_{proj}} \times \bf{\hat{e}_{\gamma, h}}.
\end{aligned}
\end{equation}

\begin{figure}
\centering
  \begin{tabular}[b]{c}
\hspace{-0.3in}    
\includegraphics[width=.5\linewidth]{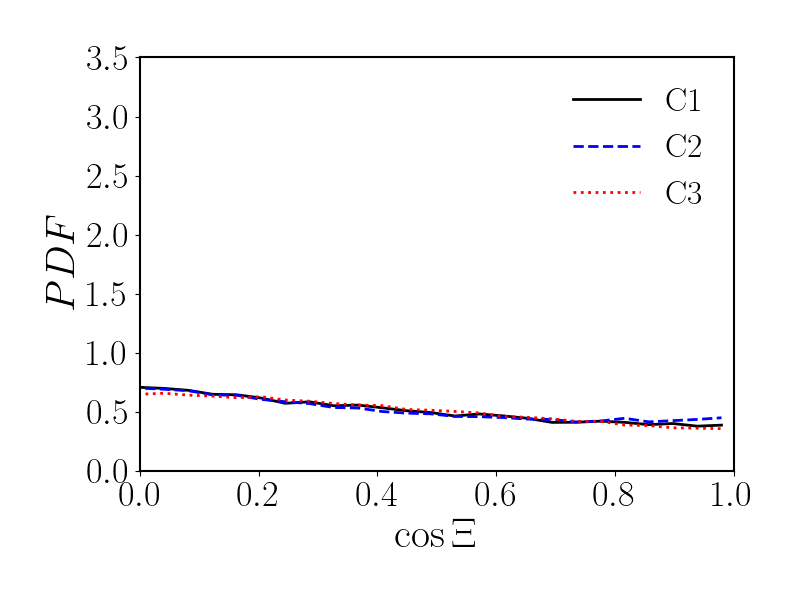} \\
    \small (a)
  \end{tabular} \qquad
  \begin{tabular}[b]{c}
  \hspace{-0.4in}
\includegraphics[width=.5\linewidth]{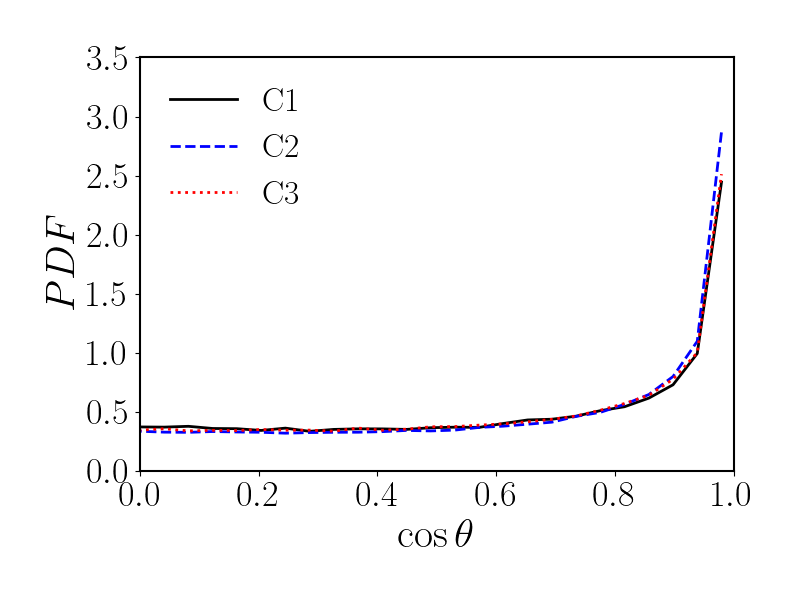}\\
    \small (b)
  \end{tabular}
  \begin{tabular}[b]{c}
    \hspace{-0.1in}
\includegraphics[width=.5\linewidth]{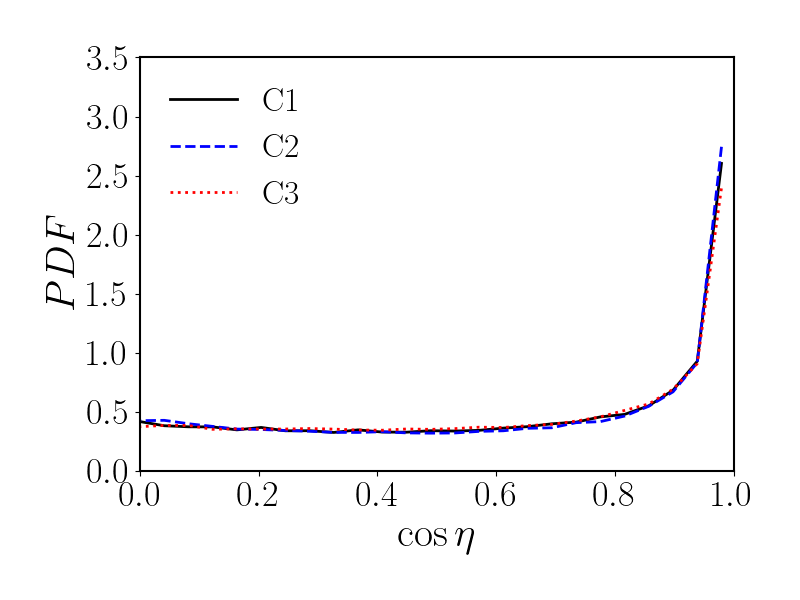} \\
    \small (c)
  \end{tabular} \qquad
   \caption{PDFs of cosine of Euler angles between the eigenvectors of $\boldsymbol{\widehat{\varsigma_{s}}}$ tensor with the eigenvectors of $\boldsymbol{\widehat{H^{'}}}$  (\ref{eq:nondimhijp}) tensor using DNS database of simulation cases C1-C3 included in Table \ref{tab:DNS_cases}: \textbf{(a) $\cos{\Xi}$, (b) $\cos{\theta}$, (c) $\cos{\eta}$}.}
    \label{fig:dns_Euler_angles_h_varsigma_s}
\end{figure}

In figure \ref{fig:dns_Euler_angles_h_varsigma_s}, we present the PDFs of the cosine of three Euler angles: $\Xi$, $\theta$, and $\eta$ (\ref{eq:cosine}) using the DNS data of all three cases C1-C3 (as given in Table.\ref{tab:DNS_cases}
).
The PDF statistics of the three Euler angles reveal consistent trends across different datasets (C1-C3), as reflected in the probability density functions (PDFs) shown in figure~\ref{fig:dns_Euler_angles_h_varsigma_s}. 
Peaks of the PDFs near unity indicate a preferential parallel or anti-parallel alignment, while peaks near zero correspond to orthogonality between the eigenvectors. Importantly, these alignment characteristics remain nearly identical across all three simulations (C1-C3), despite differences in the initial Mach and Reynolds numbers. This invariance suggests a degree of universality in the alignment between the $\boldsymbol{\widehat{\varsigma_{s}}} $ and $\boldsymbol{\widehat{H^{'}}}$  (\ref{eq:nondimhijp}) tensors, indicating that their mutual orientation is largely unaffected by variations in the initial turbulent Mach number and Reynolds number. Therefore, $\boldsymbol{\widehat{H^{'}}}$  (\ref{eq:nondimhijp}) can serve as the symmetric tensor employed in constructing the tensor bases set (\ref{eq:tensor_basis_general}).

\begin{figure}
   \centering
     \begin{tabular}[b]{c}
\hspace{-0.5in}    
\includegraphics[width=.5\linewidth]{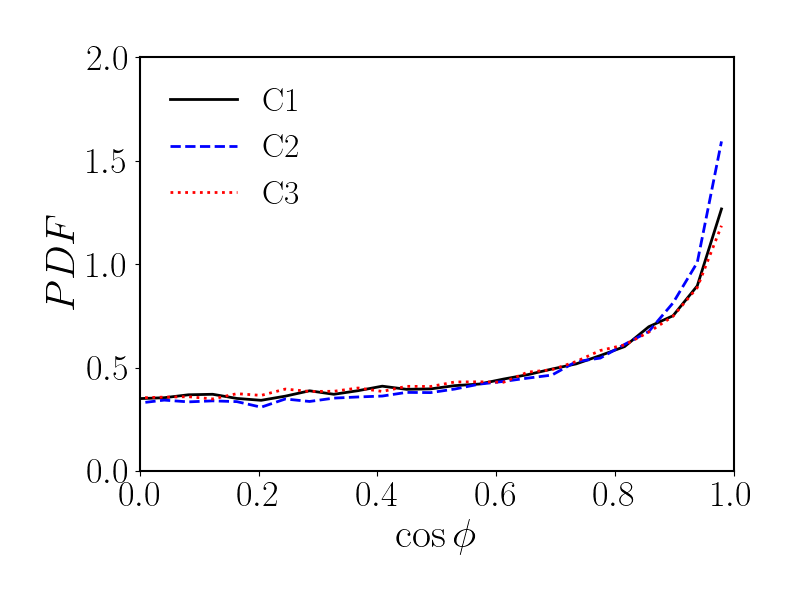} \\
  \end{tabular} \qquad
\caption{PDFs of cosine of angles between the locally defined vectors $\bf{\zeta}$ (\ref{eq:axial_varsigma}) and $\boldsymbol{\omega}$ (\ref{eq:vorticity}) computed using the DNS database of simulation cases C1-C3 included in Table \ref{tab:DNS_cases}.}
    \label{fig:dns_eigendirection_omega_zeta}
\end{figure}

Figure~\ref{fig:dns_eigendirection_omega_zeta} shows the PDFs of the direction cosines between $\boldsymbol{\omega}$ (\ref{eq:vorticity})  and $\boldsymbol{\zeta}$ (\ref{eq:axial_varsigma}) across simulations C1-C3.
Our analysis reveals that (i) the vorticity vector exhibits clear alignment preferences with the $\boldsymbol{\zeta}$ vector, as indicated by a distinct peak at one in the corresponding PDF distributions, and (ii) the PDF distributions remain consistent across all simulations C1--C3. 
This consistency suggests a universal tendency in the alignment behavior between the vorticity vector and the $\boldsymbol{\zeta}$ vector in flow fields with different initial turbulent Mach numbers and Reynolds numbers. 
\newpage
The tensor bases $\mathbf{T}^{(n)}$  and invariants $\lambda_{i}$ are constructed as functions of the tensors $\boldsymbol{\widehat{H}'}$ and $\boldsymbol{W}$  (\ref{eq:vorticity}). The explicit forms of $\mathbf{T}^{(n)}$ are defined as follows:

 \begin{equation}
\begin{aligned}
\label{eq:tensor_basis}
\textbf{T}^{(1)}=\boldsymbol{\widehat{H^{'}}W}, & \hspace{2cm} 
\textbf{T}^{(2)}=\boldsymbol{W\widehat{H^{'}}}^{2},\\
\textbf{T}^{(3)}=\boldsymbol{W^{2}\widehat{H^{'}}}, & \hspace{2cm}
\textbf{T}^{(4)}=\boldsymbol{W\widehat{H^{'}}W^{2}},\\
\textbf{T}^{(5)}= \boldsymbol{\widehat{H^{'}}}\boldsymbol{W}\boldsymbol{\widehat{H^{'}}}^{2},
& \hspace{2cm} \textbf{T}^{(6)}=\boldsymbol{W^{2}\widehat{H^{'}}^{2}},\\
\textbf{T}^{(7)}=\boldsymbol{W}\boldsymbol{\widehat{H^{'}}^{2}}\boldsymbol{W^{2}}, & \hspace{2cm} 
\textbf{T}^{(8)}=\boldsymbol{\widehat{H^{'}}}\boldsymbol{W^{2}}\boldsymbol{\widehat{H^{'}}^{2}}.\\
\end{aligned}
\end{equation}
Similarly, the invariants $\lambda_{i}s$ are defined as:
\begin{equation}
\begin{aligned}
\label{eq:invariants_HW}
\lambda_{1}=Tr(\boldsymbol{W}^{2}),& \hspace{0.5cm}
\lambda_{2}=Tr(\boldsymbol{\widehat{H^{'}}}),  & \hspace{0.5cm}
\lambda_{3}=Tr(\boldsymbol{W^{2}\widehat{H^{'}}}),& \hspace{0.5cm}
\lambda_{4}=Tr(\boldsymbol{W^{2}\widehat{H^{'}}^{2}}).& \hspace{0.5cm}
\end{aligned}
\end{equation}

Next, to incorporate the property that the $\boldsymbol{\widehat{\varsigma}}$ tensor vanishes as the flow approaches the incompressible limit, we design the invariant layers accordingly. This construction ensures that the primary inputs to the neural network can capture variations in local compressibility limits. Earlier, \cite{suman2013velocity} identified the most influential local parameters of compressibility as:  
(i) the normalized dilatation rate, denoted by $a_{ii}$ (\ref{eq:delta_aii}), and  
(ii) the normalized rate of change of the dilatation rate, denoted by $\delta$ (\ref{eq:delta_aii}).  
Both these local parameters ($a_{ii}~\&~\delta$) become zero in regions of strictly incompressible flow. Furthermore, we modify the proposed list of scalar invariants (\ref{eq:invariants_HW}) such that all invariants approach zero in the incompressible limit.  Consequently, the final set of invariants is defined as follows:
\begin{equation}
\begin{aligned}
\label{eq:final_invariants}
\lambda_{1}=Tr(\boldsymbol{a})Tr(\boldsymbol{W}^{2}),& \hspace{0.5cm}
\lambda_{2}=Tr(\boldsymbol{\widehat{H^{'}}}),  & \hspace{0.5cm}
\lambda_{3}=Tr(\boldsymbol{a})Tr(\boldsymbol{W}^{2}\boldsymbol{\widehat{H^{'}}}),\\
\lambda_{4}=Tr(\boldsymbol{a})Tr(\boldsymbol{W}^{2}\boldsymbol{\widehat{H^{'}}}^{2}),& \hspace{0.5cm}
\lambda_{5}=Tr(\boldsymbol{a}),& \hspace{0.5cm}
\lambda_{6}=\delta.
\end{aligned}
\end{equation}
In incompressible flow regions, $a_{ii}=0$, $\hat{H^{'}_{ii}}=0$ and $\delta=0$, which consequently brings all invariants to zero, leads to $\boldsymbol{\widehat{\varsigma}} \rightarrow 0$ which in turns leads to $B_{ij} \rightarrow 0$ .
The notional mapping of the tensor bases and invariants with the $\boldsymbol{\widehat{\varsigma}}$ tensor, as implemented in the TBNN architecture, is expressed as: 
\begin{equation}    \label{eq:model_varsigmahat}  
\boldsymbol{\widehat{\varsigma}}^{IBBM}=\sum_{n=1}^{8}{[g_{ibbm}^{(n)}(\lambda_{1},....\lambda_{6})]\textbf{T}^{(n)}},
\end{equation}
here, $\boldsymbol{\widehat{\varsigma}}^{IBBM}$ denotes the closure for the $\boldsymbol{\widehat{\varsigma}}$ tensor. In (\ref{eq:model_varsigmahat}), the coefficients (denoted by $g_{ibbm}^{n}$) are the non linear functions of six invariants (\ref{eq:final_invariants}). These coefficients are computed using a deep neural network.

\begin{figure}
    \centering
    \includegraphics[width=0.6\linewidth]{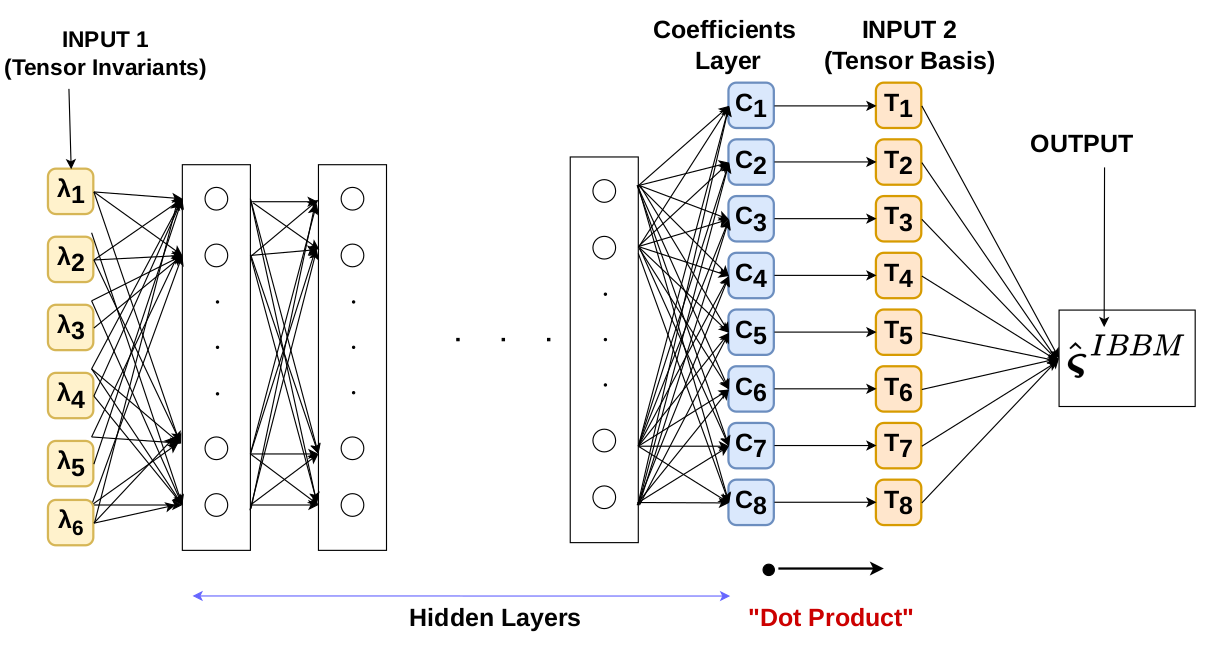}
    \caption{\label{fig:TBNN}Schematic of $\boldsymbol{\widehat{\varsigma}^{IBBM}}$ predicting neural network architecture (\ref{eq:model_varsigmahat}).}    
\end{figure}

Figure~\ref{fig:TBNN}, illustrates the architecture of the neural network developed to predict the tensor $\boldsymbol{\widehat{\varsigma}}$.  
The model consists of two input layers.  
The first layer has six nodes, which receive the scalar invariants $\lambda_i$ (\ref{eq:final_invariants}) as inputs.  
This layer is followed by five hidden layers with [50, 120, 120, 50, 8] neurons, respectively. The eight neurons in the final hidden layer are interpreted as scalar coefficients associated with the eight tensor bases elements (\ref{eq:tensor_basis}). The second input layer has eight neurons, each corresponding to one of the tensor bases elements $\mathbf{T}^{(n)}$, where the tensor bases ($\mathbf{T}^{(n)}$) are defined in (\ref{eq:tensor_basis}).  
The final output of the network is constructed by forming a weighted sum of these basis tensors, with the eight coefficients generated in the last hidden layer. This operation yields a nine-component tensor (corresponding to the components of a second-order ($3 \times 3$) tensor), which represents the nine components of the predicted output $\boldsymbol{\widehat{\varsigma}}^{IBBM}$.

The training efficiency of a neural network is highly influenced by the design of its loss function \citep{shikha2023augmented}. A well-defined loss function enables the network to capture the desired physical characteristics embedded within the input data. In a recent study \cite{shikha2023augmented}, an orientation-based loss function was introduced for modeling second-order tensors. Their approach demonstrated improved training convergence and predictive accuracy compared to traditional loss functions, which are designed based on the component-wise mean errors of the target tensor. In the present study, we used a similar custom loss function inspired by the directional alignment tendencies of the $\boldsymbol{\widehat{\varsigma}}$ tensor with respect to the input tensors  $\boldsymbol{\widehat{H^{'}}}$  (\ref{eq:nondimhijp}) and  $\boldsymbol{W}$  (\ref{eq:vorticity}). Using the values of the cosine of angles (\ref{eq:cosine}), we formulate an optimal loss function ($J_1$), which drives the neural network-based predicted tensor $\boldsymbol{\widehat{\varsigma}}^{IBBM}$ to align its orientation with that of the DNS-based $\boldsymbol{\widehat{\varsigma}}$ tensor.
\begin{equation}
\begin{aligned}
\label{eq:TBNN_Loss}
J_{1}= L_{1}+L_{2}+L_{3}+L_{4},\\
L_{1}= \frac{ \sum_{j=1}^{m}(|\cos{\Xi}_{DNS}|  - |\cos{\Xi}_{model}| )_{j}^{2}}{ \sum_{j=1}^{m}|\cos{\Xi}_{DNS}|_{j}^{2}},\\
L_{2}=  \frac{ \sum_{j=1}^{m}(|\cos{\theta}_{DNS}|  - |\cos{\theta}_{model}| )_{j}^{2}}{ \sum_{j=1}^{m}|\cos{\theta}_{DNS}|_{j}^{2}},\\
L_{3}= \frac{ \sum_{j=1}^{m}(|\cos{\eta}_{DNS}|  - |\cos{\eta}_{model}| )_{j}^{2}}{ \sum_{j=1}^{m}|\cos{\eta}_{DNS}|_{j}^{2}},\\
L_{4}= \frac{ \sum_{j=1}^{m}(|\cos{\phi}_{DNS}|  - |\cos{\phi}_{model}| )_{j}^{2}}{ \sum_{j=1}^{m}|\cos{\phi}_{DNS}|_{j}^{2}},\\
\end{aligned}
\end{equation}
where $m$ denotes the number of samples used to compute the loss value $J_{1}$. The loss $J_{1}$ is minimized during the training of the neural network using the Adamax optimizer \citep{goodfellow2016deep}.

The network is trained on 1,00,000 data points extracted from the DNS database, while an additional 20,000 data points are reserved for validation. Both training and validation sets are randomly sampled from a fixed time snapshot of the DNS simulation case C1 (see Table~\ref{tab:DNS_cases}). The batch size and learning rate are tuned via Bayesian optimization \citep{snoek2012practical}, yielding optimal values of 1,00,000 and $1 \times 10^{-2}$, respectively. The model is trained for 1000 epochs, with the network state (weights and biases) saved every 50 epochs. Evaluation of the saved states indicates that the model trained for 650 epochs provides the best predictive performance. Accordingly, the network state at the 650\textsuperscript{th} epoch is adopted for the final prediction of the tensor $\boldsymbol{\widehat{\varsigma}}^{IBBM}$.

\subsection{Model of \texorpdfstring{$\Psi$}{Psi}}

To model the magnitude of the $\boldsymbol{\varsigma}$ tensor, denoted by $\Psi$ (\ref{eq:magnitude}), we employ a separate deep neural network (DNN) framework. In order to ensure rotational invariance of the model output, all input quantities are chosen as scalar invariants of the coordinate system. Prior to defining the input set, we conducted a detailed DNS-based investigation of the characteristic variations of $\Psi$ conditioned on various flow parameters. 

\begin{figure}
   \centering
     \begin{tabular}[b]{c}
\hspace{-0.2in}    
\includegraphics[width=.5\linewidth]{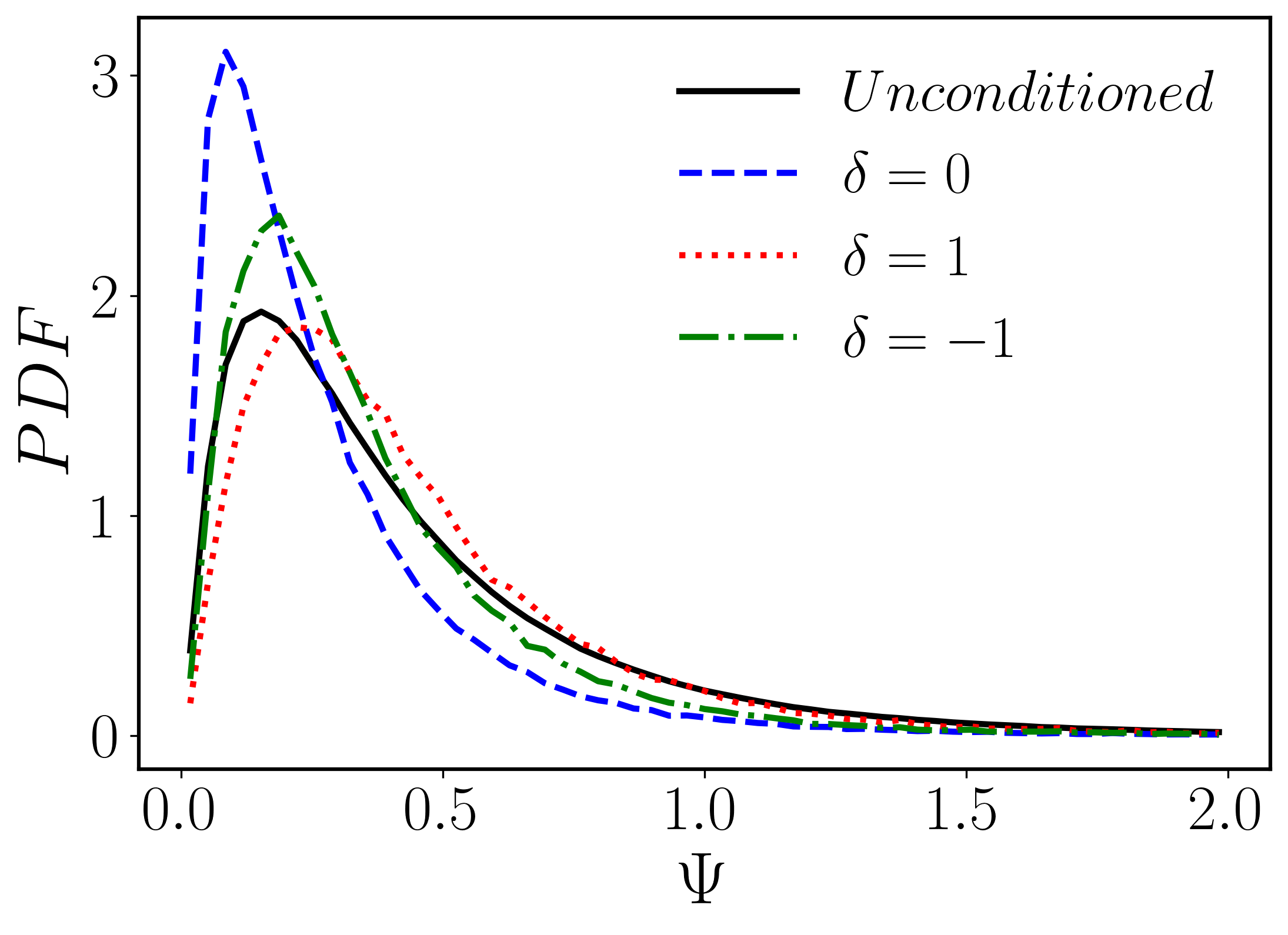} \\
    \small (a)
  \end{tabular} \qquad
  \begin{tabular}[b]{c}
  \hspace{-0.2in}
\includegraphics[width=.5\linewidth]{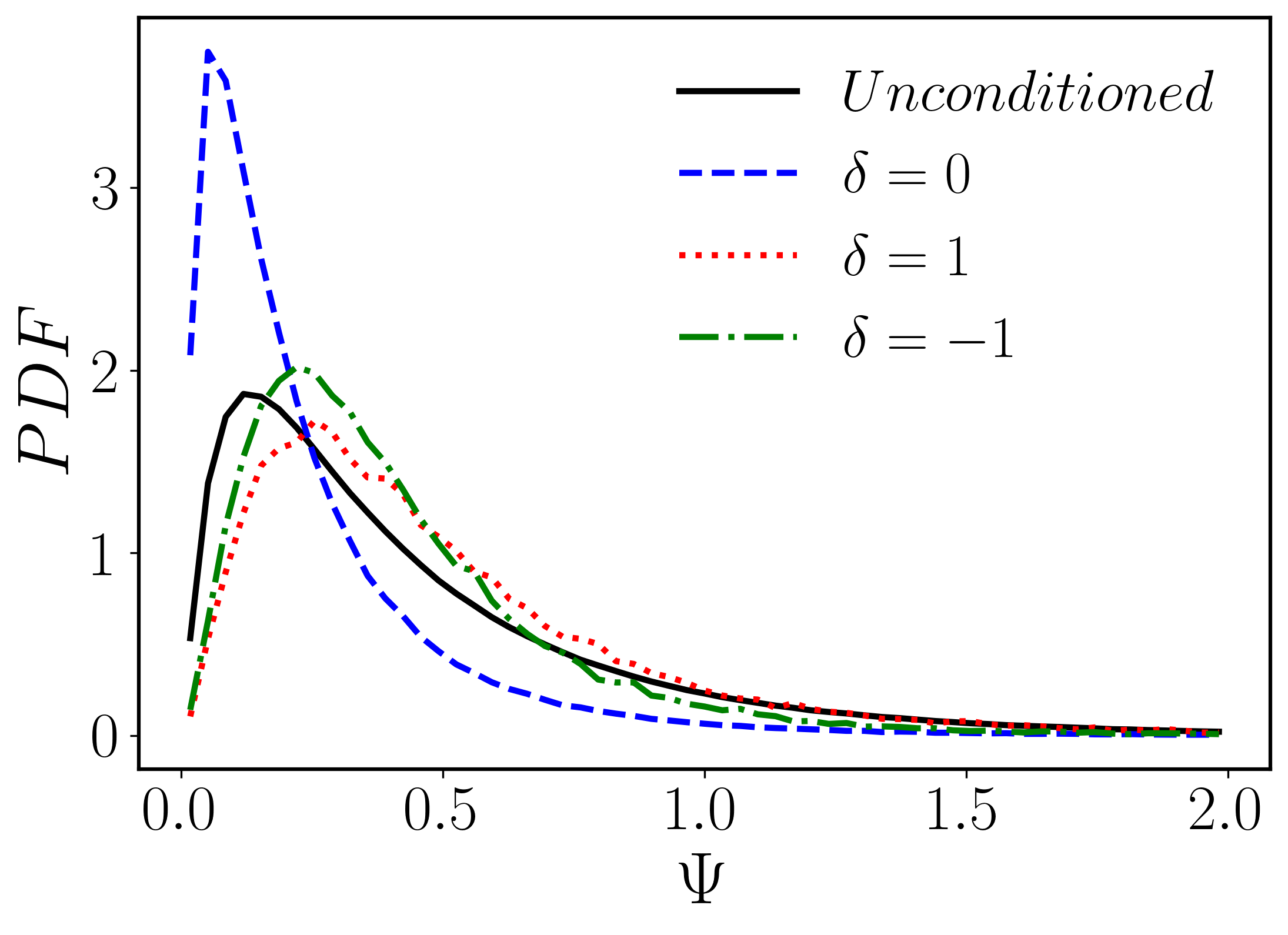} \\
    \small (b) 
    \end{tabular} \qquad
      \begin{tabular}[b]{c}
  \hspace{-0.2in}
\includegraphics[width=.5\linewidth]{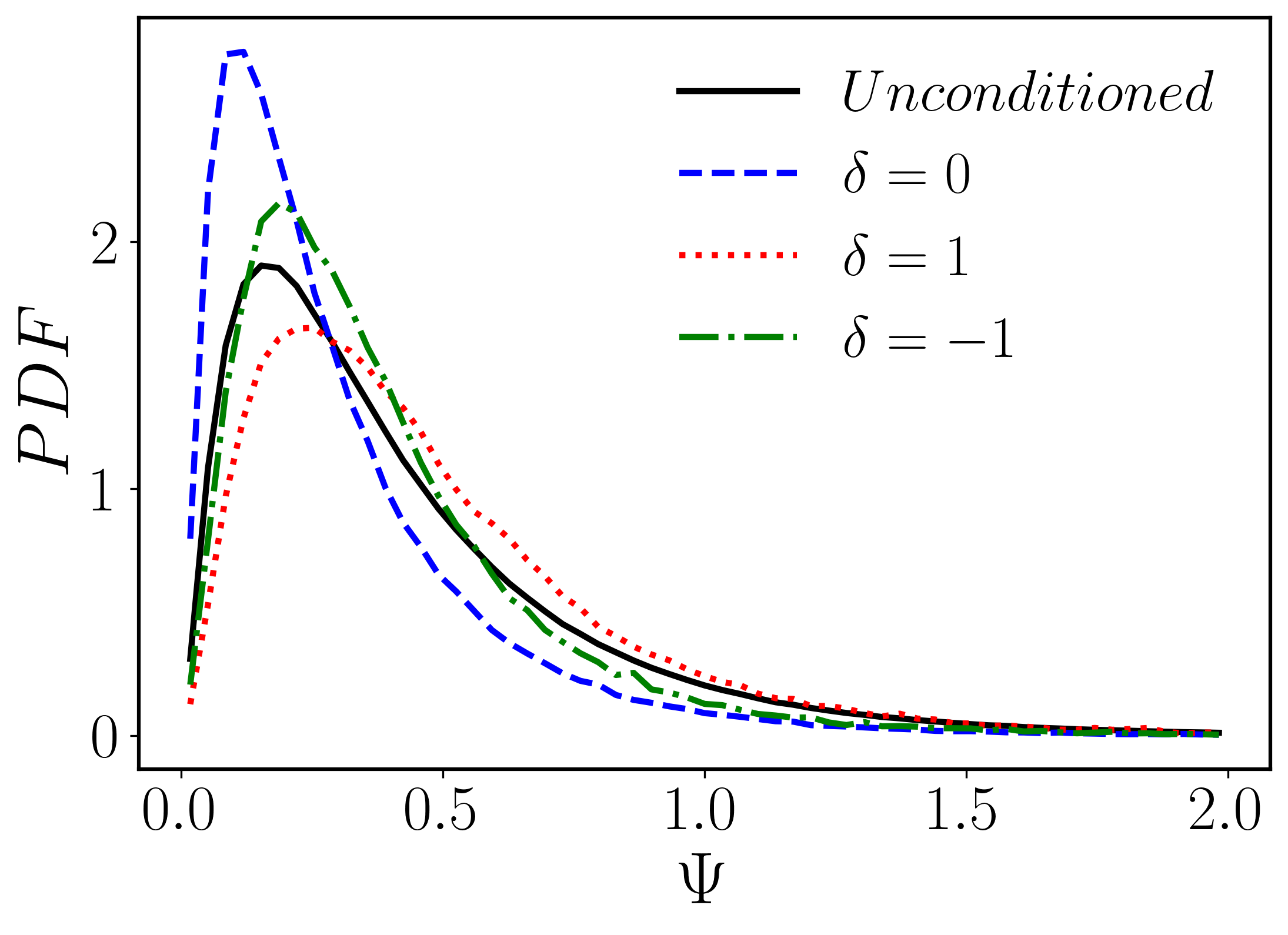} \\
    \small (c) 
    \end{tabular} \qquad
\caption{PDFs of $\Psi$ (\ref{eq:magnitude}) conditioned on $\delta$ computed using the DNS database of simulation cases: (a) C1, (b) C2, (c) C3. }
    \label{fig:delta_condition}
\end{figure}

Following the approach of~\cite{suman2013velocity}, we investigate the variation of $\Psi$ by examining its probability density functions (PDFs) conditioned on two local compressibility parameters: (i) $a_{ii}$ (\ref{eq:delta_aii}), and (ii) $\delta$ (\ref{eq:delta_aii}). 
To examine the influence of $\delta$ on $\Psi$, we evaluate its PDFs using the DNS datasets under three distinct $\delta$-conditioned datasets: (i) $\delta = 0 \pm 0.5$, (ii) $\delta = 1 \pm 0.05$, and (iii) $\delta = -1 \pm 0.05$. Figures~\ref{fig:delta_condition} (a), (b), and (c) display the resulting PDFs of $\Psi$ obtained from these $\delta$ conditioned datasets for simulations C1, C2, and C3, respectively (see Table~\ref{tab:DNS_cases}).
From Figs.~\ref{fig:delta_condition}(a)–(c), we observe that the $\delta$-conditioned PDFs differ significantly from the unconditional PDFs. Moreover, the variation trends of the unconditional PDFs and the corresponding $\delta$-conditioned PDFs remain consistent across all simulation cases. These observations clearly indicate that the local value of $\delta$ has a direct influence on the statistics of $\Psi$ (\ref{eq:magnitude}).

\begin{figure}
   \centering
     \begin{tabular}[b]{c}
\hspace{-0.2in}    
\includegraphics[width=.5\linewidth]{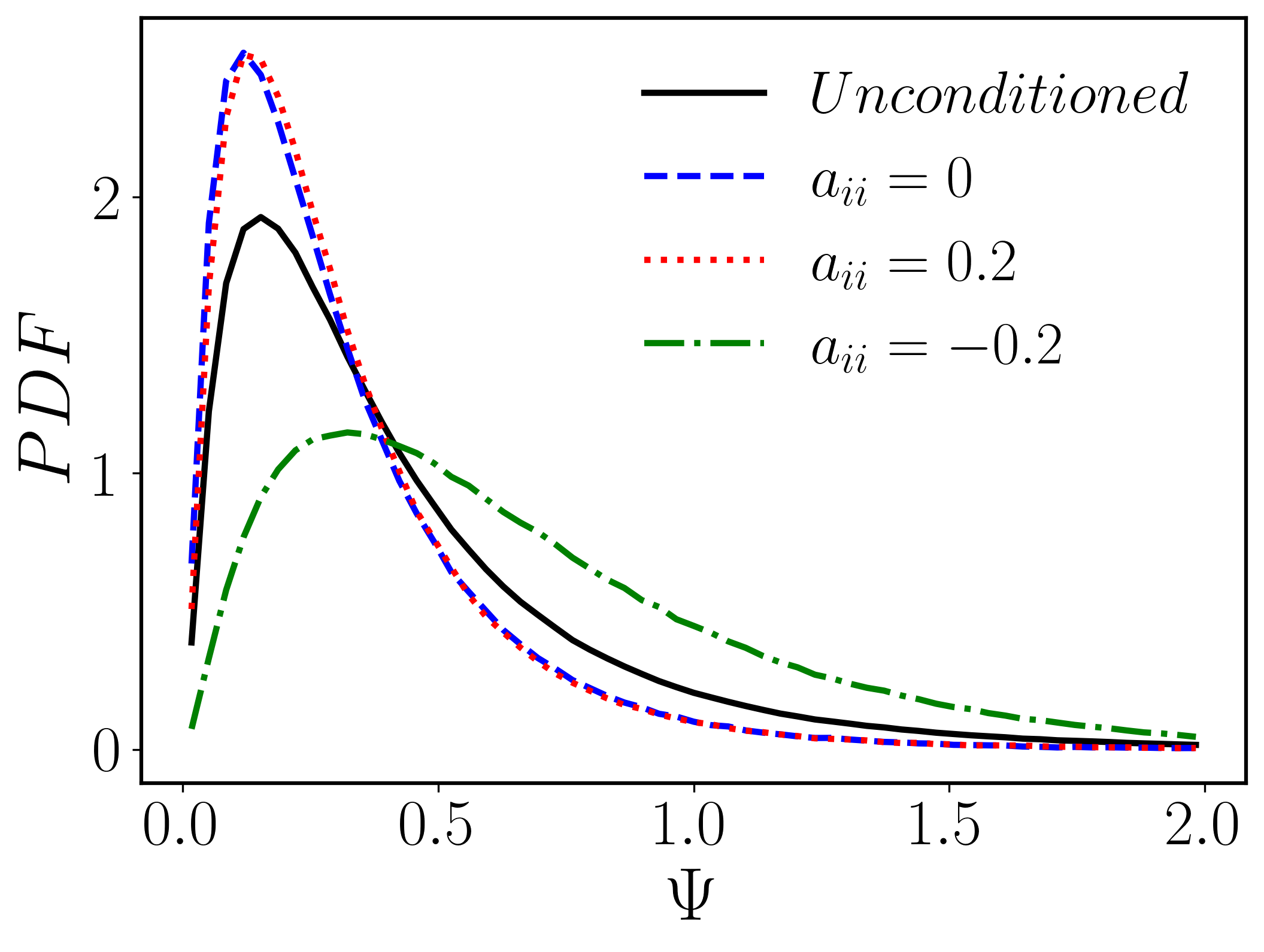} \\
    \small (a)
  \end{tabular} \qquad
  \begin{tabular}[b]{c}
  \hspace{-0.2in}
\includegraphics[width=.5\linewidth]{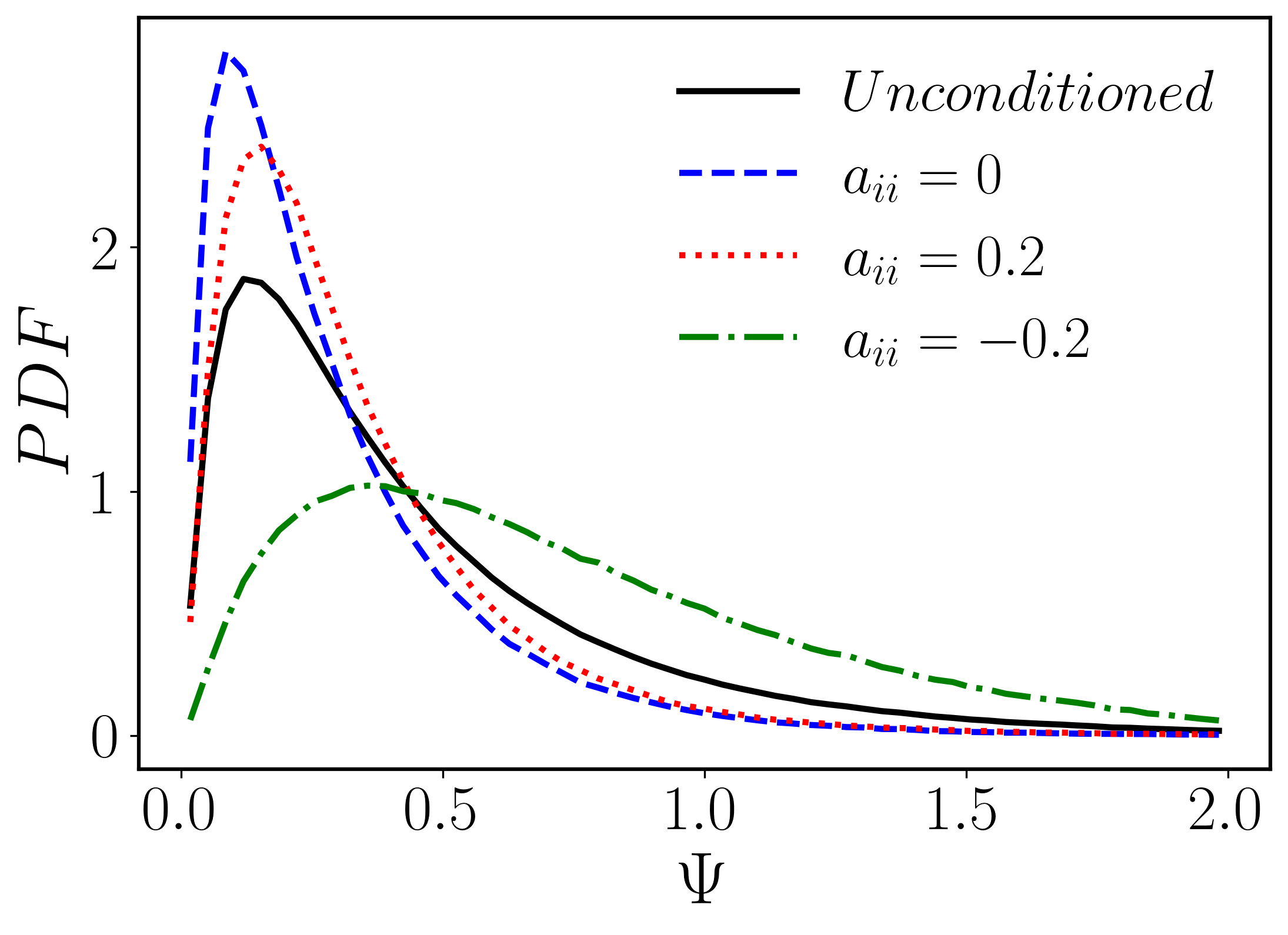} \\
    \small (b) 
    \end{tabular} \qquad
      \begin{tabular}[b]{c}
  \hspace{-0.2in}
\includegraphics[width=.5\linewidth]{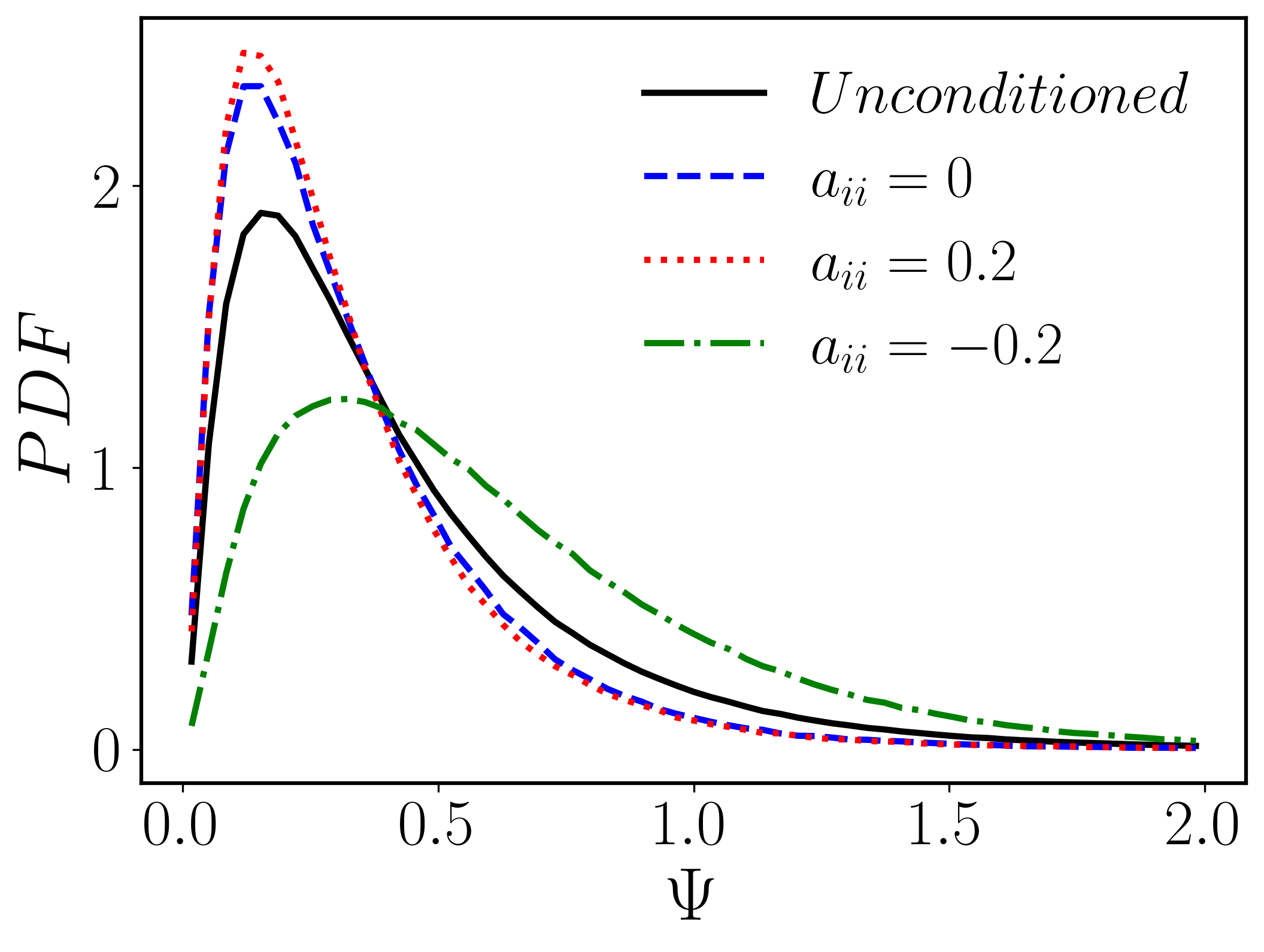} \\
    \small (c) 
    \end{tabular} \qquad
\caption{PDFs of $\Psi$ (\ref{eq:magnitude}) conditioned on $a_{ii}$ computed using the DNS database of simulation cases: (a) C1, (b) C2, (c) C3. }
    \label{fig:aii_condition}
\end{figure}

We next examine the dependence of $\Psi$ on the normalized dilation rate ($a_{ii}$) by computing its PDFs from DNS datasets conditioned on three distinct ranges: $a_{ii} = 0 \pm 0.05$, $a_{ii} = -0.2 \pm 0.05$, and $a_{ii} = 0.2 \pm 0.05$. Figure~\ref{fig:aii_condition} presents the resulting PDFs of $\Psi$ (\ref{eq:magnitude}) conditioned on $a_{ii}$ using the DNS datasets from the simulations cases C1-C3 (see Table~\ref{tab:DNS_cases}).
We observe that the PDFs of $\Psi$ are strongly influenced by the local values of $a_{ii}$. 
Furthermore, the trends in the PDFs of $\Psi$, conditioned on specific values of $a_{ii}$, remain consistent across all simulation cases (C1–C3), as shown in Figs.~\ref{fig:aii_condition}(a)–(c).

These results highlight the importance of including $a_{ii}$ and $\delta$ as input features when developing a DNN-based model for $\Psi$. Since both quantities are ultimately derived from the velocity field and its gradients, it is also necessary to account for pressure-related effects (\ref{eq:nondimenV}). In particular, incorporating an appropriate representation of the tensor $\boldsymbol{\widehat{H^{'}}}$  (\ref{eq:nondimhijp}) into the input set is crucial for improving the predictive capability of the model for $\Psi$. 

To model $\Psi$, which is a scalar quantity, we need a scalar representation of the tensor $\boldsymbol{\widehat{H^{'}}}$  (\ref{eq:nondimhijp}). 
To preserve the rotational invariance of the model predictions, the input set needs to be in the form of the scalar invariants of $\boldsymbol{\widehat{H^{'}}}$  (\ref{eq:nondimhijp}). A similar strategy has been employed in the study of~\cite{shikha2024modeling}, where the magnitude of the vibrational non-equilibrium tensor was modeled using the first and third principal invariants of the self-normalized form of symmetric pressure Hessian tensor ($\boldsymbol{H}$) as input quantities. 
Following this approach, we include the invariants of $\boldsymbol{\widehat{H^{'}}}$  (\ref{eq:nondimhijp}) as the input set for modeling $\Psi$ (\ref{eq:magnitude}). 
The three invariants ($I_1,\, I_2,\, I_3$) of $\boldsymbol{\widehat{H^{'}}}$  (\ref{eq:nondimhijp}) are defined as:
\begin{subequations}
\begin{equation}
\label{eq:I1}
I_{1} \equiv Tr[\boldsymbol{\widehat{H^{'}}}],
\end{equation} 
\begin{equation}
\label{eq:I2}
I_{2} \equiv \frac{1}{2}(I_{1}^{2}-Tr[\boldsymbol{\widehat{H^{'}}}^2]),
\end{equation} 
\begin{equation}
\label{eq:I3}
I_{3} \equiv \frac{1}{3}(-I_{1}^{3}+3I_{1}I_{2}-Tr[\boldsymbol{\widehat{H^{'}}}^{3}]),
\end{equation}
\end{subequations}
where $\operatorname{Tr}$ denotes the trace of the tensor. For a self-normalized tensor, $I_2$ is not an independent invariant, since  
\[
I_{2} \equiv \tfrac{1}{2}\left(I_{1}^{2}-1\right).
\]  
Therefore, only $I_1$ and $I_3$ are included in the set of input quantities for the modeling of $\Psi$.
\begin{figure}
   \centering
     \begin{tabular}[b]{c}
\hspace{-0.2in}    
\includegraphics[width=.4\linewidth]{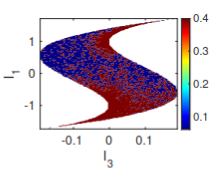} \\
    \small (a)
  \end{tabular} \qquad
  \begin{tabular}[b]{c}
  \hspace{-0.2in}
\includegraphics[width=.4\linewidth]{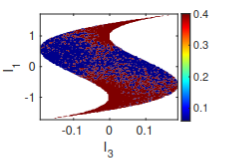} \\
    \small (b) 
    \end{tabular} \qquad
      \begin{tabular}[b]{c}
  \hspace{-0.2in}
\includegraphics[width=.4\linewidth]{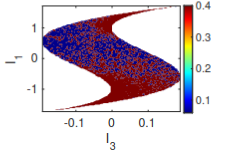} \\
    \small (c) 
    \end{tabular} \qquad
\caption{PDFs of $\Psi$ conditioned on $I_{1}$ and $I_{3}$ computed using the DNS database of simulation cases: (a) C1, (b) C2, (c) C3. }
\label{fig:I1_I3_contours}
\end{figure}
The choice of $I_1$ and $I_3$ as input quantities is further supported by DNS-based investigations. Specifically, we examine the distribution of $\Psi$ conditioned on local values of $I_{1}$ and $I_{3}$, using DNS data from all simulation cases C1–C3 (Table~\ref{tab:DNS_cases}). The resulting contours of $\Psi$, conditioned on $I_{1}$ and $I_{3}$, are shown in figure~\ref{fig:I1_I3_contours} for these simulations.

A comparison of the contour distributions in figure~\ref{fig:I1_I3_contours} across simulations C1–C3 reveals several key features. In all cases, the higher values of $\Psi$ are concentrated in regions with relatively large magnitudes of $I_{1}$ and $I_{3}$, as indicated by the clustering of red colors at the edges of the contours. However, some differences are visible in the middle regions ($I_{1} \in [-1,1]$). For example, in C3, there are comparatively more red spots in the intermediate regions, while C2 shows the fewest, indicating low values of $\Psi$ (\ref{eq:magnitude}). Despite these differences, the overall distributional trends remain consistent across all simulations, reflecting the dependence of $\Psi$ on $I_{1}$ and $I_{3}$. Based on these observations, we propose the following notional mapping for $\Psi^{IBBM}$:
\begin{equation}
    \label{eq:general_phi}
    \Psi^{IBBM} = \mathcal{G}(a_{ii}, \delta, I_{1}, I_{3}),
\end{equation}
where $\mathcal{G}$ denotes the nonlinear mapping obtained via a deep neural network and $\Psi^{IBBM}$ denotes the closure for the magnitude of $\boldsymbol{\varsigma}$ tensor .

\begin{figure}
\centering
\includegraphics[width=0.5\linewidth]{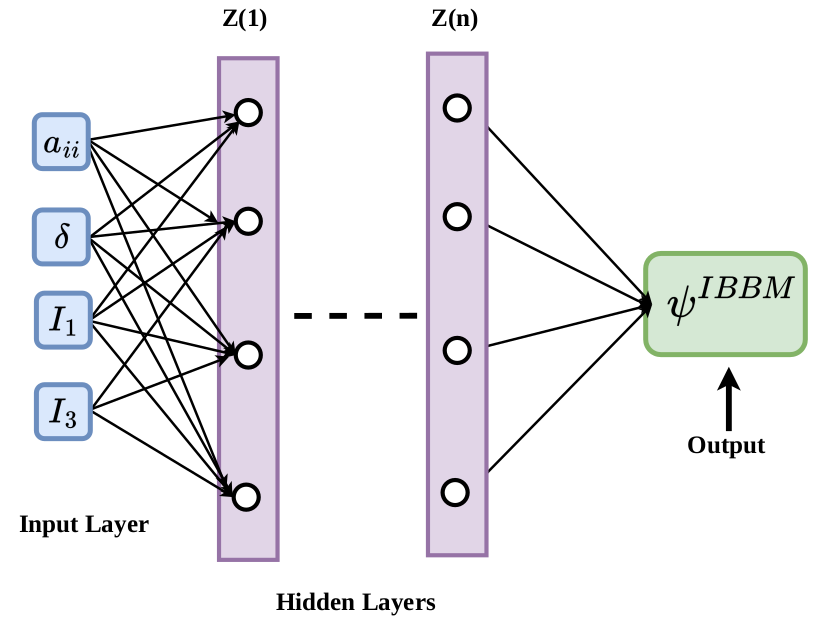}
\caption{\label{fig:Phi_DNN} Schematic of the neural network architecture for predicting $\Psi^{IBBM}$ (\ref{eq:magnitude}).}
\end{figure}

The schematic of the $\Psi^{IBBM}$ (\ref{eq:magnitude}) predicting neural network is shown in figure~\ref{fig:Phi_DNN}.
The neural network architecture consists of one input layer with four neurons corresponding to the input features 
$\left[a_{ii},~\delta,~I_{1},~I_{3}\right]$. This is followed by three hidden layers, each containing 50 neurons, 
and a single-node output layer that predicts $\Psi^{IBBM}$ (\ref{eq:magnitude}).

The network is trained by minimizing a custom loss function, $J_{2}$, defined as the relative mean square error (RMSE) 
between the DNS-computed and model-predicted values of $\Psi$ (\ref{eq:magnitude}):
\begin{equation}
\label{eq:15}
J_{2} = \frac{\sum_{j=1}^{m} \left( \Psi_{\mathrm{DNS},j} - \Psi_{\mathrm{model},j} \right)^{2}}
              {\sum_{j=1}^{m} \left( \Psi_{\mathrm{DNS},j} \right)^{2}},
\end{equation}
where $m$ denotes the number of training samples used in the loss evaluation.

The training dataset comprises $3 \times 10^{5}$ DNS samples obtained from the compressible decaying turbulence case C1 (Table~\ref{tab:DNS_cases}). At the end of each epoch, model performance is validated 
using $5\times 10^{4}$ randomly selected validation samples. Bayesian optimization identifies the optimal batch size 
and learning rate as $50,000$ and $1\times 10^{-3}$, respectively. The network is trained for 500 epochs, with weights 
and biases saved every 50 epochs. Testing across the saved states indicates that the best predictions are obtained at 
the $150^{\mathrm{th}}$ epoch, which is subsequently used for $\Psi$ (\ref{eq:magnitude}) prediction in the performance evaluations.

By combining the outputs of the two deep neural networks (\ref{eq:model_varsigmahat} and~\ref{eq:general_phi}), the model predicts the \(\boldsymbol{\varsigma}\) tensor. The combined architecture formed by these two networks is referred to as the \emph{invariants-based baroclinic mechanism} (IBBM) model. The IBBM prediction of the \(\boldsymbol{\varsigma}\) tensor is expressed as
\begin{equation}
    \label{eq:ibnm_varsigma}
    \varsigma^{\text{IBBM}}_{ij}
    = \widehat{\varsigma}^{\text{IBBM}}_{ij}\,
    \Psi^{\text{IBBM}} .
\end{equation}
Using the locally available tensors \(A_{ij}\) and \(P_{ij}=H_{ij}-B_{ij}\), the expression for \(V_{M_{ij}}\) is written as
\begin{equation}
    \label{eq:ibnm_VMij}
    V_{M_{ij}}
    = -\,\varsigma^{\text{IBBM}}_{ij}
    \sqrt{A_{pq}A_{pq}}\,
    \sqrt{P_{mn}P_{mn}} .
\end{equation}
With this closure, the evolution equation for the $\boldsymbol{B}$ tensor becomes
\begin{multline}  
\label{eq:Bij_modeled}
\frac{dB_{ij}}{dt}
= -A_{kj}B_{ik}
  -A_{ki}B_{kj}
  -(n-1)A_{kk}B_{ij}-B_{ij}\frac{c}{L}
  -\varsigma^{\text{IBBM}}_{ij}
   \sqrt{A_{pq}A_{pq}}\,
   \sqrt{P_{mn}P_{mn}}
 \\ -\frac{C_{pq}C_{pq}}{3\tau_{p}}\, B_{ij}.
\end{multline}
Before incorporating the IBBM closure into the full dynamical system, we first assess its predictive capability through an \emph{a priori} evaluation. In this assessment, the IBBM model is evaluated at fixed time instants using input tensors extracted directly from the DNS datasets listed in Table~\ref{tab:DNS_cases}.

\subsection{\label{subsec:a-priori}A-priori evaluation of the IBBM model}
In this section, we evaluate the performance of the IBBM model directly using $H_{ij},A_{ij}$ from a one-time instant DNS dataset. Using these $H_{ij},~ A_{ij}$ components, the IBBM model is used to obtain $\varsigma_{ij}^{IBBM}$ values. We use the DNS database case C1 (details given in Table \ref{tab:DNS_cases}). Following the decomposition of the $\boldsymbol{\varsigma}$ tensor (\ref{eq:nondimenV}), we assess the model performance in two distinct aspects: (i) the statistics of eigendirection alignments tendencies of the $\boldsymbol{\varsigma}$ tensor: represented as $\boldsymbol{\widehat{\varsigma}}$ (\ref{eq:varsigmahat}), and (ii) the statistics of its magnitude: represented as $\Psi$ (\ref{eq:magnitude}). To evaluate the model performance, we utilize 3,000,000 data points extracted directly from the DNS simulation of interest. The statistics of the $\boldsymbol{\varsigma}$ tensor, computed using the model predictions, are compared against the corresponding statistics obtained from the DNS datasets. For assessing the eigendirection alignment statistics, we further decompose the $\boldsymbol{\widehat{\varsigma}}$ tensor in its symmetric $\boldsymbol{\widehat{\varsigma_{s}}}$ and antisymmetric $\boldsymbol{\widehat{\varsigma_{a}}}$ parts (\ref{eq:varsigma_decomp}). We then compare the PDFs of the cosines of the three Euler angles (\ref{eq:cosine}) defined between the eigendirections of the $\boldsymbol{\widehat{\varsigma_{s}}}$ (\ref{eq:varsigma_decomp}) and $\boldsymbol{\widehat{H^{'}}}$  (\ref{eq:nondimhijp}) tensors. From the antisymmetric part of the $\boldsymbol{\widehat{\varsigma_{a}}}$ tensor, the local vector $\zeta$ (\ref{eq:axial_varsigma})  is extracted. The model performance is subsequently quantified by comparing the PDFs of the cosine of the angle between the locally computed vectors $\zeta$ (\ref{eq:axial_varsigma}) and $\boldsymbol{\omega}$ (\ref{eq:vorticity}): denoted as $\cos{\phi}$ (\ref{eq:cosine}).

In this process, the $\boldsymbol{\widehat{H^{'}}}$  (\ref{eq:nondimhijp}) tensor and $\boldsymbol{\omega}$ (\ref{eq:vorticity})  vector are obtained directly from the corresponding DNS datasets (as described in Table~\ref{tab:DNS_cases}). We denote the set of eigenvectors of the $\boldsymbol{\widehat{\varsigma_{s}}}$ tensor as $\boldsymbol{\hat{e}}_{\alpha,{\varsigma_{s}}}, \boldsymbol{\hat{e}}_{\beta,{\varsigma_{s}}}, \boldsymbol{\hat{e}}_{\gamma,{\varsigma_{s}}}$, where the corresponding eigenvalues are sorted such that $\alpha_{\varsigma_{s}} > \beta_{\varsigma_{s}} > \gamma_{\varsigma_{s}}$. Similarly, the eigenvectors of $\boldsymbol{\widehat{H^{'}}}$  (\ref{eq:nondimhijp}) are denoted as $\boldsymbol{\hat{e}}_{\alpha,{h}}, \boldsymbol{\hat{e}}_{\beta,{h}}, \boldsymbol{\hat{e}}_{\gamma,{h}}$, with eigenvalues ordered as $\alpha_{h} > \beta_{h} > \gamma_{h}$. Using the corresponding local eigendirections $\boldsymbol{\hat{e}}_{\alpha,{\varsigma_{s}}}, \boldsymbol{\hat{e}}_{\beta,{\varsigma_{s}}}, \boldsymbol{\hat{e}}_{\gamma,{\varsigma_{s}}}$ and $\boldsymbol{\hat{e}}_{\alpha,{h}}, \boldsymbol{\hat{e}}_{\beta_{h}}, \boldsymbol{\hat{e}}_{\gamma,{h}}$, we compute the PDFs of the cosines of three Euler angles: $\Xi,~ \theta,~ \eta$ (\ref{eq:cosine}). Once the reference statistics are obtained, the same procedure is repeated to compute the PDFs of the direction cosines between the eigendirections of the IBBM model-predicted $\boldsymbol{\widehat{\varsigma_{s}}}$ tensor and those of the $\boldsymbol{\widehat{H^{'}}}$  (\ref{eq:nondimhijp}) tensor sourced from the same DNS simulation used for model evaluation.

For the evaluation of the antisymmetric part of the $\boldsymbol{\widehat{\varsigma}}$ tensor, the local values of the corresponding axial vector $\zeta$ (\ref{eq:axial_varsigma}) and the vorticity vector $\boldsymbol{\omega}$ (\ref{eq:vorticity}) are obtained using the DNS datasets (Table\ref{tab:DNS_cases}). At each spatial location, the PDF of $\cos{\phi}$ (\ref{eq:cosine}) is calculated using the DNS data as the reference statistics. The same procedure is then applied to the $\zeta$ vector extracted from the IBBM model-predicted $\boldsymbol{\widehat{\varsigma}}$ tensor, and the resulting PDFs are compared against the DNS reference to assess the model performance. 

To assess the performance of the proposed model, we further evaluate it based on the magnitude statistics of the predicted $\boldsymbol{\widehat{\varsigma}}$ tensor. Specifically, we compare the PDFs of $\Psi$ (\ref{eq:magnitude}). In addition to this global evaluation, we further scrutinize the model by examining the conditional distributions of $\Psi$ (\ref{eq:magnitude}) with respect to $I_{1} \& I_{3}$ (\ref{eq:I1})-(\ref{eq:I3}). This two-level evaluation provides both an overall assessment of prediction accuracy and a detailed investigation of how the model captures the local distribution of $\Psi$ in $I_{1} ~\&~ I_{3}$ space.


The IBBM model is trained using DNS data from simulation C1 ($M_{t}=1.0$). Therefore, at this stage, we evaluate the model within the same flow configuration but at spatial locations different from those used in the training and validation of the neural networks.

\begin{figure}
\centering
  \begin{tabular}[b]{c}
\hspace{-0.3in}    
\includegraphics[width=.5\linewidth]{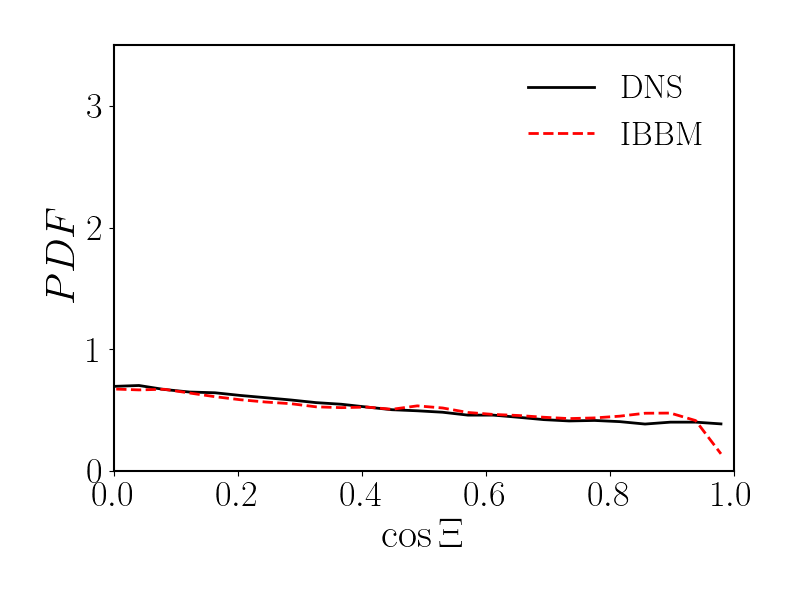} \\
    \small (a)
  \end{tabular} \qquad
  \begin{tabular}[b]{c}
  \hspace{-0.4in}
\includegraphics[width=.5\linewidth]{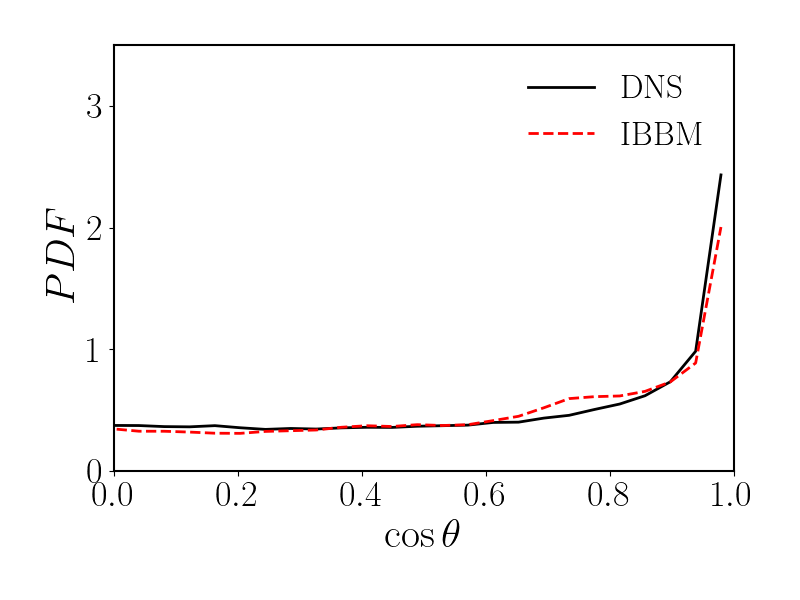}\\
    \small (b)
  \end{tabular}
  \begin{tabular}[b]{c}
    \hspace{-0.1in}
\includegraphics[width=.5\linewidth]{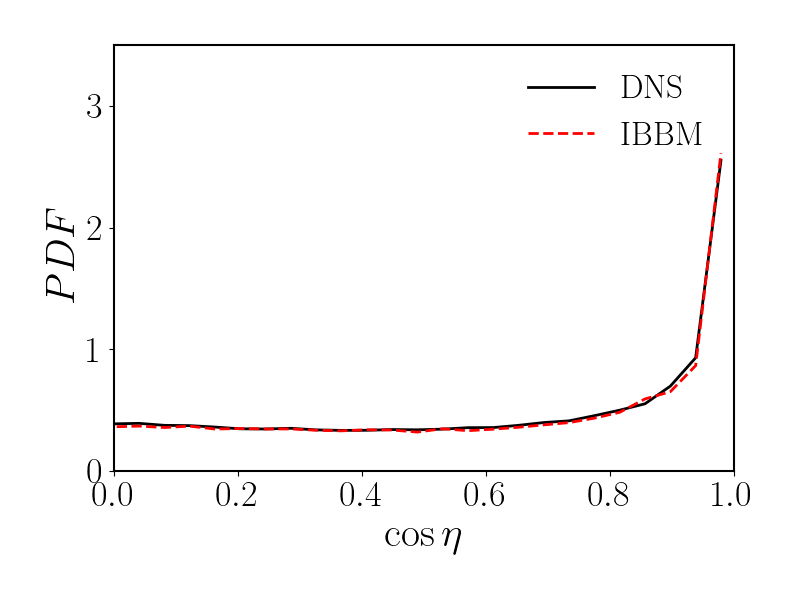} \\
    \small (c)
  \end{tabular} \qquad
    \caption{\label{fig:Mt1p0_align} PDFs of $\cos{\Xi},~\cos{\theta},~\cos{\eta}$ (\ref{eq:cosine}) using: (i) the DNS data of case C1 (details available in Table \ref{tab:DNS_cases}), (ii) the IBBM Model.}
\end{figure}

Figure~\ref{fig:Mt1p0_align} shows the PDFs of the cosine of three Euler angles: $\Xi,~ \theta, ~ \eta $ (\ref{eq:cosine}) defined between the eigendirections of the $\boldsymbol{\widehat{H^{'}}}$ tensor and those of the $\boldsymbol{\widehat{\varsigma_{s}}}$ (\ref{eq:varsigma_decomp}) tensor. 
Subfigure~\ref{fig:Mt1p0_align}(a) shows the comparison of the PDF of $\cos{\Xi}$ obtained from the DNS data of case C1, while Subfigure~\ref{fig:Mt1p0_align}(b) presents the corresponding comparison for $\cos{\theta}$, and Subfigure~\ref{fig:Mt1p0_align}(c) for $\cos{\eta}$. In each case, we also plot the PDFs predicted using the $\boldsymbol{\widehat{\varsigma}_{s}}$ tensor from the IBBM model. From these comparisons, we observe that, in the DNS data, the PDF of $\cos{\Xi}$ peaks at zero, whereas the PDFs of the other two Euler angles peak at one. The IBBM-based predictions successfully reproduce these features of the DNS-based PDFs and also capture their overall distributions. Hence, the IBBM model accurately reproduces the alignment characteristics between $\boldsymbol{\widehat{H}^{'}}$ and $\boldsymbol{\widehat{\varsigma}_{s}}$ tensors for the $M_t=1.0$ case.

\begin{figure}
  \centering
  \begin{tabular}[b]{c}
\hspace{-0.6in}     
\includegraphics[width=.5\linewidth]{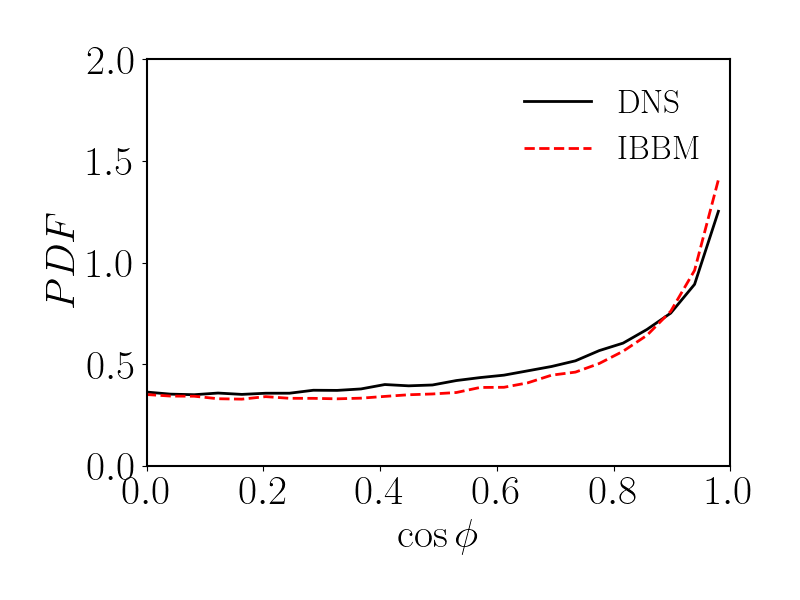} \\
     \end{tabular} \qquad
   \caption{\label{fig:vorticity_Mt1p0} PDFs of $\cos{\phi}$ (\ref{eq:cosine}) using: (i) the DNS data of case C1 (details available in Table \ref{tab:DNS_cases}), (ii) the IBBM Model.}
\end{figure}

Figure~\ref{fig:vorticity_Mt1p0} presents the PDF of $\cos{\phi}$ (\ref{eq:cosine}), where $\phi$ is the angle between the locally defined vectors $\zeta$ (\ref{eq:axial_varsigma}) and $\boldsymbol{\omega}$ (\ref{eq:vorticity}). Here, $\zeta$ represents the axial vector  extracted from the $\boldsymbol{\widehat{\varsigma_{a}}}$ (\ref{eq:varsigma_decomp}) tensor.
The PDF of $\cos{\phi}$ obtained using the DNS data of case C1 ($M_{t}=1$) exhibits a peak at one. The IBBM model not only captures this peak but also reproduces the overall PDF distribution observed in the DNS results. Overall, the model accurately represents the alignment tendencies between the locally defined vectors $\boldsymbol{\omega}$ and $\boldsymbol{\zeta}$ for most data points.

\begin{figure}
  \centering
  \begin{tabular}[b]{c}
\hspace{-0.6in}     
\includegraphics[width=.5\linewidth]{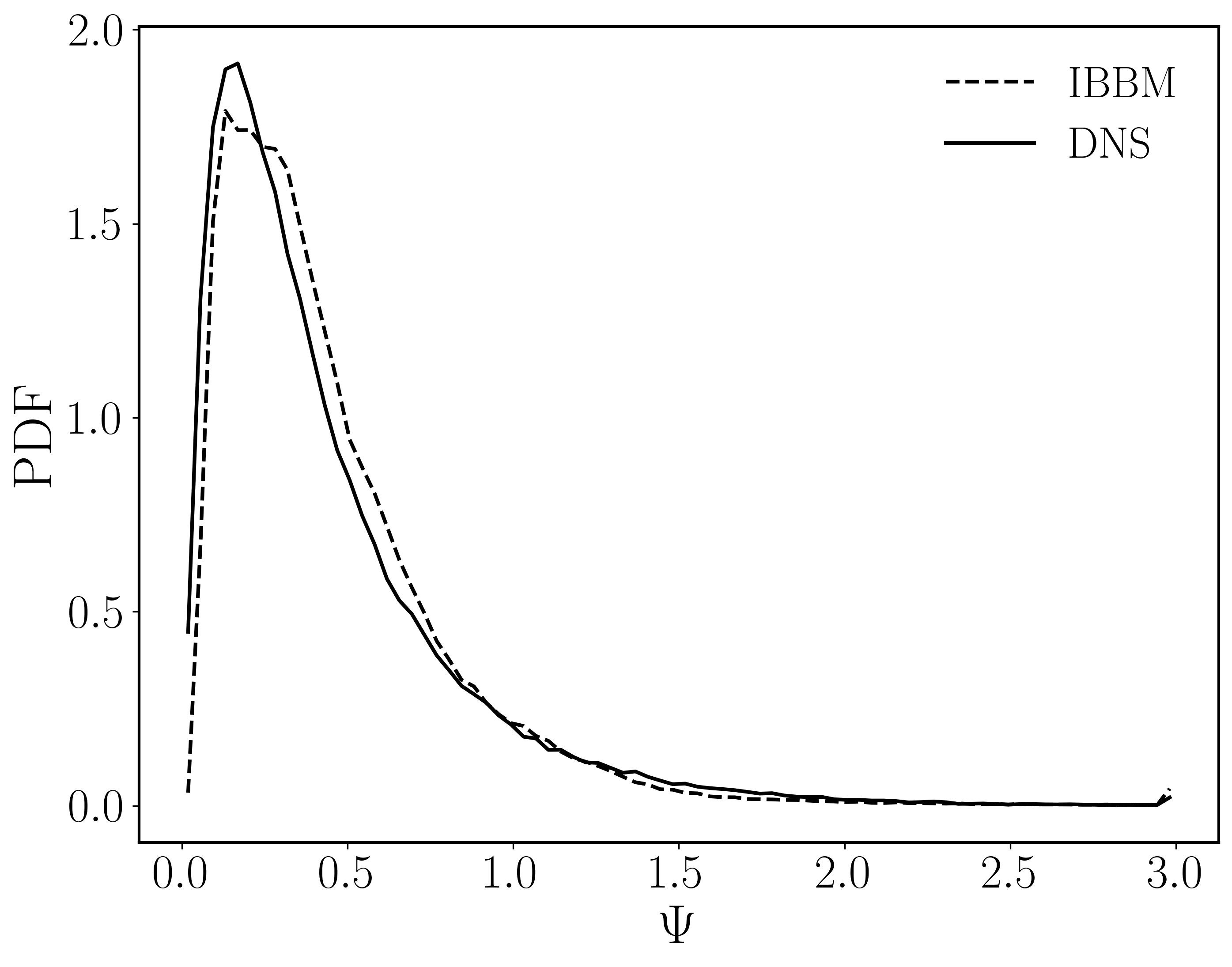} \\
     \end{tabular} \qquad
   \caption{\label{fig:Mt1p0_psi}  PDFs of the $\Psi$ (\ref{eq:magnitude}) using: (i) the DNS data of case C1 (details available in Table \ref{tab:DNS_cases}), (ii) the IBBM model.}
\end{figure}

Next, we evaluate the IBBM model based on the model’s capability to reproduce the magnitude of the $\boldsymbol{\varsigma}$ tensor. For this, we first evaluate the PDF of $\Psi$ (\ref{eq:magnitude}). Figure~\ref{fig:Mt1p0_psi} shows the PDF of $\Psi$ obtained using the DNS simulation case C1 ($M_t = 1$, table \ref{tab:DNS_cases}). For better comparison, the IBBM model predicted PDF is also plotted in the same figure.
The DNS distribution spans $(0,3)$, with the probability density increasing up to a pronounced peak at ($\Psi \approx 0.15$) and subsequently decreasing up to zero. The probability of $\Psi~\geq 1.5$ is nearly zero. This characteristic trend is closely reproduced by the IBBM model. Moreover, the predicted distribution captures both the peak location ($\Psi \approx 0.15 $) and the overall spread of the DNS curve, with only minor discrepancies in the peak magnitude.

\begin{figure}
\centering
  \begin{tabular}[b]{c}
\hspace{-0.3in}    
\includegraphics[width=.5\linewidth]{Figures/Rough_figure/Model_evauluation/mag/Mt1R250/DNS_I1I3.png} \\
    \small (a)
  \end{tabular} \qquad
  \begin{tabular}[b]{c}
  \hspace{-0.4in}
\includegraphics[width=.5\linewidth]{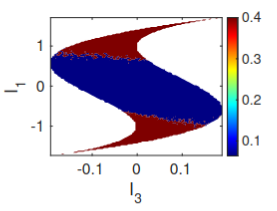}\\
    \small (b)
  \end{tabular}
    \caption{\label{fig:Mt0p1_psi_I23} The contours of $\Psi$ (\ref{eq:magnitude}) conditioned on $I_{1}-I_{3}$ using: (a) the DNS data of case C1 (details available in Table \ref{tab:DNS_cases}), (b) the IBBM model.}
\end{figure}

We further, extend the model evaluation based on it's capabilities to capture the distribution of $\Psi$ (\ref{eq:magnitude}) conditioned over the values of $I_{1}$–$I_{3}$ (\ref{eq:I1}) and (\ref{eq:I3}), where $I_{1}$ and $I_{3}$ denote the first and third invariants of the $\boldsymbol{\widehat{H^{'}}}$  (\ref{eq:nondimhijp}) tensor. Figures~\ref{fig:Mt0p1_psi_I23} (a,b) presents the contours of $\Psi$ (\ref{eq:magnitude}) in the $I_{1}$–$I_{3}$ plotted using the DNS data of case C1 ($M_t = 1$, table \ref{tab:DNS_cases}) and the IBBM model predictions, respectively.
The DNS data exhibit a characteristic \textit{S}-shaped distribution in the invariant space, with higher $\Psi$ (\ref{eq:magnitude}) values concentrated near the outer boundaries. The model predictions capture the overall shape of the DNS distribution and the regions of elevated $\Psi$ (\ref{eq:magnitude}), although the contours are smoother and miss certain high-$\Psi$ locations near the central region. This suggests that, while the model reproduces the overall distribution trends in the $I_{1}$–$I_{3}$ space, some localized high-magnitude features observed in the central regions of the DNS contours are less pronounced in the model predictions. The performance of the IBBM model has also been evaluated using the DNS datasets corresponding to cases C2 and C3 (Table~\ref{tab:DNS_cases}). In both cases, the model successfully reproduces the key trends observed in the DNS data, and its predictive performance remains as good as it is shown for the case C1. Since the results obtained for cases C2 and C3 are similar to those presented for case C1, they have not been included in the manuscript in order to avoid the repetition of figures.

Based on this comparison, we conclude that the proposed IBBM model effectively captures both the eigendirection alignments and the magnitude distributions of the $\boldsymbol{\varsigma}$ tensor, yielding predictions that remain in close agreement with DNS results. 
The authors would like to clarify that the validation of the proposed model is currently restricted to the range of turbulent Mach numbers ($0.8 \leq M_t \leq 1.0$) and Taylor-scale Reynolds numbers ($60 \leq Re_{\lambda} \leq 250$) represented by the available DNS datasets. This limitation arises primarily from the limited availability of high-fidelity DNS databases for compressible turbulence (the computational resource available to the authors is
limited, and to the best of authors' knowledge, no such database is available in public
domain). As additional DNS datasets become available, the proposed framework can be evaluated over an wider range of flow conditions.

\section{\label{VII} Summary of the enhanced hybrid model for velocity-gradient dynamics}
In this section, we summarize the complete set of closed differential equations governing the evolution of the velocity-gradient tensor in compressible turbulence, including the effects of vibrational non-equilibrium.
The resulting closed-form evolution equations for the velocity-gradient dynamics are written as follows:

\begin{equation}
\label{eq:Aij_modeled_HEHEE}
\frac{dA_{ij}}{dt}=-A_{ik}A_{kj}-H_{ij}+B_{ij}-\frac{C_{pq}C_{pq}}{3\tau_{v}}\left(A_{ij}+\frac{A_{kk}}{3C_{pq}C_{pq}}C_{ri}C_{rj} \right),
\end{equation}
\begin{multline}
\label{eq:Hij_modeled_HEHEE}
    \frac{dH_{ij}}{dt}=-\left( H_{ij} +\frac{A_{lm}A_{ml}}{C_{pq}C_{pq}} C_{ri}C_{rj}\right ) \frac{c}{L} -\frac{C_{pq}C_{pq}}{3\tau_{p}} H_{ij}\\+\frac{(1-{\xi_{m}}_{0})}{(1+\xi_{m})} \Pi_{ij}^{IBVM}\sqrt{P_{mn}P_{mn}} \sqrt{A_{pq}A_{pq}},  
\end{multline}
\begin{multline}
\label{eq:Bij_modeled_HEHEE}
\frac{dB_{ij}}{dt}= -A_{kj}B_{ik} -A_{ki}B_{kj} -(n-1)A_{kk}B_{ij} -B_{ij}\frac{c}{L}-\widehat{\varsigma_{ij}}^{IBBM}\Psi^{IBBM}\sqrt{A_{pq}A_{pq}}\sqrt{P_{mn}P_{mn}} \\ -\frac{C_{pq}C_{pq}}{3\tau_{p}} B_{ij},
\end{multline}
\begin{equation}
\label{eq:14}
\textbf{M} \equiv\textbf{C}^{-1},
\end {equation}
\begin{equation}
\label{eq:dmij_modeled_HEHEE}
\frac{dM_{ij}}{dt}= A_{ik}M_{kj},  
\end {equation}
\begin{equation}
\label{eq:dxi_modeled_HEHEE}
\frac{d\xi_{m}}{dt}=-\frac{\xi_{m}-1}{\tau_{\text{vib}}},
\end {equation}

\begin{equation}
\label{eq:L_modeled_HEHEE}
L=max \left [ \frac{L_{o}}{\sqrt{\frac{C_{pq}C_{pq}}{3}}},\frac{L_{o}}{\sqrt{\tau_{\nu}A}} \right ].
\end{equation}

Here, $c$ denotes the speed of sound and $L$ represents a characteristic fluid length scale, with $L_{0}\equiv L(t=0)$. The tensor $\boldsymbol{C}$ (\ref{eq:deformation_gradient_tensor}) represents the deformation-gradient tensor.
The proposed framework (\ref{eq:Aij_modeled_HEHEE}--\ref{eq:L_modeled_HEHEE}) constitutes a closed system of 34 ordinary differential equations, which can be solved by direct numerical integration. This closed system is hereafter referred to as the \textit{hybrid enhanced homogenized Euler equations} (H-EHEE). To the best of our knowledge, this is the first closed set of ODEs that combines both phenomenological and DNN-based models, describing the evolution of velocity gradient dynamics in compressible turbulence with the presence of vibrational non-equilibrium.

\section{\label{VIII}Non-dimensional parameters of the H-EHEE model}
The H-EHEE model is comprised of several mechanisms governing the evolution of the $\boldsymbol{A}$ and $\boldsymbol{P}$ tensors, including inertial, thermodynamic, viscous and  vibrational-relaxation processes. The relative importance of these mechanisms is expected to vary significantly with the flow conditions. To (i) clearly understand the relative importance of these mechanisms in the evolution history of the $\boldsymbol{A}$  and $\boldsymbol{P}$ tensors, and also (ii) to identify the role of initial conditions on this evolution, we perform a formal non-dimensionalization of the equation set (\ref{eq:Aij_modeled_HEHEE}- \ref{eq:L_modeled_HEHEE}) and identify the important non-dimensional numbers which control the behavior of the modeled equations. 

Following a procedure similar to what has been adopted earlier by   \citet{suman2011dynamical}, the following normalized variables are introduced:
\begin{align}
L = \bar{L}\,L^{*}, \qquad
t = \frac{\bar{t}}{A^{*}}, \qquad
A_{ij} = \bar{A}_{ij}\,A^{*}, \qquad
P_{ij} = \bar{P}_{ij}\,P^{*}, \qquad
M_{ij} = \bar{M}_{ij}\,M^{*}.
\end{align}
The symbol $L^{*}$ denotes the characteristic length scale, while $A^{*}$, $P^{*}$ and $M^{*}$ represent the characteristic magnitudes of the tensors $A_{ij}$, $P_{ij}$ and $M_{ij}$, respectively.
The H-EHEE system is non-dimensionalized as:
\begin{multline}
\frac{d\bar{A}_{ij}}{d\bar{t}} =-\bar{A}_{ik}\bar{A}_{kj}-\phi_m\left(\bar{H}_{ij}-\bar{B}_{ij}\right)
-\frac{1}{\mathrm{Re}_m}
\left(
\bar{A}_{ij}
+ \frac{\bar{A}_{kk}}{3\,C_{pq}C_{pq}}\,C_{ri}C_{rj}
\right),
\end{multline}
\begin{multline}
\label{eq:norm_Hij_eq}
\frac{d\bar{H}_{ij}}{d\bar{t}}
=
-\frac{1}{M_m}
\left[
\bar{H}_{ij}
+
\frac{1}{\phi_m}
\frac{\bar{A}_{lm}\bar{A}_{ml}}
     {C_{pq}C_{pq}}
\,C_{ri}C_{rj}
\right]
\frac{1}{\bar{L}}
-\frac{1}{\theta_m}\bar{H}_{ij}
+\chi_m \,\hat{\Pi}^{\mathrm{IBVM}}_{ij},
\end{multline}
\begin{multline}
\frac{d\bar{B}_{ij}}{d\bar{t}}
 =
-\bar{A}_{kj}\bar{B}_{ik}
-\bar{A}_{ki}\bar{B}_{kj}
-(n-1)\bar{A}_{kk}\bar{B}_{ij}
-\frac{1}{M_m\bar{L}}\,\bar{B}_{ij}
-\hat{\varsigma}^{\mathrm{IBNM}}_{ij}
-\frac{1}{\theta_m}\bar{B}_{ij},
\end{multline}
\begin{align}
\frac{d\bar{M}_{ij}}{d\bar{t}}
=
\bar{A}_{ik}\bar{M}_{kj},
\qquad
\bar{M}_{ij}M^{*}_{ij}
=
{C}_{ij}^{-1},
\end{align}
\begin{align}
\bar{L}
=\max
\left[
\frac{\bar{L}_0}
{\sqrt{C_{pq}C_{pq}/3}},
\;
\frac{\bar{L}_0}
{\sqrt{\tau_v A^{*}}}
\right],
\end{align}
\begin{align}
\frac{d{\xi}_m}{d\bar{t}}
=
-D_m({\xi}_m-1).
\end{align}
The non-dimensionalized H-EHEE system is governed by six model parameters. The complete set of non-dimensional model parameters is summarized as:
\begin{multline}
\label{eq:non_dimen_para}
M_{m}= \frac{A^{*}L^{*}}{c}, \qquad
Re_{m}=\frac{3\tau_{\nu}A^{*}}{C_{pq}C_{pq}}, \qquad
\theta_{m}=\frac{3\tau_{p}A^{*}}{C_{pq}C_{pq}}=Re_{m}Pr, \qquad
\phi_{m}= \frac{P^{*}}{A^{*2}},\\
D_{m}= \frac{1/A^{*}}{\tau_{vib}}, \qquad 
\xi_{m}= \frac{e_{\nu}}{e_{\nu}^{*}}.
\end{multline}
The parameter $\chi_{m}$ used in (\ref{eq:norm_Hij_eq}) is defined as
\begin{equation}
    \chi_{m}= \frac{1-\xi_{m0}}{1+\xi_{m}},
\end{equation}
where $\xi_{m0}=\xi_{m}(t=0)$. It should be noted that $\chi_{m}$ is not an independent dimensionless parameter, since it depends on the evolving non-equilibrium index $\xi_{m}$ and its initial value $\xi_{m0}$. Consequently, $\chi_{m}$ is not included in the set of independent model parameters listed in (\ref{eq:non_dimen_para}). 
The first four model parameters listed in (\ref{eq:non_dimen_para}), are the same as those identified in \cite{suman2011dynamical}, while two additional parameters arise due to the inclusion of vibrational nonequilibrium physics. 
The parameters $M_{m}$ and $Re_{m}$ are referred to as the model Mach number and the model Reynolds number, respectively. The symbols $D_{m}$ and $\xi_{m}$ denote the two additional model parameters, referred to as the model Damk\"ohler number and the model non-equilibrium index. The symbol $Pr$ denotes the Prandtl number, while $c$ and $L$ represent the speed of sound and the local fluid length scale. Additional details regarding the definition and physical interpretation of these model non-dimensional parameters can be found in \cite{suman2011dynamical}. 

\subsection{\label{VIIA} Comparison of H-EHEE parameters with DNS}
Following the analysis of \cite{suman2011dynamical}, the model non-dimensional parameters $M_{m}$ and $Re_{m}$ exhibit proportional trends with their corresponding quantities defined in DNS datasets, namely the turbulent Mach number: $M_{t}$ (\ref{eq:mach_number}) and the Taylor-microscale Reynolds number: $Re_{\lambda}$ (\ref{eq:reynolds_number}).
\begin{equation}
M_{m} \propto M_{t}, \qquad Re_{m} \propto Re^{a}_{\lambda},
\end{equation}
where $a<1$.
However, due to differences in their definitions and underlying modeling assumptions, an exact one-to-one mapping between these parameters cannot be established. The parameter $\phi_{m}$ reflects both the intensity of thermodynamic fluctuations and their spatial distribution. Its characteristic magnitude is therefore expected to vary with the Mach-number regime. In the low-Mach-number limit, where pressure fluctuations are strongly constrained by the velocity field through the Poisson equation, $\phi_{m}$ is expected to remain of order unity.


The model Damk\"ohler number ($D_{m}$) represents the relative magnitude of the characteristic flow time scale to the vibrational relaxation time scale. In direct numerical simulations (DNS) of isotropic decaying compressible turbulence with vibrational non-equilibrium effects, the vibrational Damk\"ohler number $D$ is defined as \citep{khurshid2019decaying, srivastava2024influence} the ratio of the large-eddy turnover time ($\tau_{\text{LE}}$) to the vibrational relaxation time ($\tau_{\text{vib}}$)
\begin{equation}
    \label{eq:damkohler_number}
    D = \frac{\tau_{\text{LE}}}{\tau_{\text{vib}}},
\end{equation}
In the H-EHEE model (\ref{eq:Aij_modeled_HEHEE}--\ref{eq:L_modeled_HEHEE}), the quantities required to explicitly evaluate $\tau_{\text{LE}}$ are not available. Consequently, a direct quantitative relationship between the model parameter: $D_{m}$ and the DNS parameter: $D$ cannot be established. Nonetheless, both $D_{m}$ and $D$ characterize the relative magnitude of the flow time scale compared to the vibrational relaxation time scale. On this basis, we infer that:
\begin{equation}
    D_{m} \propto D.
\end{equation}

Another important non-dimensional parameter associated with vibrational non-equilibrium physics in DNS cases is the ratio of the mean vibrational temperature to the mean translational temperature, denoted by $\xi$, and defined as \citep{khurshid2019decaying, srivastava2024influence}:
\begin{equation}
    \label{eq:xi}
    \xi = \frac{\langle T_{\nu} \rangle}{\langle T \rangle},
\end{equation}
where $\langle \cdot \rangle$ denotes spatial averaging and $T_{\nu}$ represents the vibrational temperature.
In the present modeling framework, the vibrational and rotational temperatures are not explicitly available. As a result, the parameter $\xi$ cannot be directly evaluated. Instead, we introduce a modeled quantity, denoted by $\xi_{m}$ (\ref{eq:xi_m}). By definition, $\xi_{m}$ represents the ratio of vibrational non-equilibrium energy to the vibrational mode energy corresponding to the equilibrium state with the local translational–rotational temperature.
Although both $\xi$ and $\xi_{m}$ provide measures of the departure from the vibrational equilibrium state, they are defined in terms of fundamentally different variables, and therefore no direct relationship exists between them. For instance, $\xi_{m}=1$ corresponds to a state of local vibrational equilibrium, whereas $\xi=1$ does not necessarily imply local vibrational equilibrium. In the latter case, the mean vibrational and translational temperatures may be equal, while their respective fluctuations may still differ.

It is important to emphasize that the proposed H-EHEE framework (\ref{eq:Aij_modeled_HEHEE})--(\ref{eq:L_modeled_HEHEE}) can be evaluated across different compressibility and vibrational-relaxation regimes by varying the model parameters, listed in (\ref{eq:non_dimen_para}). The effects of compressibility and vibrational non-equilibrium are represented explicitly through parameters such as $M_m$, $D_m$, and $\xi_m$, which enter the governing equations and their associated closure terms. The H-EHEE model solution is computed using a range of [$5,000$--$60,000$] random initial conditions (fluid elements), with the exact number determined by the convergence of the simulation statistics. Each fluid element comprises initial velocity gradient tensor components ($A_{ij}$) generated as uniformly distributed random numbers between [-1,1]. The initial magnitude of $\textbf{A}$ is set to unity by normalizing it with the initial magnitude: $ \sqrt{A_{ij}A_{ij}} $. All initial $\textbf{M}$ tensor is set as the identity tensor. The $\textbf{H}$ tensor is initially chosen to be isotropic $(H_{ij}=-A_{lm}A_{ml} \frac{\delta_{ij}}{3})$, note that, $\delta_{ij}$ is the symbol of "Kronecker delta" function. The $\textbf{B}$ tensor is initially set to be zero. The value of $\tau_{v}$ is chosen according to the desired initial model Reynolds number, $Re_{m}$.The initial value of $\chi_{m}$ is set to zero, corresponding to $\xi_{m 0} = 1$, for vibrational equilibrium flows, and to $0.33$, corresponding to $\xi_{m0} = 0.5$, for vibrational non-equilibrium flow cases.
 In all evaluation cases, the initial values (at $t=0$), of  $A^{*},~\& ~L^{*}$ are taken unity. The values of Prandtl number and specific heat ratio are taken as 0.7 and 1.4, respectively. 
The solution is integrated up to a dimensionless time $tA_{o}=10$ (unless specified otherwise), where $A_{o}=\sqrt{A_{ij}A_{ij}}$ at t=0. We check all velocity gradient statistics at the time after which the ensemble-averaged $<A_{ij}A_{ij}>$ reaches its peak value; by this time, the model has generated sufficient length scales and all nonlinear and viscous processes are activated.

\section{\label{IX}Model evaluation}
The performance of the H-EHEE model is evaluated by numerically integrating the complete set of modeled equations, (\ref{eq:Aij_modeled_HEHEE})--(\ref{eq:L_modeled_HEHEE}). The resulting dynamical solutions are then compared with the corresponding DNS statistics and predictions from the existing dynamical model for compressible flows proposed by~\cite{danish2014direct} (the N-EHEE model), wherever such reference results are available.

The evaluation is conducted systematically across various flow regimes. We first examine the behavior of the H-EHEE model in the limit of vanishingly small Mach numbers, corresponding to the incompressible regime. This is followed by an assessment of the model at intermediate Mach number flows ($M_{t}=[0.1-1.2]$), with and without having vibrational non-equilibrium effects. 

\subsection{Low Mach number: incompressible flow limit}

In the limit of small model Mach number ($M_{m}$), the sound speed is much larger than the characteristic fluid velocity scale, leading to
\begin{equation}
    \frac{c}{L} \gg A .
\end{equation}
In this regime, the local compressibility-related parameters $a_{ii}$ and $\delta$ approach zero. Consequently, the modeled inviscid mechanism, $\varsigma_{ij}^{IBBM}$ (\ref{eq:ibnm_varsigma}), becomes negligible, which leads to $\textbf{B} \rightarrow 0$. Moreover, the $\textbf{H}$ tensor relaxes rapidly toward its incompressible limit governed by the Poisson equation,
\begin{equation}
    Z \approx -A_{lm}A_{ml},
\end{equation}
where $Z$ denotes the isotropic part of the $\textbf{H}$ tensor. Under these conditions, (\ref{eq:Aij_modeled_HEHEE}) reduces to
\begin{equation}
    \frac{dA_{ij}}{dt}
    =-A_{ik}A_{kj}-Q_{ij}
    +\frac{1}{3}A_{lm}A_{ml}\,\delta_{ij}
    +\frac{C_{pq}C_{pq}}{3\tau_{\nu}}A_{ij},
\end{equation}
which corresponds to the Vieillefosse restricted Euler model augmented with an anisotropic pressure-Hessian contribution ($Q_{ij}$) and viscous effects. 

To numerically examine the incompressible limit of the H-EHEE model, the initial model parameters are chosen as:
\begin{equation}
    M_{m}=0.05, \qquad Re_{m}=12, \qquad  \phi_{m} \approx 1, \qquad \chi_{m}=0,\qquad D_{m}=10,000 .
\end{equation}
The model performance is assessed in terms of both structural properties, such as eigenvector alignment tendencies, and statistical measures associated with the magnitude of the velocity gradient tensor, $\sqrt{A_{mn}A_{mn}}$. Let $(\alpha_{s},\beta_{s},\gamma_{s})$ denote the eigenvalues of the normalized strain-rate tensor $\boldsymbol{s}$. The normalized strain-rate tensor $\boldsymbol{s}$ is obtained from the self-normalized velocity-gradient tensor $\boldsymbol{a}=\boldsymbol{A}/|\boldsymbol{A}|$, as:
\begin{equation}    \label{eq:strain_rate}     \boldsymbol{s}=\frac{\boldsymbol{a}+\boldsymbol{a}^{T}}{2}.
\end{equation}

The eigenvalues $(\alpha_{s},\beta_{s},\gamma_{s})$
are ordered as: $\alpha_{s}>\beta_{s}>\gamma_{s}$ \citep{vieillefosse1982local}. The corresponding eigenvectors are denoted by $\hat{\boldsymbol{e}}_{\alpha_{s}}$, $\hat{\boldsymbol{e}}_{\beta_{s}}$, and $\hat{\boldsymbol{e}}_{\gamma_{s}}$.
For comparison across realizations, the eigenvalues are further normalized as
\begin{equation}
\label{eq:31}
\alpha_{s}^{*}= \frac{\alpha_{s}}{\sqrt{\alpha_{s}^{2}+\beta_{s}^{2}+\gamma_{s}^{2}}}, \qquad
\beta_{s}^{*}= \frac{\beta_{s}}{\sqrt{\alpha_{s}^{2}+\beta_{s}^{2}+\gamma_{s}^{2}}}, \qquad
\gamma_{s}^{*}= \frac{\gamma_{s}}{\sqrt{\alpha_{s}^{2}+\beta_{s}^{2}+\gamma_{s}^{2}}}.
\end{equation}
The normalized eigenvalues $(\alpha_{s}^{*},\beta_{s}^{*},\gamma_{s}^{*})$ directly quantify the relative magnitudes of the strain-rate eigenvalues.

\begin{figure}
\centering
  \begin{tabular}[b]{c}
\hspace{-0.3in}    
\includegraphics[width=.5\linewidth]{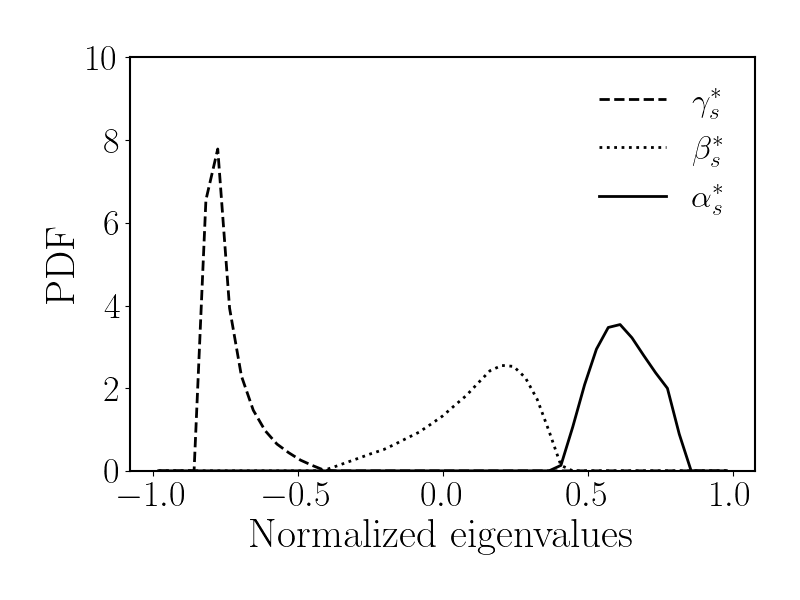} \\
    \small (a)
  \end{tabular} \qquad
  \begin{tabular}[b]{c}
  \hspace{-0.4in}
\includegraphics[width=.5\linewidth]{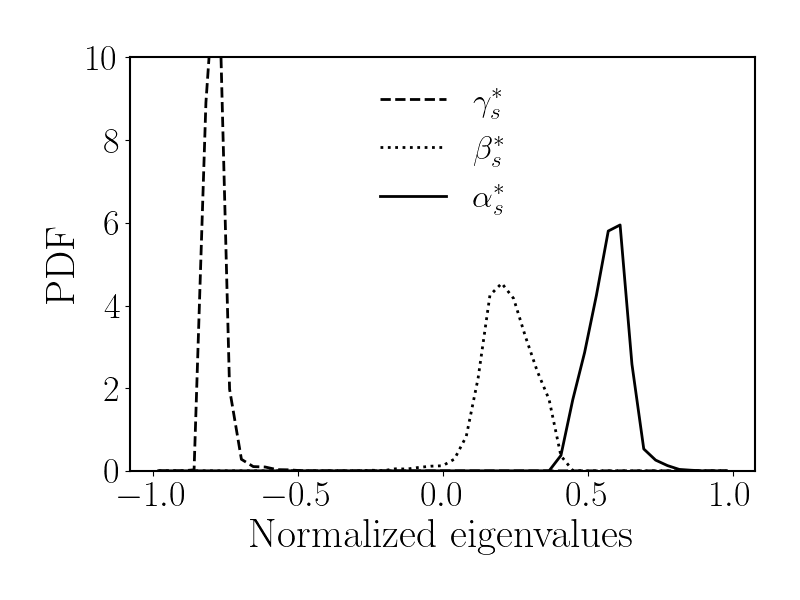}\\
    \small (b)
  \end{tabular}
  \begin{tabular}[b]{c}
 \hspace{-0.4in}
\includegraphics[width=.5\linewidth]{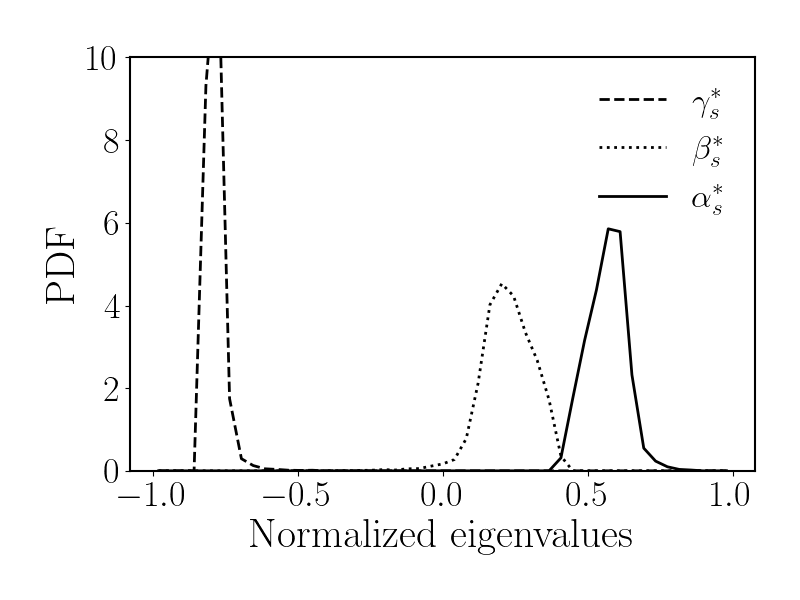}\\
    \small (c)
  \end{tabular}
  \caption{\label{fig:strain_rate_eigenvalues}The PDFs of the normalized strain-rate eigenvalues using: (a) DNS dataset \citep{WinNT}, (b) the H-EHEE model, (c) the N-EHEE model \citep{danish2014direct}.}
  \end{figure}

We start our comparison with the PDFs of the self-normalized eigenvalues (\ref{eq:31}) of the strain rate tensor. We utilize the homogeneous isotropic incompressible DNS data at $Re_{\lambda}=433$) \citep{perlman2007data,WinNT} to produce the incompressible DNS behavior.
In figure~\ref{fig:strain_rate_eigenvalues} (a), \ref{fig:strain_rate_eigenvalues} (b), and \ref{fig:strain_rate_eigenvalues} (c), we present the results from DNS, the H-EHEE model, and the N-EHEE model, respectively. In figure~\ref{fig:strain_rate_eigenvalues} (a), we observe that, most likely, two of the normalized eigenvalues of the strain rate tensor take positive values, whereas one takes a negative value, as the plot has two peaks at positive values and one peak at a negative value. The peaks of these three eigenvalues are known to exist with a ratio of 3:1:-4. Both models, the proposed H-EHEE model and the N-EHEE model (compare figures \ref{fig:strain_rate_eigenvalues} (b) and \ref{fig:strain_rate_eigenvalues} (c), respectively), successfully recover this characteristic behavior observed in DNS. Moreover, the PDFs predicted by the H-EHEE model closely match those obtained using the N-EHEE model.

The local vorticity vector (\ref{eq:vorticity}) alignment with the strain rate eigenvectors signifies the extent of the vortex stretching phenomenon. The vorticity alignment with the strain rate eigenvector corresponding to positive eigenvalues indicates vortex stretching, while the vorticity alignment with the strain rate eigenvector corresponding to negative eigenvalues indicates vortex compression. Therefore, the PDF of the vorticity alignment with the strain rate eigenvectors is a crucial quantity for evaluating a velocity gradient dynamics model. 

\begin{figure}
\centering
  \begin{tabular}[b]{c}
\hspace{-0.3in}    
\includegraphics[width=.5\linewidth]{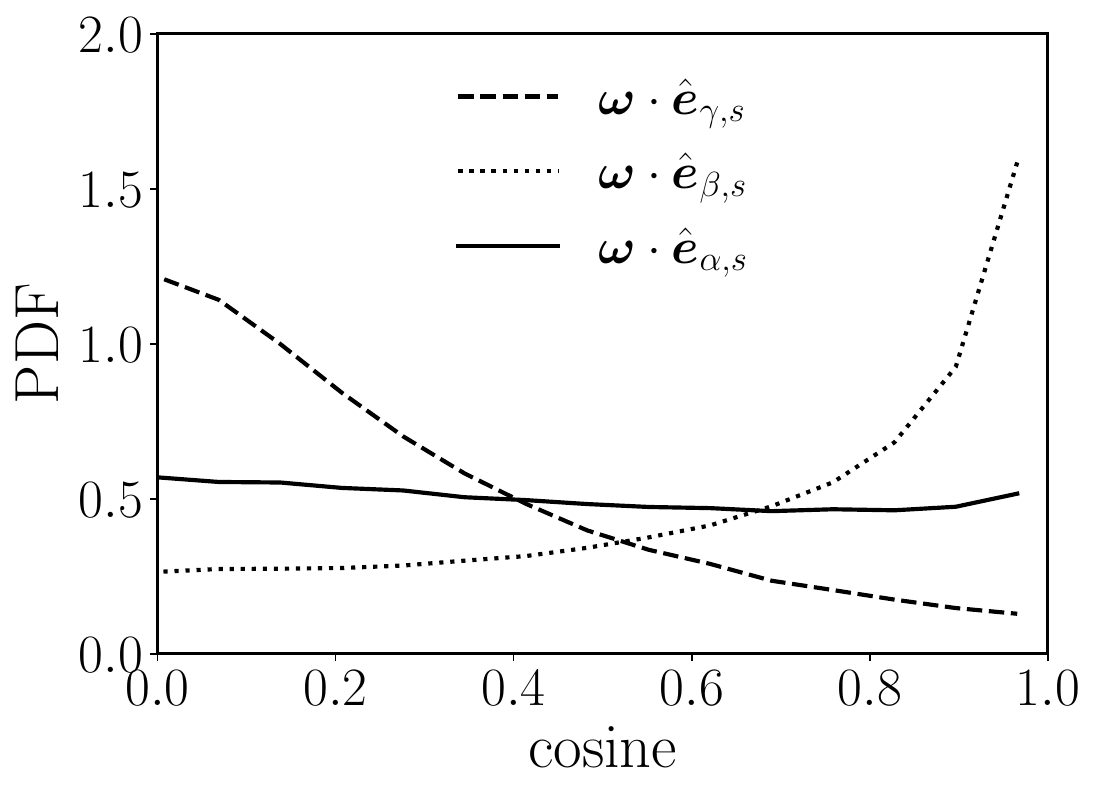} \\
    \small (a)
  \end{tabular} \qquad
  \begin{tabular}[b]{c}
  \hspace{-0.3in}
\includegraphics[width=.5\linewidth]{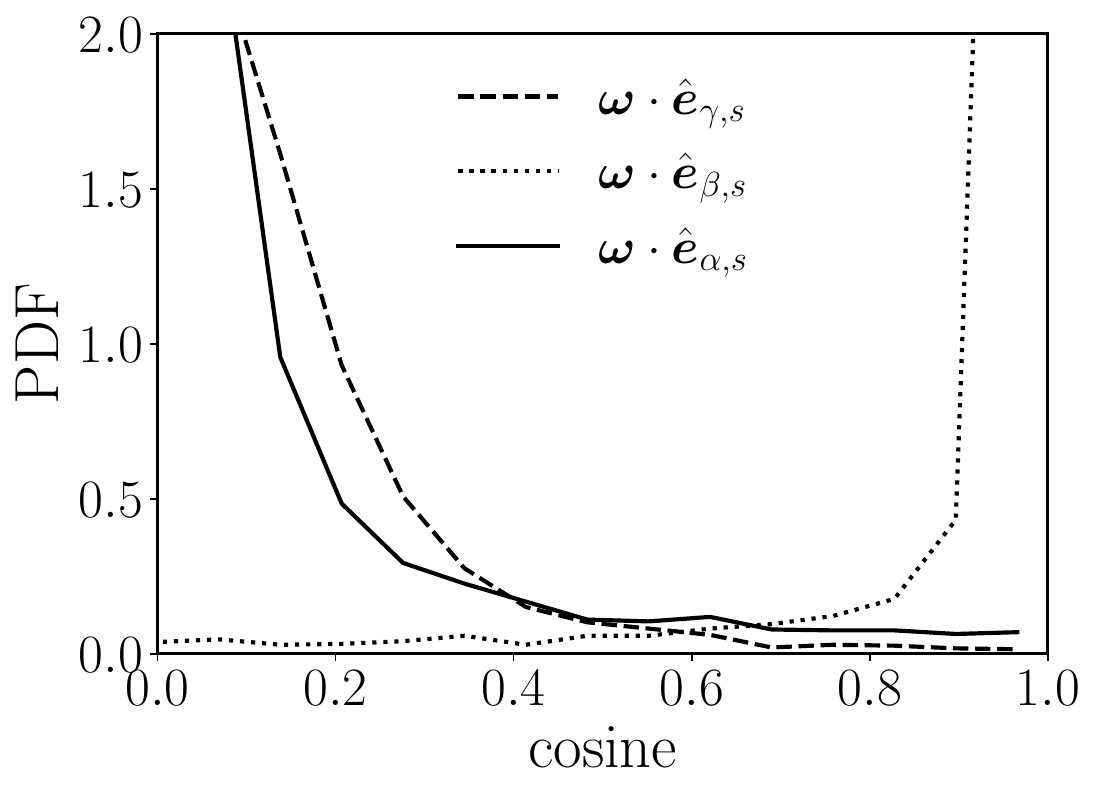}\\
    \small (b)
  \end{tabular}
  \begin{tabular}[b]{c}
 \hspace{-0.3in}
\includegraphics[width=.5\linewidth]{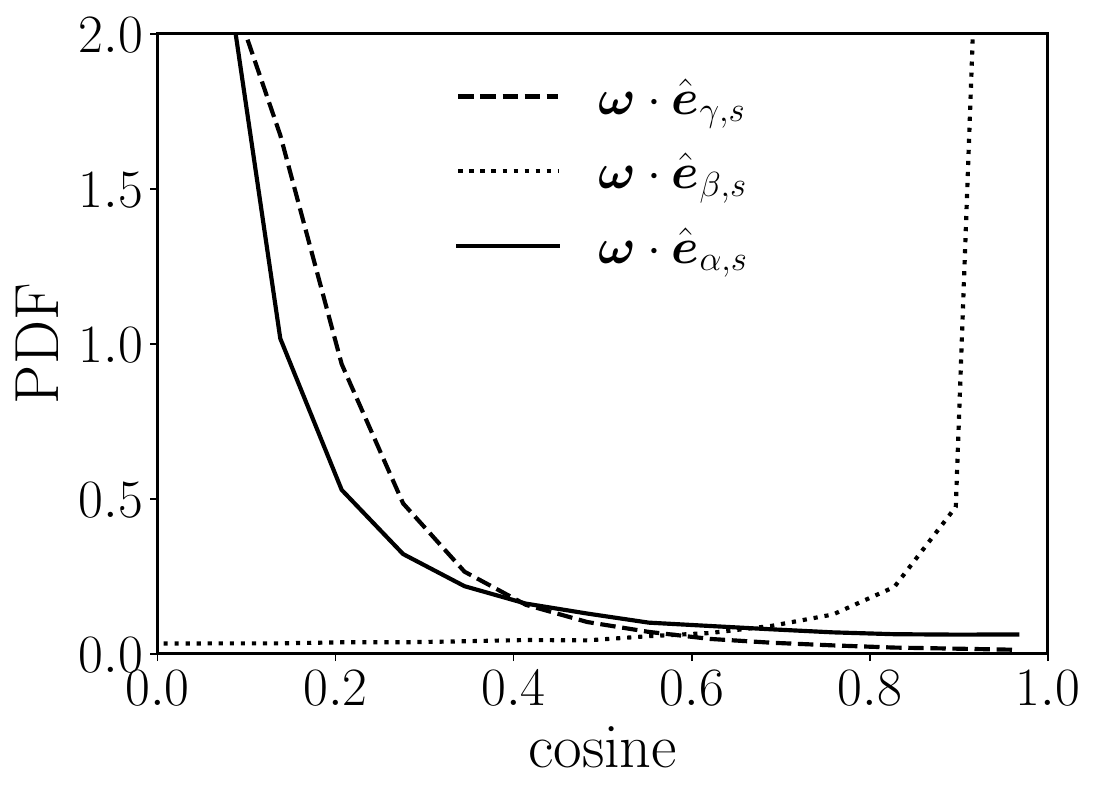}\\
    \small (c)
  \end{tabular}
  \caption{\label{fig:s_omega}The PDFs of the cosine of angles between the local strain-rate eigendirections with respect to the local vorticity vector using: (a) DNS dataset \citep{WinNT}, (b) the H-EHEE model, (c) the N-EHEE model \citep{danish2014direct}.}
  \end{figure}

Next, we evaluate our model based on the alignment tendencies of the strain-rate eigendirections with the vorticity vector. In figure~\ref{fig:s_omega}, we present the alignments of different eigendirections of the strain rate tensor with the local vorticity vector.
DNS results as shown in figure~\ref{fig:s_omega} (a), show the strongest alignment of the vorticity vector along with the intermediate eigendirection ($\widehat{e_{\beta_{s}}}$) and orthogonal alignment with the most negative eigendirection ($\widehat{e_{\gamma_{s}}}$). The most positive eigendirection ($\widehat{e_{\alpha_{s}}}$) does not show any preferential alignment with the vorticity vector. 
Both models, the H-EHEE and the N-EHEE model  (as shown in figures~\ref{fig:s_omega} (b), and \ref{fig:s_omega} (c), respectively), capture the essential alignment characteristics except the non-preferential alignment feature of $\widehat{e_{\alpha_{s}}}$ eigenvector. However, the models seem to overpredict the extent of these tendencies ( represented by sharp rises in the curves near 0 and 1). 

Another important feature of the velocity gradient tensor is its set of invariants. \cite{chong1990general}, and \cite{suman2010velocity} demonstrate that the information about the local topology associated with the fluid element can be inferred by examining the invariants of the locally normalized velocity gradient tensor ($a_{ij}$). The three invariants of $\boldsymbol{a}=\boldsymbol{A}/|\boldsymbol{A}|$ are defined as:
\begin{equation}
\label{eq:33}
m \equiv -a_{ii} \equiv 0,
\end {equation}
\begin{equation}
\label{eq:34}
q \equiv \frac{1}{2}(m^{2}-s_{ij}s_{ji}-w_{ij}w_{ji}),
\end {equation}
\begin{equation}
\label{eq:35}
r \equiv \frac{1}{3}(-m^{3}+3mq-s_{ij}s_{jk}s_{ki}-3w_{ij}w_{jk}s_{ki}),
\end {equation}
where $m$, $q$, and $r$ are the first, second, and third invariants of the $\textbf{a}$ tensor, respectively.

\begin{figure}
\centering
  \begin{tabular}[b]{c}
\hspace{-0.3in}    
\includegraphics[width=.5\linewidth]{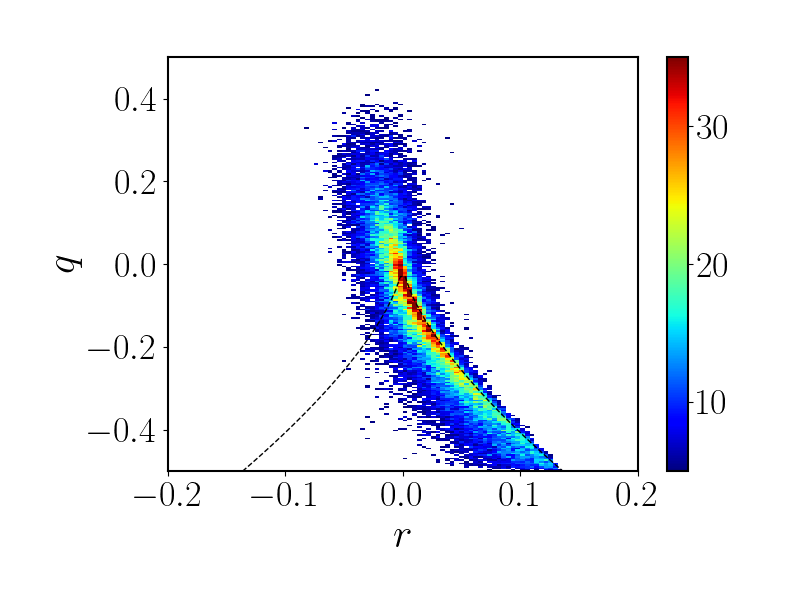} \\
    \small (a)
  \end{tabular} \qquad
  \begin{tabular}[b]{c}
  \hspace{-0.4in}
\includegraphics[width=.5\linewidth]{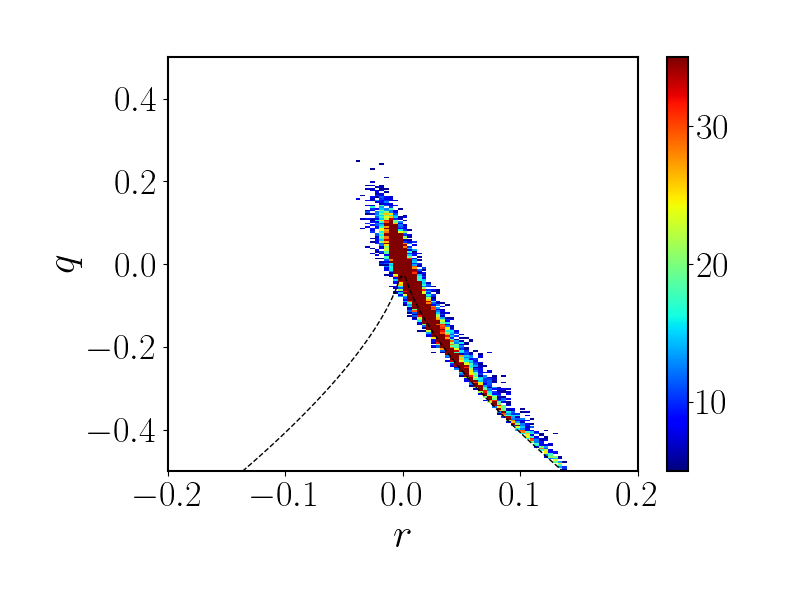}\\
    \small (b)
  \end{tabular}
  \begin{tabular}[b]{c}
 \hspace{-0.4in}
\includegraphics[width=.5\linewidth]{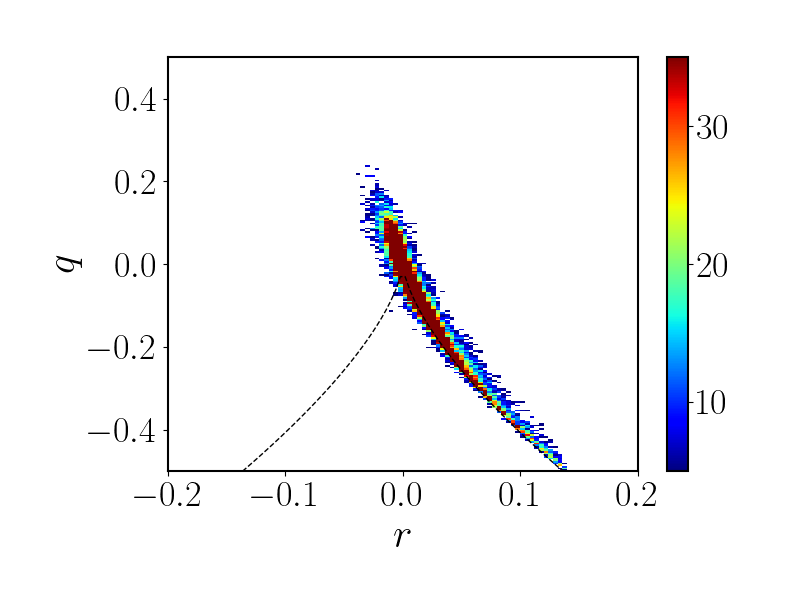}\\
    \small (c)
  \end{tabular}
  \caption{\label{fig:joint_PDF}The Joint-PDFs of $q~ \& ~r$ using: (a) DNS dataset \citep{WinNT}, (b) the H-EHEE model, (c) the N-EHEE model \citep{danish2014direct}.}
  \end{figure}

We evaluate the H-EHEE model in terms of the joint PDF of the second and third invariants (\ref{eq:33}-\ref{eq:35}) of the normalized velocity gradient tensor.
Figure~\ref{fig:joint_PDF} shows the joint-PDF of the second and third invariants. Figure~\ref{fig:joint_PDF}(a) presents the joint-PDF obtained from the DNS data ($m=0$ plane). In the DNS distribution, the majority of the probability mass is concentrated in the lower-right quadrant, while there is substantially less spread observed in the upper-right and lower-left quadrants. The distribution in the upper-left quadrant appears nearly uniform. The region of maximum probability density, indicated by the red contours near the origin ($q=0$, $r=0$), extends along the Vieillefosse line $\left(q=-\sqrt[3]{27r^2/4}\right)$ into the fourth quadrant.
Figures~\ref{fig:joint_PDF}(b) and \ref{fig:joint_PDF}(c) show the results obtained from the H-EHEE and N-EHEE models, respectively. The two models exhibit very similar joint PDF distributions and successfully reproduce the key qualitative features observed in the DNS data. However, compared to the DNS results, the models predicted joint-PDFs show a noticeably narrower distribution in all quadrants.

Overall, in the low-Mach-number regime, the H-EHEE model shows performance comparable to that of the N-EHEE model. In this regime, the $\boldsymbol{B}$ tensor should vanish, hence the $P_{ij}$ tensor can be approximated as: $P_{ij} \equiv H_{ij}$. The recovery of N-EHEE-like behavior indicates that, within the H-EHEE framework, the $\boldsymbol{B}$ tensor is effectively suppressed. This occurs because the neural-network-based closure for the inviscid mechanism $V_{ij}$ (\ref{eq:Pressure_Hessian}) identifies regions where $a_{ii}$ becomes small and consequently predicts negligible contributions. As a result, the H-EHEE model correctly recovers the limiting behavior of the $P_{ij}$ tensor subjected to the incompressible flow conditions.
\subsection{Intermediate Mach number \texorpdfstring{($M_t < 0.6$)}{(Mt < 0.6)}}
In this section, the proposed model is evaluated for intermediate Mach numbers ($M_{t}=[0.1-0.6]$) flows under conditions with and without vibrational non-equilibrium effects.

\subsubsection{Fluid flows without vibrational non-equilibrium effects}
First, we present the model evaluations in flow fields without having vibrational non-equilibrium effects. These model simulations are performed with $D_m = 10{,}000$ and $\chi_m = 0$. For this evaluation, the H-EHEE model predictions are compared with DNS results reported by \cite{lee1991eddy} and \cite{sarkar1991analysis}, as well as with the N-EHEE model of \cite{danish2014direct}. Since an exact quantitative relation between the DNS parameters ($M_{t}, Re_{\lambda}$) and the model parameters ($M_{m}, Re_{m}$) cannot be established (as discussed in subsection \ref{VIIA}, the comparisons presented here are primarily qualitative in nature. 

\begin{figure}
\centering
  \begin{tabular}[b]{c}
\hspace{-0.3in}    
\includegraphics[width=.5\linewidth]{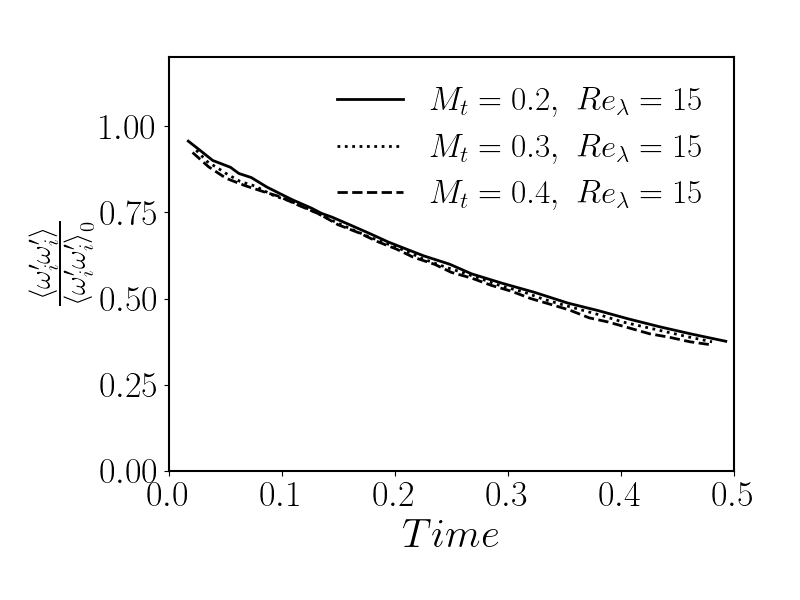} \\
    \small (a)
  \end{tabular} \qquad
  \begin{tabular}[b]{c}
  \hspace{-0.4in}
\includegraphics[width=.5\linewidth]{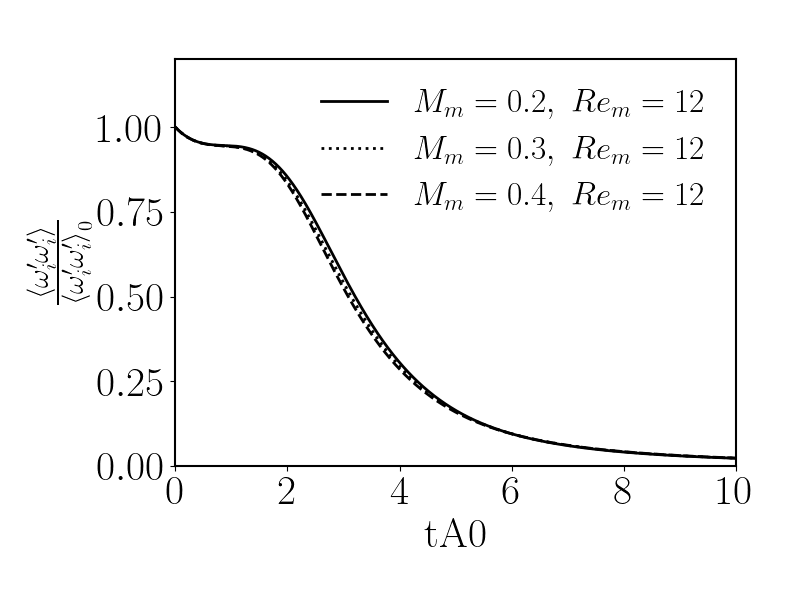}\\
    \small (b)
  \end{tabular}
  \begin{tabular}[b]{c}
 \hspace{-0.4in}
\includegraphics[width=.5\linewidth]{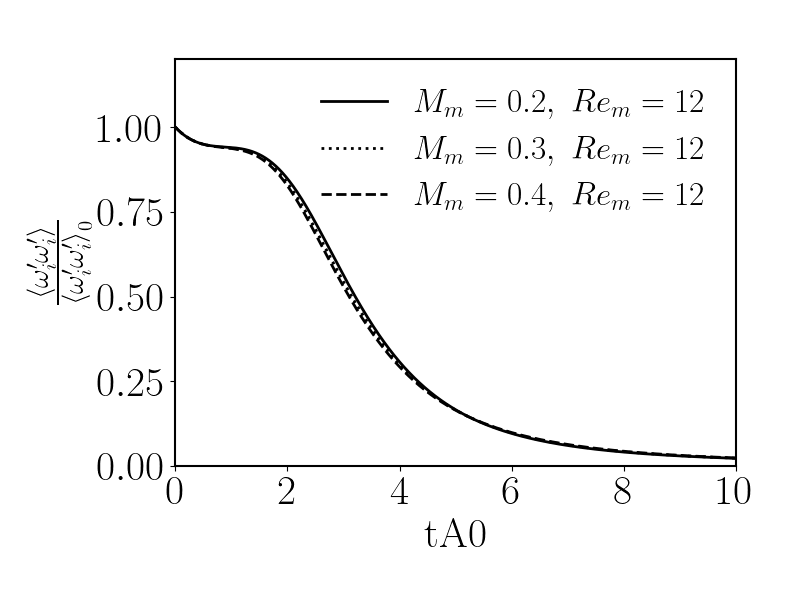}\\
    \small (c)
  \end{tabular}  \caption{\label{fig:inter_mach_solenoidal}The evolution of solenoidal dissipation $\frac{\langle \omega_{i}^{'}
\omega_{i}^{'}\rangle} {\langle\omega_{i}^{'}
\omega_{i}^{'}\rangle_{0}}$ with different initial Mach numbers using: (a) DNS dataset \citep{sarkar1991analysis}, (b) the H-EHEE model, (c) the N-EHEE model \citep{danish2014direct}}
  \end{figure}

\begin{figure}
\centering
  \begin{tabular}[b]{c}
\hspace{-0.3in}    
\includegraphics[width=.5\linewidth]{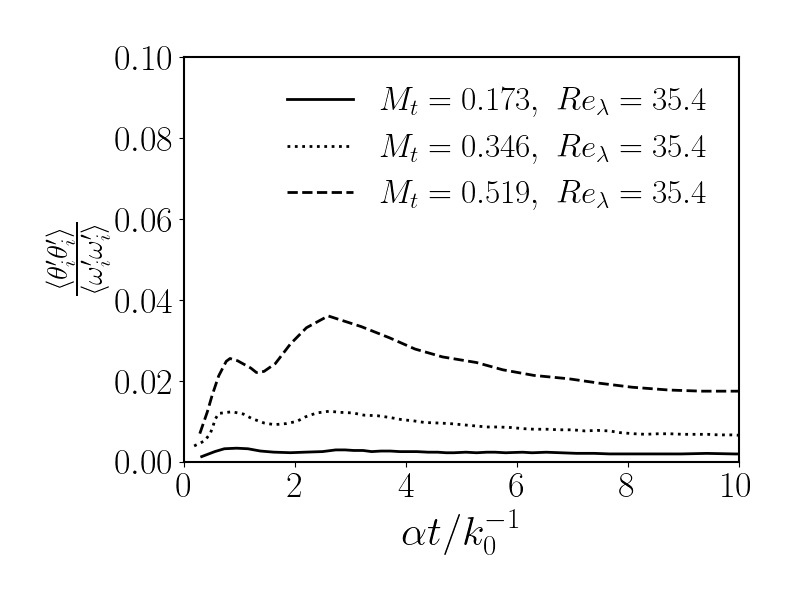} \\
    \small (a)
  \end{tabular} \qquad
  \begin{tabular}[b]{c}
  \hspace{-0.4in}
\includegraphics[width=.5\linewidth]{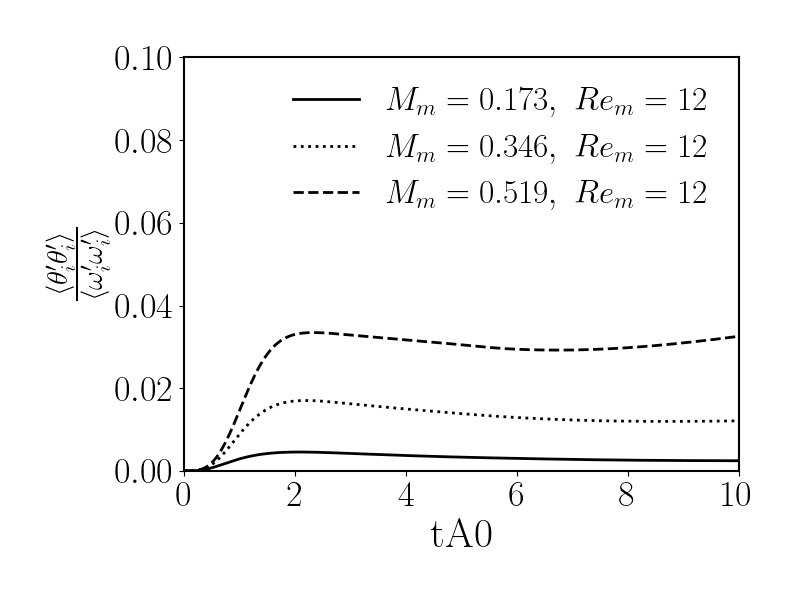}\\
    \small (b)
  \end{tabular}
  \begin{tabular}[b]{c}
 \hspace{-0.4in}
\includegraphics[width=.5\linewidth]{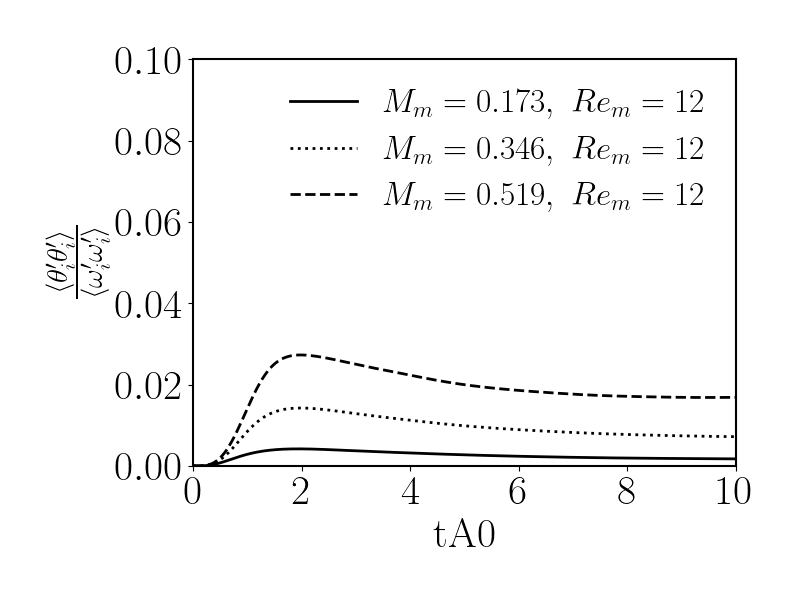}\\
    \small (c)
  \end{tabular}
  \caption{\label{fig:inter_mach_dilatational}Effect of Mach number on the evolution of $\frac{\langle \theta_{i}^{'}
\theta_{i}^{'}\rangle} {\langle\omega_{i}^{'}
\omega_{i}^{'}\rangle}$ with different initial Mach numbers using: (a) DNS \citep{lee1991eddy}, (b) the H-EHEE model, (c) the N-EHEE model \citep{danish2014direct}.}
  \end{figure}

To assess the ability of the proposed model to capture Mach-number-dependent compressibility effects, we focus on its predictions of the statistical behavior of  
(i) the root-mean-square (rms) vorticity ($\omega'$), and  
(ii) the rms dilatation rate ($\theta'$),  
defined as
\begin{equation}    \label{eq:theta_and_omega_prime}
    \omega' = \langle \omega_i \omega_i \rangle^{1/2}, 
    \quad
    \theta' = \left\langle \left( \frac{\partial u_i}{\partial x_i} \right)^2 \right\rangle^{1/2}.
\end{equation}
For constant-viscosity simulations, the temporal evolution of ${\omega'}^{2}$ and ${\theta'}^{2}$ is directly related to the solenoidal dissipation rate ($\epsilon_s$) and the dilatational dissipation rate ($\epsilon_d$) of kinetic energy, respectively, as defined:
\begin{equation}
\label{eq:solenoidal_dilatational_dissipation}
\epsilon_s = \langle \mu \, \omega_i \omega_i \rangle, 
\quad
\epsilon_d = \left\langle \frac{4\mu}{3} \left( \frac{\partial u_i}{\partial x_i} \right)^2 \right\rangle.
\end{equation}
Figures~\ref{fig:inter_mach_solenoidal} and \ref{fig:inter_mach_dilatational} show the temporal evolution of the normalized solenoidal dissipation rate ($\langle \omega_i' \omega_i' \rangle / \langle \omega_i' \omega_i' \rangle_0$), and the ratio of the dilatational dissipation rate to the solenoidal dissipation rate ($\langle \theta_i' \theta_i' \rangle / \langle \omega_i' \omega_i' \rangle$), respectively, for different initial Mach numbers, where subscript $_{0}$ denotes the value of the quantity at $t=0$.
The DNS results, together with the theoretical analysis of \cite{sarkar1991analysis,lee1991eddy}, indicate that in the intermediate Mach-number range ($M_t = 0.1$--$0.6$), Mach-number effects on solenoidal dissipation are negligible (as demonstrated in figure~\ref{fig:inter_mach_solenoidal} a). In contrast, the dilatational dissipation exhibits a strong dependence on the initial turbulent Mach number (figure~\ref{fig:inter_mach_dilatational} a).
Figures~\ref{fig:inter_mach_solenoidal}(b) and \ref{fig:inter_mach_solenoidal}(c) present the evolution of $\langle \omega_i' \omega_i' \rangle / \langle \omega_i' \omega_i' \rangle_0$ obtained using the H-EHEE and N-EHEE models, respectively. Both models correctly reproduce the insensitivity of $\langle \omega_i' \omega_i' \rangle$ to Mach number, which is consistent with the DNS behavior.

Figures~\ref{fig:inter_mach_dilatational}(b) and \ref{fig:inter_mach_dilatational}(c) show the evolution of $\langle \theta_i' \theta_i' \rangle / \langle \omega_i' \omega_i' \rangle$ predicted by the H-EHEE and N-EHEE models, respectively. For all simulations, the ratio initially increases with time, reaches a peak, and subsequently decays toward a statistically steady state. Moreover, simulations with higher initial Mach numbers $M_{m}$ exhibit both higher peak values and higher late-time steady-state levels. These trends are in qualitative agreement with the DNS results shown in figure~\ref{fig:inter_mach_dilatational}(a).

 \begin{figure}
\centering
  \begin{tabular}[b]{c}
\hspace{-0.3in}    
\includegraphics[width=.5\linewidth]{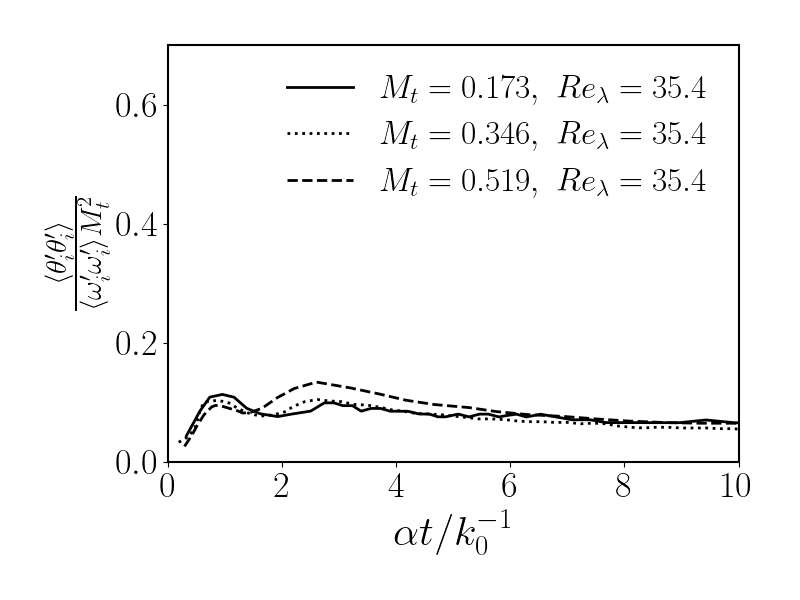} \\
    \small (a)
  \end{tabular} \qquad
  \begin{tabular}[b]{c}
  \hspace{-0.4in}
\includegraphics[width=.5\linewidth]{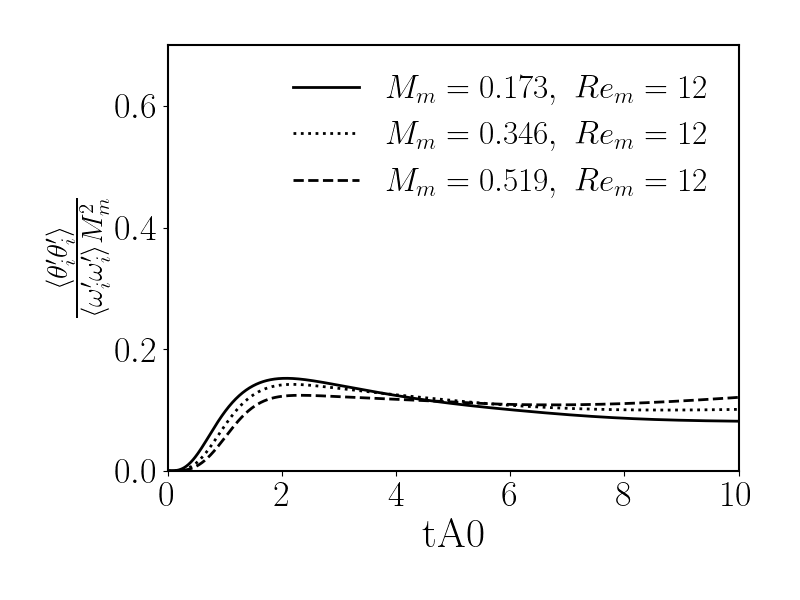}\\
    \small (b)
  \end{tabular}
  \begin{tabular}[b]{c}
 \hspace{-0.4in}
\includegraphics[width=.5\linewidth]{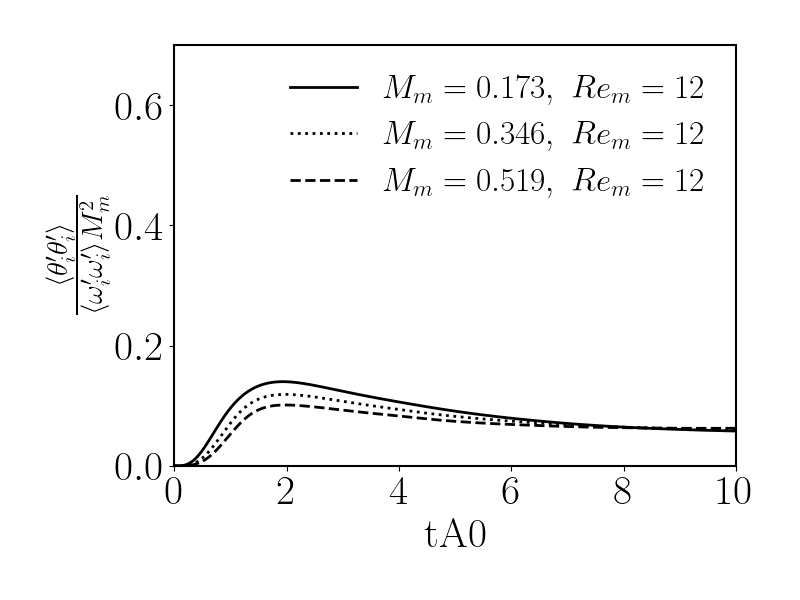}\\
    \small (c)
  \end{tabular}
  \caption{\label{fig:inter_mach_dilatational_norm} Effects of Mach number on the evolution of the ratio of dilatation to solenoidal dissipation normalized with the initial value of Mach number $\left(\frac{\langle \theta_{i}^{'}
\theta_{i}^{'}\rangle} {\langle\omega_{i}^{'}
\omega_{i}^{'}\rangle M_{m}^{2}}\right )$ using datasets from: (a) DNS \citep{lee1991eddy}, (b) the H-EHEE model, (c) the N-EHEE model \citep{danish2014direct}.}
  \end{figure}

Finally, figure~\ref{fig:inter_mach_dilatational_norm} examines the scaling of the ratio of the rms dilatation rate to rms vorticity normalized with the initial Mach numbers. After normalization, results from different Mach-number simulations collapse onto a single curve, indicating an approximate $M_{t}^{2}$ scaling of the dilatational dissipation. Figures~\ref{fig:inter_mach_dilatational_norm}(b) and \ref{fig:inter_mach_dilatational_norm}(c) show the corresponding results obtained using the H-EHEE and N-EHEE models. Similar to the DNS, the normalized curves collapse to a narrow band when scaled with $M_m^{2}$, demonstrating that both models does capture the essential compressibility scaling. However, at later times, the collapse of the H-EHEE model is somewhat inferior to that of the N-EHEE model. This aspect will be examined further in our future work.

\subsubsection{Fluid flows with vibrational non-equilibrium effects} 
Next, we evaluate the H-EHEE model based on its predictions of vibrational non-equilibrium effects. Equation~(\ref{eq:model_vibnon}) represents the closure for vibrational non-equilibrium in velocity-gradient dynamics. The corresponding dynamical closure incorporates predictions from the IBVM model, as discussed in section~\ref{V}. The IBVM model was previously evaluated at single time instants by \cite{shikha2024modeling}, using input features extracted from DNS datasets of isotropic decaying compressible turbulence with vibrational non-equilibrium effects \citep{srivastava2024influence}. These datasets are characterized by four non-dimensional parameters: (i) the turbulent Mach number $M_t$ (\ref{eq:mach_number}), (ii) the Taylor Reynolds number $Re_{\lambda}$ (\ref{eq:reynolds_number}), (iii) the vibrational Damk\"ohler number $D$ (\ref{eq:damkohler_number}), and (iv) the initial ratio of the mean vibrational temperature to the mean translational temperature, denoted by $\xi$ (\ref{eq:xi}). 

An \textit{a priori} evaluation of the IBVM model was conducted by \cite{shikha2024modeling} using DNS datasets spanning a range of initial values of $D$, $\xi$, $Re_{\lambda}$, and $M_t$, demonstrating satisfactory performance at a single time instant. In the present work, we extend this analysis by evaluating the IBVM predictions in a fully dynamical equations set. We compare the Hybrid -EHEE model results with DNS results from simulation cases of \cite{srivastava2024influence} initialized with $M_t = 0.5$, $Re_{\lambda} = 60$, $\xi = 0.5$, and $D=[0.25-4]$. Note that, since the N-EHEE model does not account for the effects of vibrational non-equilibrium, it does not qualify for this evaluation.

\cite{srivastava2024influence} have investigated the impact of vibrational non-equilibrium on the key statistical quantities, namely: (i) the solenoidal dissipation rate: $\omega '^{2}$ and (ii) the dilatational dissipation rate: $\theta'^{2}$, as defined in (\ref{eq:theta_and_omega_prime}). Their study examined the influence of varying $D$ on the temporal evolution of $\omega'$ and $\theta'$.  We evaluate the H-EHEE model based on its capabilities to reproduce DNS trends across different initial model Damk\"ohler numbers. Because an explicit quantitative relationship between the DNS-based Damk\"ohler number ($D$) and the model parameter ($D_m$) cannot be established, we select values of $D_m$ such that the trend of the influence of varying $D_m$ is clearly observable. The H-EHEE model simulations are performed for multiple initial model Damk\"ohler numbers, as summarized in Table~\ref{tab:vib_non_eqlb_cases}.
\begin{table}
    \centering
    \begin{tabular}{c c c c c}
        \hline
        Simulation & $M_m$ & $Re_m$ & $D_m$ & ${\xi_m}_{0}$ \\ \hline
        R1    & 0.5 & 60 & 2 & 0.5 \\
        R2    & 0.5 & 60 & 4 & 0.5 \\
        R3    & 0.5 & 60 & 8 & 0.5 \\
        novib & 0.5 & 60 & -- & 1   \\
        \hline
    \end{tabular}
    \caption{The H-EHEE model simulation cases used to evaluate vibrational non-equilibrium effects.}
    \label{tab:vib_non_eqlb_cases}
\end{table}
For the \textit{novib} case, the setting ${\xi_m}_{0} = 1$ effectively disables the vibrational non-equilibrium mechanism using multiplier as $\left[(1-{\xi_m}_{0})/(1+{\xi_m})\right]$ in (\ref{eq:model_vibnon}), thereby reducing the model to the vibrational equilibrium limit.
\begin{figure}[h!]
\centering
\begin{tabular}[b]{c}
\includegraphics[width=0.5\linewidth]{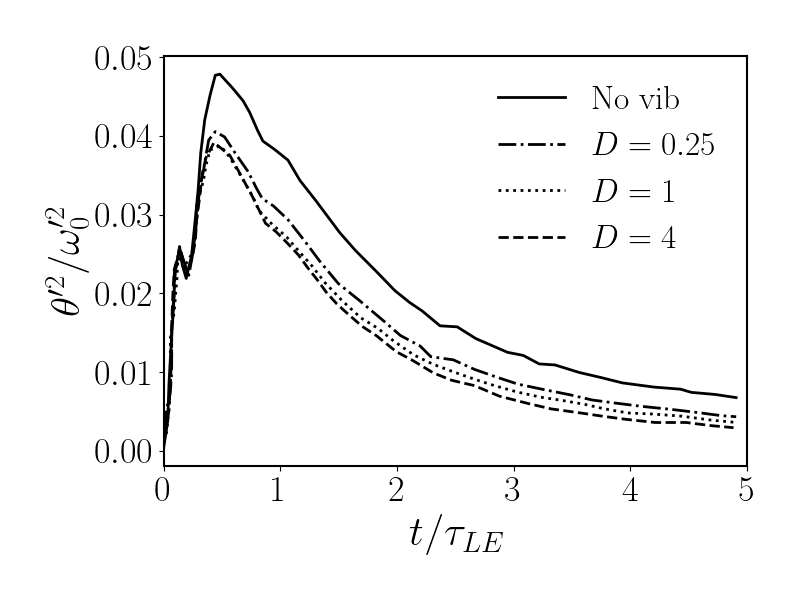} \\
(a)
\end{tabular} \quad
\hspace{-0.4in}
\begin{tabular}[b]{c}
\includegraphics[width=0.5\linewidth]{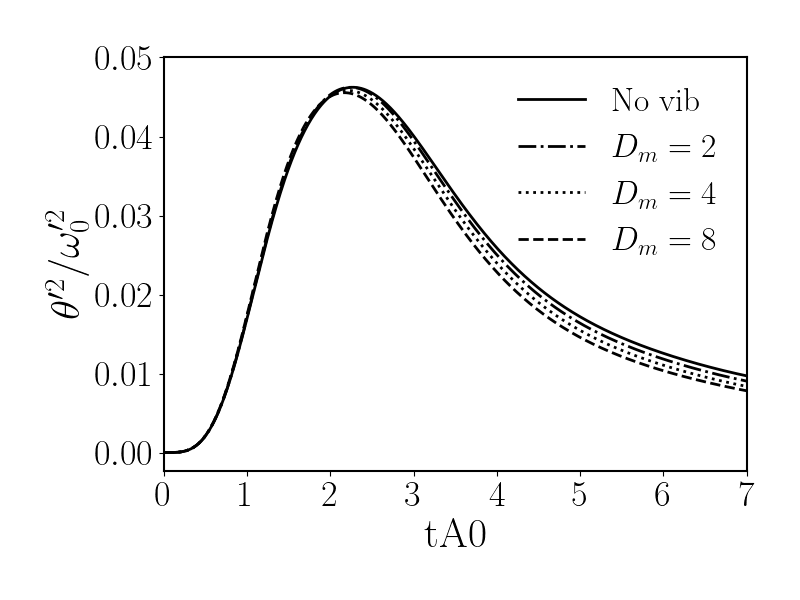} \\
(b)
\end{tabular}
\caption{Effects of initial Damk\"ohler number on the temporal evolution of ${\theta'}^2/{\omega'_0}^2$ using: (a) DNS datasets \citep{srivastava2024influence}, (b) the H-EHEE model.}
\label{fig:D_theta_prime}
\end{figure}

\begin{figure}
\centering
\begin{tabular}[b]{c}
\includegraphics[width=0.5\linewidth]{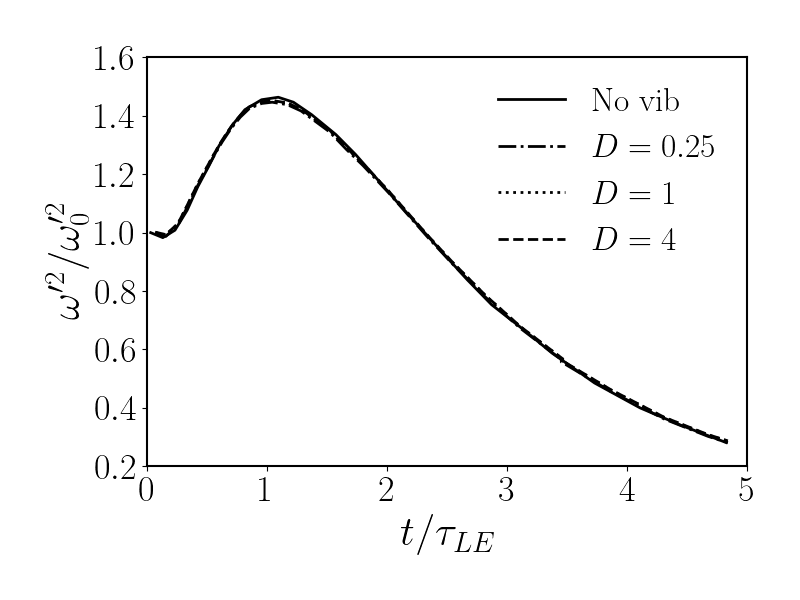} \\
(a)
\end{tabular} \quad
\hspace{-0.4in}
\begin{tabular}[b]{c}
\includegraphics[width=0.5\linewidth]{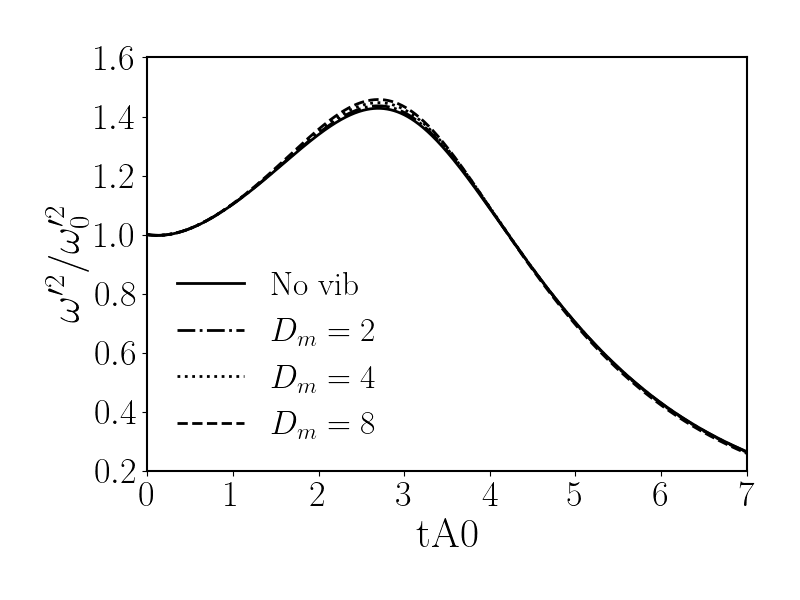} \\
(b)
\end{tabular}
\caption{Effects of initial Damk\"ohler number on the temporal evolution of ${\omega'}^2/{\omega'_0}^2$ using: (a) DNS datasets \citep{srivastava2024influence}, (b) the H-EHEE model.}
\label{fig:D_omega_prime}
\end{figure}
Figures~\ref{fig:D_theta_prime}(a)–(b) compare the temporal evolution of the normalized dilatation dissipation rate ${\theta'}^{2}/{\omega'_0}^{2}$ obtained from DNS datasets \citep{srivastava2024influence} and the H-EHEE model for different initial Damk\"ohler numbers. Similarly, Figures~\ref{fig:D_omega_prime}(a)–(b) present the corresponding evolution of the normalized solenoidal dissipation rate ${\omega'}^{2}/{\omega'_0}^{2}$. The DNS results show that increasing $D$ leads to a systematic reduction in ${\theta'}^{2}$, while ${\omega'}^{2}$ remains largely insensitive to variations in $D$. The H-EHEE model does qualitatively reproduce the observed trends in the DNS.

Based on the results of this subsection, we summarize that, the H-EHEE model accurately captures the influence of vibrational non-equilibrium effects on the temporal evolution of both solenoidal and dilatational fluctuations. The close qualitative agreement with DNS results demonstrates the capability of the proposed framework to represent vibrational non-equilibrium physics in compressible turbulence.


\subsection{Higher Mach number
\texorpdfstring{($M_t=1.2$)}{(Mt=1.2)}:
based on \texorpdfstring{$V_{ij}$}{Vij} mechanism predictions}

In higher Mach number flows ($M_{t}=1.2$), we specifically assess the H-EHEE model performance by examining the geometric characteristics of the $P_{ij}$ tensor. In the low-Mach-number limit, the $P_{ij}$ tensor behaves predominantly as a symmetric tensor. In contrast, at higher turbulent Mach numbers, a significant antisymmetric component of the $P_{ij}$ tensor is generated. The symmetric and antisymmetric parts of the $P_{ij}$ tensor are defined as
\begin{equation}
\label{eq:PH_s_as}
   P_{ij}^{s}=\frac{1}{2}\left(P_{ij}+P_{ji}\right), 
   \qquad 
   P_{ij}^{as}=\frac{1}{2}\left(P_{ij}-P_{ji}\right).
\end{equation}
In strictly incompressible or low-Mach-number turbulence, the $P_{ij}$ tensor remains symmetric at all times, implying that $P_{ij}^{as}\equiv 0$. However, at elevated Mach numbers, $P_{ij}^{as}$ become non-zero. When the initial $P_{ij}$ tensor field is symmetric, any antisymmetric contribution appearing at later times must arise from the mechanisms which are responsible for the generation and evolution of the $B_{ij}$ tensor, specifically the group $V_{ij}$ mechanism (\ref{eq:Pressure_Hessian}). In the present framework, the $V_{ij}$ mechanism is represented using the IBBM closure (\ref{eq:ibnm_varsigma}). Since all simulations are initialized with a $P_{ij}=H_{ij}$ condition, which is symmetric in nature, the emergence of antisymmetric components at later times can be directly attributed to the IBBM predictions. Consequently, a comparison of the $P_{ij}^{as}$ obtained from the H-EHEE model provides a direct assessment of the proposed IBBM based closure.
To analyze the geometry of the $P_{ij}$ tensor, we consider the normalized symmetric and antisymmetric components defined as
\begin{equation}
\label{eq:Pij_components}
    \Lambda_{ij}=\frac{P_{ij}^{s}}{\sqrt{P_{mn}P_{mn}}}, 
    \qquad 
    \Theta_{ij}=\frac{P_{ij}^{as}}{\sqrt{P_{mn}P_{mn}}}.
\end{equation}
By construction, all components of $\Lambda_{ij}$ and $\Theta_{ij}$ are bounded between $-1$ and $1$.

\begin{figure}
\centering
  \begin{tabular}[b]{c}
\hspace{-0.3in}    
\includegraphics[width=.5\linewidth]{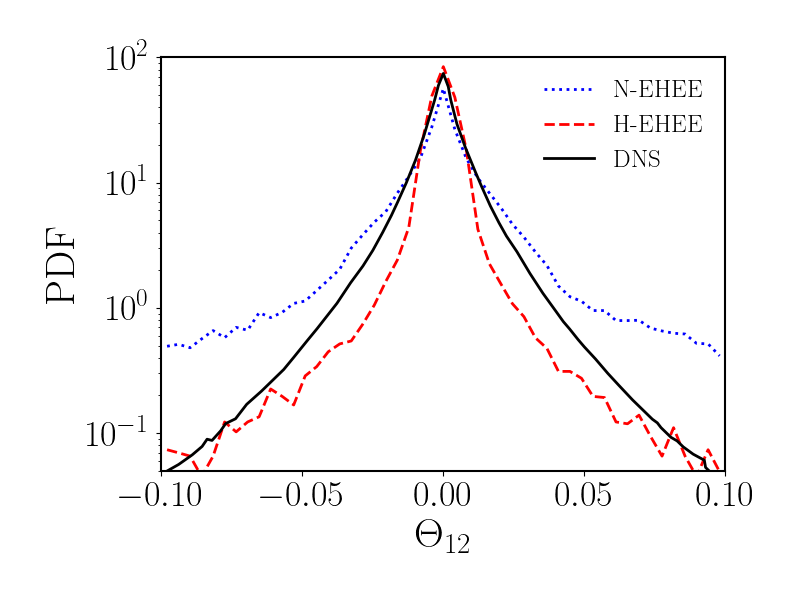} \\
    \small (a)
  \end{tabular} \qquad
  \begin{tabular}[b]{c}
  \hspace{-0.4in}
\includegraphics[width=.5\linewidth]{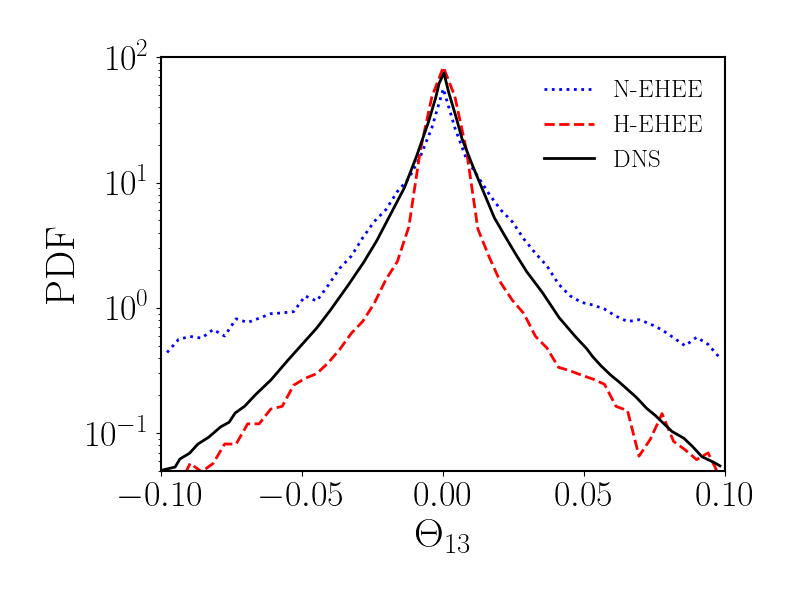}\\
    \small (b)
  \end{tabular}
  \begin{tabular}[b]{c}
 \hspace{-0.4in}
\includegraphics[width=.5\linewidth]{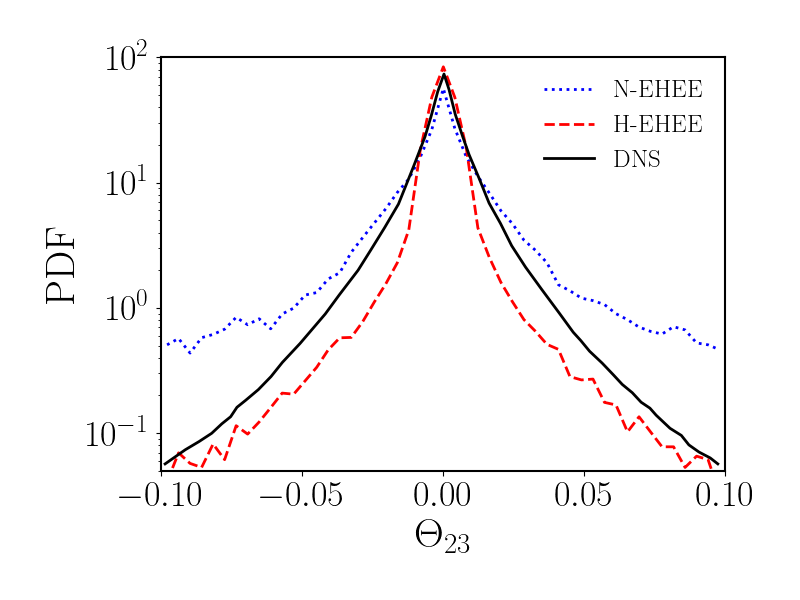}\\
    \small (c)
  \end{tabular}
  \caption{\label{fig:Theta_ij_pdf}PDFs of normalized components of anti-symmetric part ($\Theta_{ij}$) of pressure Hessian ($P_{ij}$) tensor using the DNS dataset of initial $M_{t}=1.2$ \citep{kumar2013weno}, the H-EHEE model, and the N-EHEE model \citep{danish2014direct}.}
  \end{figure}

Next, we evaluate the H-EHEE model based on its predictions of the PDFs of $\Lambda_{ij}$ and $\Theta_{ij}$ (\ref{eq:Pij_components}). Figure~\ref{fig:Theta_ij_pdf} presents the PDFs of the non-zero components of $\Theta_{ij}$.
The DNS results presented in this subsection (with initial parameters $M_t=1.2$ and $Re_{\lambda}=55.6$) are taken using the dataset of \cite{kumar2013weno}. In contrast, the H-EHEE and the N-EHEE model results are obtained at $M_m=1.2$ and $Re_m=12$. The DNS results show that all three non-zero components of $\Theta_{ij}$ exhibit a single peak at zero, followed by a monotonic decay away from zero. This qualitative behavior is captured by both the H-EHEE and the N-EHEE models. However, the H-EHEE model reproduces the overall PDF distributions much more closely to the DNS data, especially at higher non-zero values of $\Theta_{ij}$. In contrast, the N-EHEE model systematically overpredicts the PDF values in regions corresponding to high non-zero values of $\Theta_{ij}$ components.

Despite the improvements demonstrated by the H-EHEE model relative to the N-EHEE model, certain discrepancies with respect to the DNS results still remain. These discrepancies may be attributed to several factors. First, the present work primarily focuses on improving the modeling of $V_{ij}$ (\ref{eq:Pressure_Hessian}). However, additional mechanisms contributing to the evolution of the $\boldsymbol{B}$ tensor remain unclosed. In particular, viscous-heating-related contributions, such as $VII_{ij}$ (\ref{eq:Pressure_Hessian}), are neglected in the current formulation. The influence of these mechanisms is expected to become more important at higher Mach numbers, where stronger thermodynamic gradients are present. Second, the neural-network closure is trained using DNS data corresponding to an initial turbulent Mach number of $M_t=1.0$. Although the model exhibits good generalization across the available datasets spanning $0.8 \leq M_t \leq 1.0$ (see Section~\ref{subsec:a-priori}), the flow physics become progressively more complex as the Mach number increases. Consequently, the predictions obtained for the higher-Mach-number condition considered in this study involve a degree of extrapolation beyond the parameter range represented in the training data.

\begin{figure}
\centering
  \begin{tabular}[b]{c}
\hspace{-0.2in}    
\includegraphics[width=.4\linewidth]{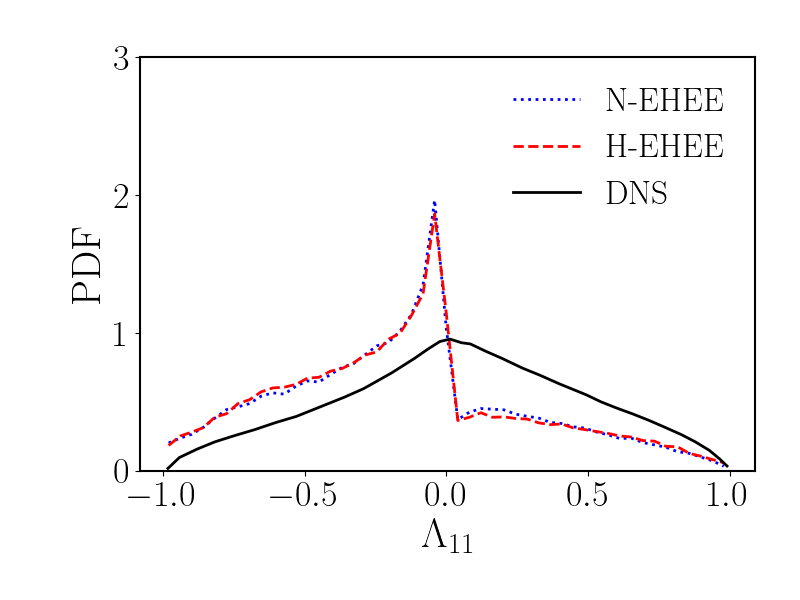} \\
    \small (a)
  \end{tabular} \qquad
  \begin{tabular}[b]{c}
  \hspace{-0.2in}
\includegraphics[width=.4\linewidth]{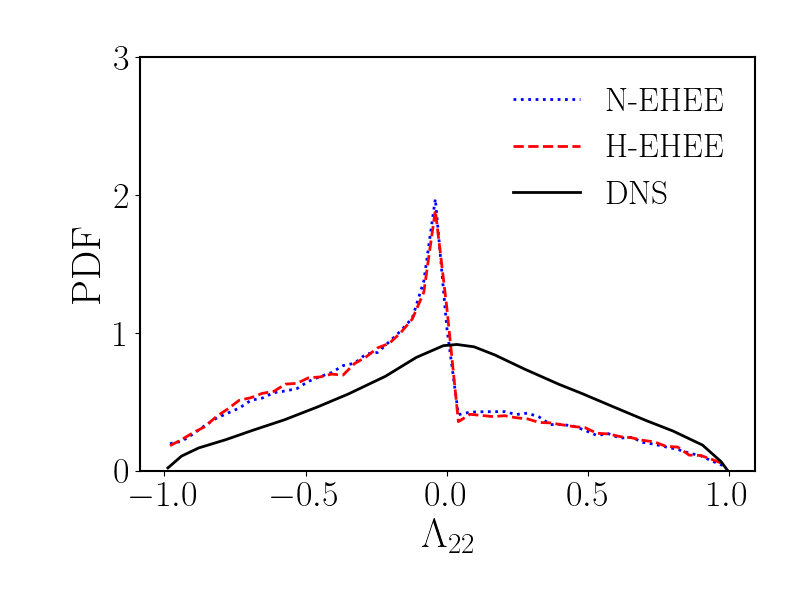}\\
    \small (b)
  \end{tabular}
  \begin{tabular}[b]{c}
 \hspace{-0.2in}
\includegraphics[width=.4\linewidth]{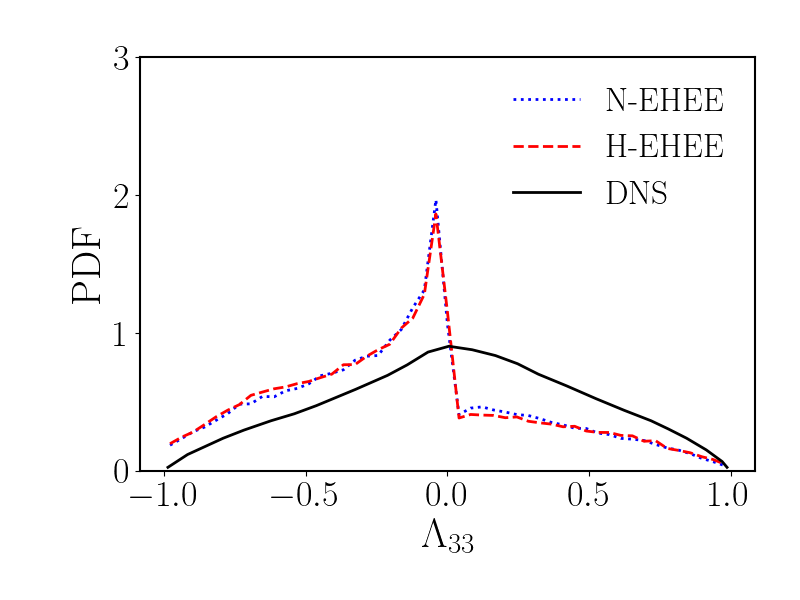}\\
    \small (c)
  \end{tabular}
   \begin{tabular}[b]{c}
\hspace{-0.2in}    
\includegraphics[width=.4\linewidth]{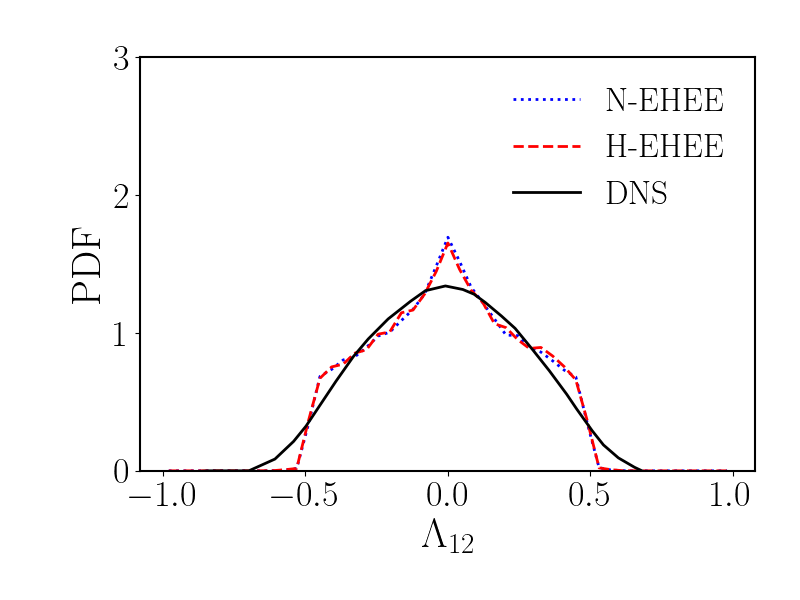} \\
    \small (d)
  \end{tabular} \qquad
  \begin{tabular}[b]{c}
  \hspace{-0.2in}
\includegraphics[width=.4\linewidth]{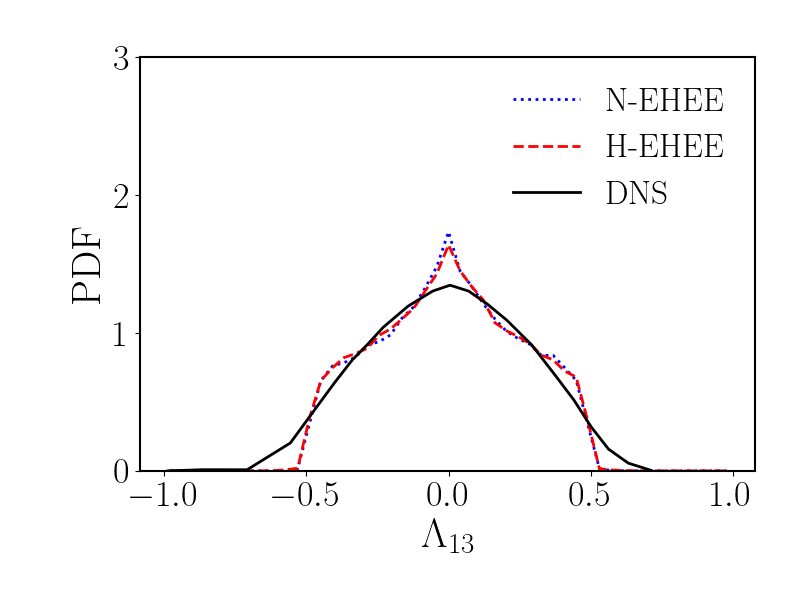}\\
    \small (e)
  \end{tabular}
  \begin{tabular}[b]{c}
 \hspace{-0.2in}
\includegraphics[width=.4\linewidth]{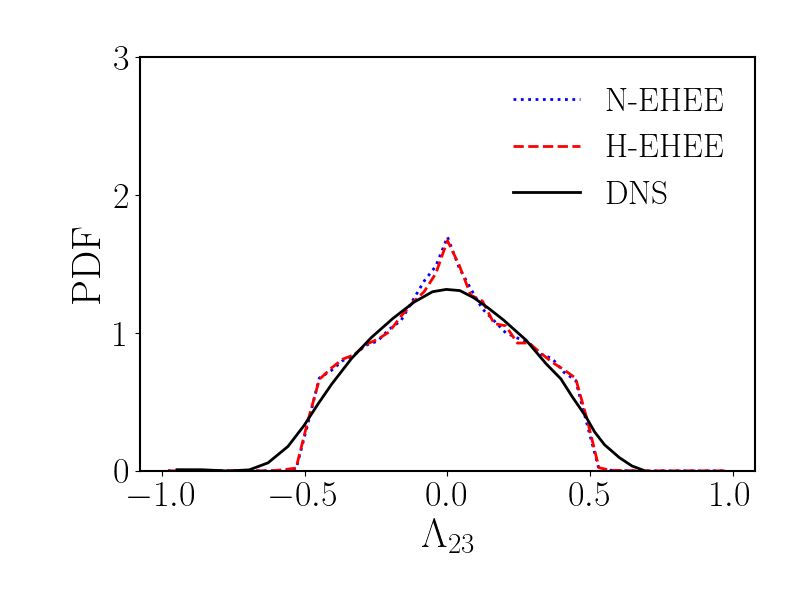}\\
    \small (f)
  \end{tabular}
  \caption{\label{fig:symmetric_Pij}PDFs of normalized components of symmetric part ($\Lambda_{ij}$) of pressure Hessian ($P_{ij}$) tensor using the DNS dataset of initial $M_{t}=1.2$ \citep{kumar2013weno}, the H-EHEE model, and the N-EHEE model \citep{danish2014direct}.}
  \end{figure}

Next, Figures~\ref{fig:symmetric_Pij}(a)--\ref{fig:symmetric_Pij}(f) show the PDFs of $\Lambda_{ij}$, obtained from DNS, the N-EHEE model, and the H-EHEE model results.
The DNS-based PDFs exhibit single-peaked distributions centered at zero for all components. Notably, the diagonal components (Figures~\ref{fig:symmetric_Pij}(a)--\ref{fig:symmetric_Pij}(c)) display broader distributions than the off-diagonal components (Figures~\ref{fig:symmetric_Pij}(d)--\ref{fig:symmetric_Pij}(f)). The H-EHEE model reproduces this qualitative behavior reasonably well. The agreement with DNS is particularly good for the off-diagonal components. For the diagonal components, the H-EHEE model slightly overpredicts the probability on the negative side and underpredicts it on the positive side. 

It is noteworthy that the predictions of the H-EHEE and N-EHEE models are very similar for these quantities. It is plausible to attribute this similarity to  the fact  the closure for the $\boldsymbol{H}$ tensor employed in the present work is taken directly from the N-EHEE model (as discussed in Section~\ref{IV}).  The limitations associated with the existing $\boldsymbol{H}$ tensor closure seems to be retained in the H-EHEE framework. 

Further, the symmetric part of the $\boldsymbol{P}$ tensor can be alternatively defined as:
\begin{equation}
\label{eq:split_sym}
P_{ij}^{s}
=
H_{ij}
-\frac{1}{2}\left(B_{ij}+B_{ji}\right).
\end{equation}
Equation \ref{eq:split_sym}  shows that $\Lambda_{ij}$ (\ref{eq:Pij_components}) receives contributions from both the $H_{ij}$ tensor and the symmetric part of the $B_{ij}$ tensor. Therefore, the discrepancies observed in Figure~\ref{fig:symmetric_Pij} may originate from inaccuracies in the prediction of either $H_{ij}$ or $B_{ij}$. Although the improved modeling of $B_{ij}$ in the H-EHEE framework leads to better agreement with DNS than the N-EHEE model, the remaining discrepancies indicate that further improvements in the prediction of the $\Lambda_{ij}$ (\ref{eq:Pij_components}) statistics will likely require more accurate closures for both the $H_{ij}$ and $B_{ij}$ tensors, particularly in highly compressible flow regimes.

Overall, at the higher turbulent Mach number $M_t = 1.2$, the H-EHEE model provides a consistent and physically meaningful representation of both the $\boldsymbol{H}$ and $\boldsymbol{B}$ tensors. In particular, the inclusion of the IBBM closure yields predictions of the $P^{as}_{ij}$ (\ref{eq:PH_s_as}) PDFs that are in better agreement with DNS compared to the N-EHEE model (especially at higher non-zero values of $\Theta_{ij}$), while simultaneously preserving the predictive performance of the original N-EHEE formulation for the $P^{s}_{ij}$ (\ref{eq:PH_s_as}) components.

\section{\label{X}Conclusions}
The evolution of the velocity gradient tensor is fundamental to understanding the dynamics of compressible turbulent flows. The primary objective of this work is to develop a dynamical modeling framework for the evolution of velocity gradients in compressible turbulence. Such a framework provides a natural basis for statistical and stochastic closure modeling of Lagrangian probability density functions methods \citep{girimaji1990material}.


In compressible turbulence, the evolution of the pressure fields on the velocity-gradient dynamics is manifested through two unclosed contributions: the pressure-Hessian tensor, which is inherently symmetric in nature, and the tensor arising from the interaction between the density-gradient and pressure-gradient fields, referred to as the baroclinic tensor. In this work, the the combined effect of these two contributions is called as the \emph{thermodynamic gradient field} (TGF) tensor. 
The baroclinic tensor effects become particularly important in shock-dominated regions. 
Accurately modeling their effects is challenging due to their inherently asymmetric nature, which necessitates a careful construction of tensor bases and invariant representations for a general second-order tensor which is not symmetric in nature. The modeling complexity further increases under vibrational non-equilibrium conditions, which commonly occur in hypersonic flows ($T \geq 800 K$) when the fluid-dynamic timescale becomes comparable to the vibrational energy relaxation timescale.

In this study, we build on the new enhanced homogenized Euler equation (N-EHEE) model \citep{danish2014direct} by incorporating a vibrational non-equilibrium mechanism. The key modifications and contributions of the proposed modeling framework are summarized as follows:
\begin{itemize}
    \item A tensor-basis- and invariants-based neural-network closure is developed for the the inviscid mechanism contributing in the baroclinic tensor generation, overcoming the limitations of the existing phenomenological model of \cite{danish2014direct}.

    \item The newly developed neural-network-based closures for the inviscid baroclinic tensor generating mechanism and the vibrational non-equilibrium mechanisms are incorporated into the dynamical velocity-gradient framework.

    \item A novel \textit{hybrid} dynamical velocity-gradient model is formulated as a closed set of ordinary differential equations, in which unclosed terms are represented using a combination of phenomenological modeling and neural-network-based closures. This framework is hereafter referred to as the \textit{hybrid enhanced homogenized Euler equation} (H-EHEE) model.

    \item The proposed framework is evaluated across a wide range of flow regimes, spanning low- to high-turbulent Mach number conditions, including regimes with active vibrational non-equilibrium effects, through systematic comparisons with direct numerical simulation data and existing models of compressible velocity gradient dynamics.
\end{itemize}

The model is assessed through a two-stage evaluation procedure. First, an \emph{a priori} assessment is performed for the newly trained neural network model, which represents the inviscid baroclinic-tensor generating mechanism. In this stage, inputs are obtained directly from DNS datasets, and the neural network predictions are compared against the corresponding statistics observed in DNS datasets. The predictions show excellent agreement with DNS trends. In the second stage, the full dynamical system is evaluated, wherein all inputs are generated by the model ODEs. The H-EHEE predictions are examined over a wide range of Mach numbers and are compared with DNS data as well as with the existing N-EHEE model \citep{danish2014direct}.

The principal conclusions drawn from the evaluation of the proposed model are summarized as follows:
\begin{enumerate}
    \item In the incompressible limit, the model reproduces the key features of the probability density functions (PDFs) of strain-rate eigenvalues, strain–vorticity alignment statistics, and the joint PDFs of the second and third invariants of the velocity-gradient tensor, showing agreement with DNS results comparable to that of the N-EHEE model.
    
    \item At intermediate Mach numbers, the model captures the DNS trends of the Mach-number dependence on both solenoidal and dilatational dissipation.
    
    \item In vibrational non-equilibrium flows, the model successfully reproduces the influence of varying initial Damk\"ohler numbers on the temporal evolution of solenoidal and dilatational dissipation components.
    
    \item At higher turbulent Mach numbers, the model provides significantly improved predictions of the PDFs of the antisymmetric components of the TGF tensor compared to the N-EHEE model. This improvement arises from the neural-network-based closure introduced for the inviscid baroclinic tensor generating mechanism, while the predictions for the symmetric TGF components remain comparable to those obtained with the earlier existing model.
\end{enumerate}

Overall, this work presents a comprehensive hybrid framework that combines phenomenological modeling with physics-assisted, data-driven closures for the velocity-gradient dynamics in compressible turbulence. The proposed H-EHEE model consistently reproduces a wide range of turbulence statistics across all flow regimes considered, including flows exhibiting vibrational non-equilibrium effects. Notably, the model demonstrates substantial improvements over the existing models in high-turbulent Mach number flows.

\section{ACKNOWLEDGMENT}
This work has been supported by the Science and Engineering Research Board (SERB), Department of Science and Technology, Government of India, through Project No. CRG/2022/002378. The high-performance computing resource required for this work has been made available by the High-Performance Computing Center (HPC), Indian Institute of Technology Delhi, India.\\

\section{Declaration of interests}
The authors report no conflict of interests.

\bibliographystyle{plainnat}  
\bibliography{references}

\end{document}